\documentclass[12pt]{article}
\pdfoutput=1
\usepackage{epsf,amsmath,amssymb,graphicx,dcolumn}
\usepackage{scalefnt,cite,hyperref}
\usepackage{cleveref,etoolbox,color,soul}
\usepackage{listings}
\usepackage[utf8]{inputenc}
\usepackage{booktabs, colortbl}
\usepackage{multirow}

\Crefname{figure}{Fig.}{Figs.}
\usepackage{placeins}
\usepackage{soul}

\graphicspath{{figs/}}

\newcommand{\citere}[1]{Ref.\,\cite{#1}}

\newcommand{\code}{\tt}

\newcommand{\HS}{{\code HiggsSignals}}
\newcommand{\how}[1]{(#1)}
\newcommand{\abbrev}{\scalefont{1}}

\newcommand{\fig}[1]{Fig.\,\ref{#1}}

\newcommand{\sct}[1]{Section~\ref{#1}}

\newcommand{\cp}{{\abbrev \ensuremath{\mathcal{CP}}}}

\newcommand{\GeV}{{\rm GeV}}
\newcommand{\TeV}{{\rm TeV}}
\newcommand{\ifb}{{\rm fb}^{-1}}

\definecolor{Darkgreen}{rgb}{0.,.8,0.25}

\newcommand{\mA}{\ensuremath{M_A}}

\newcommand{\tb}{\ensuremath{\tan\beta}}

\definecolor{darkgreen}{rgb}{0,0.5,0.1}

\newcounter{notecount}
\begin{document}
\thispagestyle{empty}
\begin{flushleft}
\end{flushleft}

\begin{flushright}
{\tt BONN-TH 2019-04},
{\tt DESY 19-093},
{\tt KA-TP-09-2019},
{\tt IFT-UAM/CSIC-19-075}
\end{flushright}

\long\def\symbolfootnote[#1]#2{\begingroup%
\def\thefootnote{\fnsymbol{footnote}}\footnote[#1]{#2}\endgroup}

\vspace{0.1cm}

\begin{center}
\Large\bf\boldmath
HL-LHC and ILC sensitivities\\[.5em] in the hunt for heavy Higgs bosons
\unboldmath
\end{center}
\vspace{0.05cm}
\begin{center}
Henning Bahl$^a$,
Philip Bechtle$^{b}$,
Sven Heinemeyer$^{c,d,e}$,\\[0.2cm]
Stefan Liebler$^f$,
Tim Stefaniak$^{a}$,
Georg Weiglein$^a$\\[0.4cm]
{\small
{\sl${}^a$DESY, Notkestra{\ss}e 85, D-22607 Hamburg, Germany}\\[0.2em]
{\sl${}^b$Physikalisches Institut der Universit\"at Bonn, Nu{\ss}allee 12, D-53115 Bonn, Germany}\\[0.2em]
{\sl${}^c$Instituto de F\'isica Te\'orica, (UAM/CSIC), Universidad Aut\'onoma de Madrid,\\
Cantoblanco, E-28049 Madrid, Spain}\\[0.2em]
{\sl${}^d$Campus of International Excellence UAM+CSIC, Cantoblanco, E-28049, Madrid, Spain}\\[0.2em]
{\sl${}^e$Instituto de F\'isica de Cantabria (CSIC-UC), E-39005 Santander, Spain}\\[0.2em]
{\sl${}^f$Institute for Theoretical Physics (ITP), Karlsruhe Institute of Technology,\\ D-76131 Karlsruhe, Germany}\\[0.2em]
}
\end{center}
\vspace*{1mm}
\begin{abstract}
\noindent

The prediction of additional Higgs bosons is one of the key features of
physics beyond the Standard Model (SM) that gives rise to an
extended Higgs sector.
We assess the sensitivity of the Large Hadron
Collider (LHC) in the high luminosity (HL) run alone and in combination
with a possible future International Linear Collider (ILC) to probe
heavy neutral Higgs bosons.  We employ the Minimal Supersymmetric
Standard Model (MSSM) as a framework and assume the light \cp-even
MSSM Higgs boson to be the Higgs boson observed at $125\,\GeV$.
We discuss the constraints on the MSSM parameter space arising from
the precision
measurements of the rates of the detected signal at $125\,\GeV$ and from
direct searches for new heavy Higgs bosons
in the $\tau^+\tau^-$, $b\bar{b}$ and di-Higgs ($hh$) final states.
A new benchmark scenario for heavy Higgs searches in the $b\bar{b}$ channel
is proposed in this context.
For the future Higgs rate measurements at the HL-LHC and ILC two different
scenarios are investigated, namely the case where the future rate
measurements agree with the SM prediction and the case where the rates agree
with the predictions of possible realizations of
the MSSM Higgs sector in nature.
\end{abstract}

\setcounter{footnote}{0}

\newpage

\pagenumbering{arabic}


\section{Introduction}
\label{sec:introduction}

The Large Hadron Collider (LHC) continues to measure the properties of the discovered Higgs
  boson~\cite{Aad:2012tfa,Chatrchyan:2012xdj} at $125\,\GeV$ with
  increasing precision. So far, given the current experimental and theoretical uncertainties,
the measurements are in good agreement with the SM predictions~\cite{Khachatryan:2016vau,ATLAS:2019slw, Sirunyan:2018sgc, Sirunyan:2018koj}.
Nevertheless, in certain parameter regions, models with extended Higgs sectors often feature a SM-like Higgs
boson that is compatible with the experimental data. Such models, often motivated by theoretical arguments
or observational puzzles (such as dark matter), are therefore equally viable in the light of the Higgs observation.

The MSSM~\cite{Nilles:1983ge,Haber:1984rc,Gunion:1984yn} is one of the best
studied models with an extended Higgs sector. In contrast to
the case of the SM, the MSSM contains two Higgs doublet fields.
This results in five physical Higgs bosons instead of the single Higgs
boson in the SM. These are (in the \cp-conserving case, which is
  assumed throughout this paper)
the light and heavy \cp-even Higgs bosons,
$h$ and $H$, the \cp-odd Higgs boson, $A$, and the charged
Higgs bosons, $H^\pm$.
At tree-level the Higgs sector is
determined by the ratio of the two vacuum expectation values, \tb,
and the mass of the \cp-odd Higgs boson, $M_A$.
In addition, the MSSM predicts two scalar
partners for all fermions as well as fermionic partners for all bosons,
which for the electroweak gauge bosons and Higgs bosons are the so-called neutralinos
and charginos, or electroweakinos.

In order to facilitate collider searches for the additional MSSM Higgs
bosons a set of new benchmark scenarios for MSSM Higgs boson searches
at the LHC has been proposed recently~\cite{Bahl:2018zmf,Bahl:2019ago}.
The scenarios are compatible --- at least over wide portions
of their parameter space --- with the most recent LHC results for the
properties of the Higgs boson at $125\,\GeV$~\cite{Khachatryan:2016vau,ATLAS:2019slw, Sirunyan:2018sgc, Sirunyan:2018koj} and the bounds on
masses and cross sections of new SUSY particles, as well as with
global fits of the phenomenological MSSM, see
e.g.~\cite{Buchmueller:2013rsa,Heinemeyer:2019vbc,Bagnaschi:2017tru}.
Each scenario contains one
\cp-even scalar with a mass around $125\,\GeV$ which is identified with
the observed Higgs boson.
Nevertheless, the scenarios differ significantly in their predictions for
the light Higgs-boson phenomenology (within the allowed uncertainties),
in particular in the non-decoupling regime, as well as in the
phenomenology of the additional, so far undetected Higgs bosons.
In many cases, such differences occur due to the presence of additional
light particles like, for example, the above-mentioned electroweakinos.

The search for the additional Higgs bosons will continue during LHC
Run~3 and subsequently at the high luminosity (HL)-LHC~\cite{Cepeda:2019klc}.
At possible future $e^+e^-$ colliders, such as the International Linear
Collider (ILC), the Compact Linear Collider (CLIC), the Future Circular
Collider (FCC)-ee or the Circular Electron Positron Collider (CEPC),
the search for extended Higgs sectors can be performed
either directly, via the search for new Higgs bosons, or indirectly, via
the precise measurements of the properties of the Higgs boson at
$125 \,\GeV$. The new benchmark scenarios, due to
their distinct phenomenology of the MSSM Higgs sector,
can be employed
to assess the reach of current and future colliders.

In this paper we present the HL-LHC and ILC%
\footnote{
We focus on the ILC as specific realization of an $e^+e^-$ Higgs factory
since it is the currently most advanced project and
anticipated precisions are based on full detector simulations.
}%
~sensitivities to heavy Higgs bosons in the $M_h^{125}$
scenario~\cite{Bahl:2018zmf},
in which all supersymmetric particles are heavy, leaving essentially a Two-Higgs-Doublet Model (THDM)-like Higgs sector
at the electroweak scale, and the $M_h^{125}(\tilde\chi)$ scenario~\cite{Bahl:2018zmf}, which features light neutralinos and charginos.
For the low $\tan\beta$ region we furthermore study the analogous scenarios $M_{h,\text{EFT}}^{125}$ and $M_{h,\text{EFT}}^{125}(\tilde\chi)$, respectively,
which employ a dedicated effective field theory (EFT) calculation and a varying SUSY mass scale~\cite{Bahl:2019ago}.
In addition, we define in this work a new benchmark scenario in which
the decays of the Higgs bosons to pairs of bottom quarks are
enhanced.
A selection of our results has already been
summarized in the report of the HL/HE-LHC Higgs working
group~\citere{Cepeda:2019klc}.

This paper is organized as follows: In \sct{sec:models} we briefly
review the MSSM benchmark models used in this study and introduce the
new benchmark scenario with enhanced Higgs-bottom-quark couplings.
In \sct{sec:expinput} we discuss the experimental input and the projections
both for the direct and indirect searches at the HL-LHC and two ILC
running scenarios. We present the exclusion reach of HL-LHC heavy Higgs
boson searches and the indirect sensitivity from HL-LHC Higgs rate
measurements in \sct{sec:reach}.
In \sct{sec:reachILC} we investigate constraints on BSM physics that can
be obtained from the HL-LHC results in combination with precision Higgs rate
measurements at the ILC. In this context we analyze both the case
where the future rate measurements agree with the SM prediction and the case
where the rates agree with the predictions of possible realizations of
the MSSM Higgs sector in nature.
We conclude in \sct{sec:conclusions}.


\section{MSSM benchmark models}
\label{sec:models}

We perform our study of the physics
potential of the HL-LHC and the ILC
in the exploration of heavy neutral Higgs bosons in the framework of
five MSSM benchmark scenarios. This allows us to compare the
sensitivity of direct searches for heavy neutral Higgs bosons with the
indirect sensitivity from precision measurements of the light Higgs
boson signal rates. The benchmark scenarios chosen here feature quite
different predictions for the considered search channels and for the
rates of the Higgs boson at $125~\GeV$, thus providing a certain variety in
the phenomenology of the two approaches and
for the different collider scenarios.

Four of the benchmark scenarios chosen here  were proposed in
Refs.~\cite{Bahl:2018zmf,Bahl:2019ago}. We shall furthermore define one
new scenario in this work in which the sensitivity of searches for
heavy Higgs boson decays to bottom quarks is enhanced (see below).
All considered scenarios are defined as two-dimensional planes
in $\mA$ and $\tb$, the two parameters that fix the MSSM Higgs-boson sector
at tree-level. We emphasize that the five benchmark scenarios considered
in this work feature the \emph{decoupling limit}~\cite{Gunion:2002zf},
i.e.\ the light \cp-even Higgs boson acquires SM-like tree-level
couplings if $\mA$ becomes large, $\mA \gg M_Z$. In scenarios where
loop corrections to the Higgs-boson mass matrix lead to an accidental
\emph{alignment without decoupling}~\cite{Gunion:2002zf,Carena:2013ooa,
Carena:2014nza,Bechtle:2016kui,Haber:2017erd},
lower bounds on $\mA$ from
the measurements of the rates of the Higgs boson at $125~\GeV$
become much weaker~\cite{Bahl:2018zmf,Bechtle:2016kui}.
While scenarios featuring at least one additional Higgs boson that is
lighter than the one at  $125~\GeV$ are tightly constrained within the
MSSM~\cite{Bahl:2018zmf} and not considered in the present paper, such a
situation can occur generically in singlet extensions of the Higgs sector,
see
e.g.~Refs.~\cite{Ellwanger:2009dp,Domingo:2015eea,Drechsel:2016jdg,Domingo:2018uim,Biekotter:2019kde,Robens:2015gla,Robens:2016xkb,Robens:2019kga}.

The theory predictions for the Higgs-boson masses, couplings and
branching ratios are obtained from \texttt{FeynHiggs} (see below for
more details on specific versions employed in this
work)~\cite{Heinemeyer:1998yj,Heinemeyer:1998np,Degrassi:2002fi,Frank:2006yh,Hahn:2013ria,Bahl:2016brp,Bahl:2017aev,Bahl:2018jom,Bahl:2018qog,Bahl:2019hmm}. For
Higgs production via
gluon fusion we make use of \texttt{SusHi} (version
1.7.0)~\cite{Harlander:2012pb, Harlander:2016hcx,
Harlander:2005rq,Harlander:2002wh,
Harlander:2002vv,Anastasiou:2014lda,Anastasiou:2015yha,
Anastasiou:2016cez,Degrassi:2010eu, Degrassi:2011vq,
Degrassi:2012vt,Actis:2008ug,Spira:1995rr}.
For
Higgs production via
bottom-quark annihilation ``matched predictions'' are employed~\cite{Bonvini:2015pxa,Bonvini:2016fgf,Forte:2015hba, Forte:2016sja}.


\subsection*{Standard scenarios: \boldmath{$M_h^{125}$} and
  \boldmath{$M_h^{125}(\tilde\chi)$ }}

The $M_h^{125}$~scenario~\cite{Bahl:2018zmf} is characterized by
relatively heavy superparticles,
such that the Higgs phenomenology at the LHC resembles that of a
THDM with MSSM-inspired Higgs couplings.
The light Higgs boson has SM-like couplings, and in the region
$\mA \lesssim 1.9 \,\TeV$ the heavy Higgs bosons can only
decay into SM particles (including the light SM-like Higgs boson).
For larger values of $\mA$
decays into electroweakinos (EWinos) are
kinematically open.
In contrast, the $M_h^{125}(\tilde{\chi})$~scenario~\cite{Bahl:2018zmf}
features light EWinos via the choice of relatively
small values of the bino, wino
and Higgsino mass parameters, $M_1$, $M_2$ and $\mu$, respectively.
This leads to sizable decay rates of the heavy Higgs bosons $H$
and $A$ into charginos and neutralinos throughout the parameter
plane, thus diminishing the event yield
of the $\tau^+\tau^-$ and $b\bar{b}$~final state signatures that are
used to search
for the additional Higgs bosons at the LHC. Furthermore, the branching
ratio of the light Higgs boson $h$ into a pair of photons is
enhanced for small values of $\tan\beta$ due to loop corrections
involving light charginos.

In these two scenarios all other SUSY masses are fixed around the TeV
scale. Consequently, in both scenarios, the light Higgs-boson mass becomes too
small (within experimental and theoretical uncertainties, as taken
over from Ref.~\cite{Bahl:2018zmf}, where a theory uncertainty of
$\pm 3\,\GeV$ was assumed~\cite{Degrassi:2002fi}),
$M_h \lesssim 122\,\GeV$, for $\tan\beta \lesssim 6$, rendering this
parameter region phenomenologically inconsistent with the observed
state at $125\,\GeV$.
For these two standard scenarios we used \texttt{FeynHiggs} version 2.14.3.
More details and a  discussion of the current constraints from
Higgs-boson searches and rate measurements in these two scenarios can be
found in Ref.~\cite{Bahl:2018zmf}.


\subsection*{Low-\boldmath{$\tan\beta$} (EFT) scenarios:
  \boldmath{$M_{h,\text{EFT}}^{125}$} and
  \boldmath{$M_{h,\text{EFT}}^{125}(\tilde\chi)$}}

The  $M_{h,\text{EFT}}^{125}$ and $M_{h,\text{EFT}}^{125}(\tilde\chi)$
scenarios~\cite{Bahl:2019ago} have similar phenomenological
properties as the standard scenarios described above, but are in
agreement with the observed Higgs-boson mass value
at lower values of \tb. A light
Higgs-boson mass $M_h \simeq 125\,\GeV$ is achieved by tuning the scalar top
masses to very
large values for every point in the parameter plane. Accordingly, for
these scenarios we employ
the THDM as effective field theory (EFT) below the SUSY scale to calculate the higher-order
corrections to the Higgs-boson masses and
self-energies~\cite{Bahl:2018jom}.
The $M_{h, {\rm EFT}}^{125}$~scenario resembles the $M_h^{125}$~scenario, while
the $M_{h, {\rm EFT}}^{125}(\tilde{\chi})$ scenario is phenomenologically similar
to the $M_h^{125}(\tilde{\chi})$~scenario, i.e.\ it features light
electroweakinos.
These scenarios are discussed in detail in Ref.~\cite{Bahl:2019ago}. For
these two EFT scenarios we used an extended version of \texttt{FeynHiggs}\,2.14.3, which
includes the THDM as EFT below the SUSY scale~\cite{Bahl:2018jom}. This feature was officially
released (alongside other changes) in version~2.16.0.


\subsection*{Large \boldmath{$H/A \to b\bar{b}$} scenario(s)}

In the MSSM the relation between the experimentally measured bottom-quark mass, $m_b$,
and the bottom-quark Yukawa coupling, $h_b$,
is affected by loop corrections which are enhanced for large $\tb$. These corrections are summarized in the parameter
$\Delta_b \propto \mu \tb$~\cite{Banks:1987iu,Hall:1993gn,Hempfling:1993kv,Carena:1994bv,Carena:1999py,Carena:2000uj}
according to
\begin{align}
\label{Deltab}
  h_b \sim \frac{m_b}{1 + \Delta_b}\,.
\end{align}
Consequently, large negative values of $\mu$ enhance the bottom-quark Yukawa
coupling and thus lead to enhanced rates of Higgs-boson decays to
bottom quarks as well as of Higgs boson production in association
with a bottom-quark pair. This enhancement is particularly strong for
the heavy non-SM Higgs bosons $H$ and $A$, while the SM-like light
Higgs boson $h$ is only affected to a much lesser extent, as these
corrections are
suppressed in the decoupling limit.%
\footnote{Corresponding $\Delta_\tau$ effects are substantially
smaller and not taken into account in a resummed way.}%
~The Higgsino mass
parameter $\mu$ can therefore via $\Delta_b$ lead
to a change in the relative
strengths of the Higgs boson couplings to bottom quarks and to
$\tau$-leptons. This, in turn, changes the relative sensitivity of
heavy Higgs boson searches in $b\bar{b}$ and $\tau^+\tau^-$ final
states.

In order
to investigate this change in sensitivity we define one new benchmark
scenario, which we denote by $M_h^{125}(\mu = -2\,\TeV)$ or the
short-hand notation $M_h^{125,\mu-}$.%
\footnote{This is similar to the change
in the $\mu$~values in the previously proposed
$m_h^{\rm mod+}$~scenario~\cite{Carena:2013ytb}, where
$\mu = \pm 200, \pm 500, \pm 1000~\GeV$ had been chosen.}%
~The SUSY input parameters are chosen to be
\begin{eqnarray}
&M_{Q_3}=M_{U_3}=M_{D_3}=1.5~\text{TeV},\quad
M_{L_3}=M_{E_3}=2~\text{TeV}, \nonumber\\[2mm]
&  \mu=-2~\text{TeV},\,\quad
M_1=1~\text{TeV},\quad M_2=1~\text{TeV},
\quad M_3=2.5~\text{TeV}, \nonumber\\[2mm]
&X_t=2.8~\text{TeV},\quad A_b=A_\tau=A_t\,.
\nonumber
\end{eqnarray}
All parameters are defined in the on-shell scheme. For the SM input parameters, we follow the recommendations of the LHC-HXSWG~\cite{deFlorian:2016spz}.

Apart from $\mu$, all parameters are chosen as in the $M_h^{125}$
scenario, in which $\mu = + 1~\text{TeV}$. In addition to this
default ``large $H/A\to b\bar{b}$ scenario'', we also investigate two
other choices of $\mu$ ($\mu = -1$ TeV and $\mu = -3$ TeV). For the
scenarios with negative $\mu$ we employ \texttt{FeynHiggs} version
2.14.4 which includes additional two-loop $\Delta_b$
corrections~\cite{Noth:2008tw,Noth:2010jy,Mihaila:2010mp} in the Higgs-boson decays.

\begin{figure}[tb!]
\begin{center}
\includegraphics[width=0.5\textwidth]{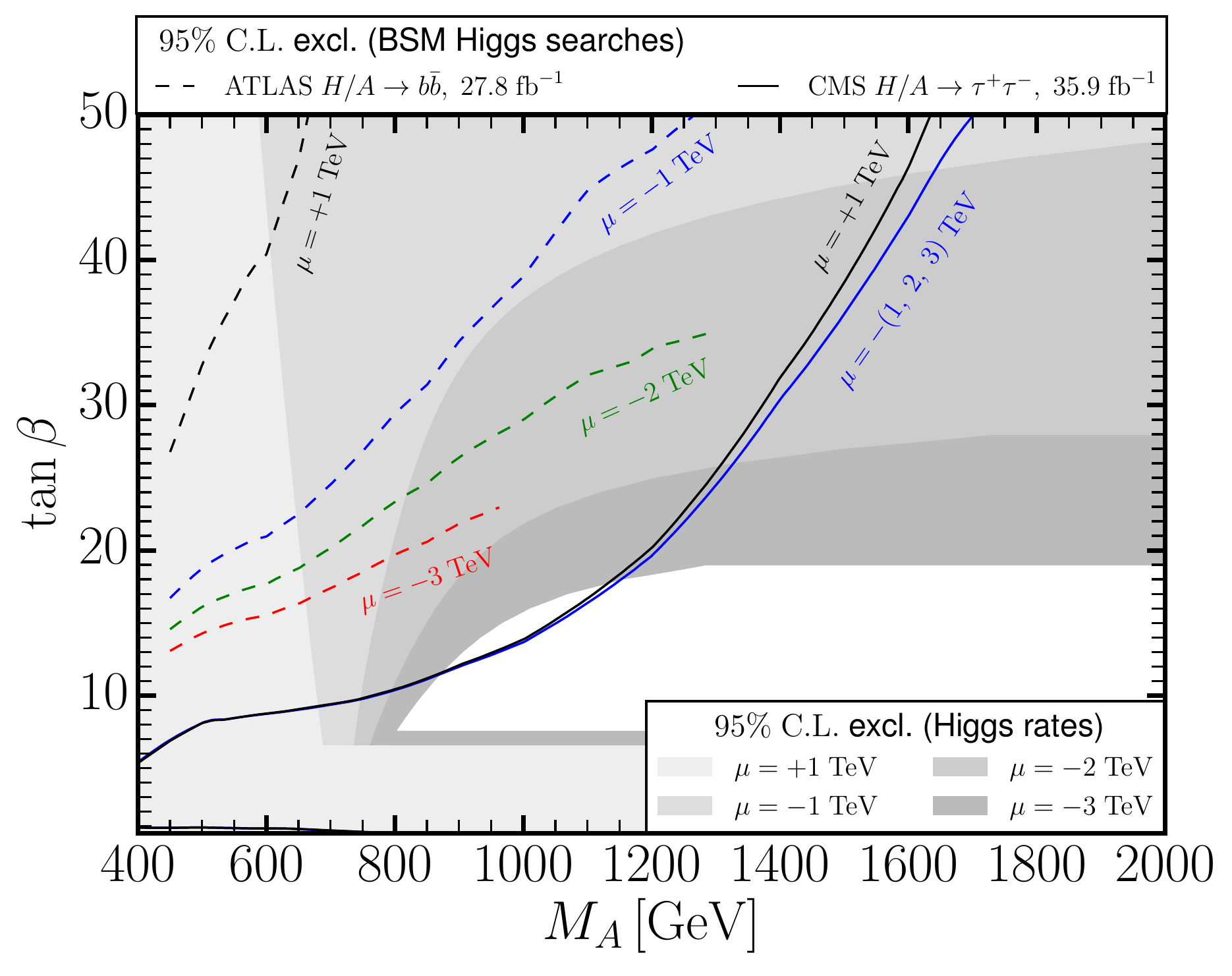}\hfill
\includegraphics[width=0.5\textwidth]{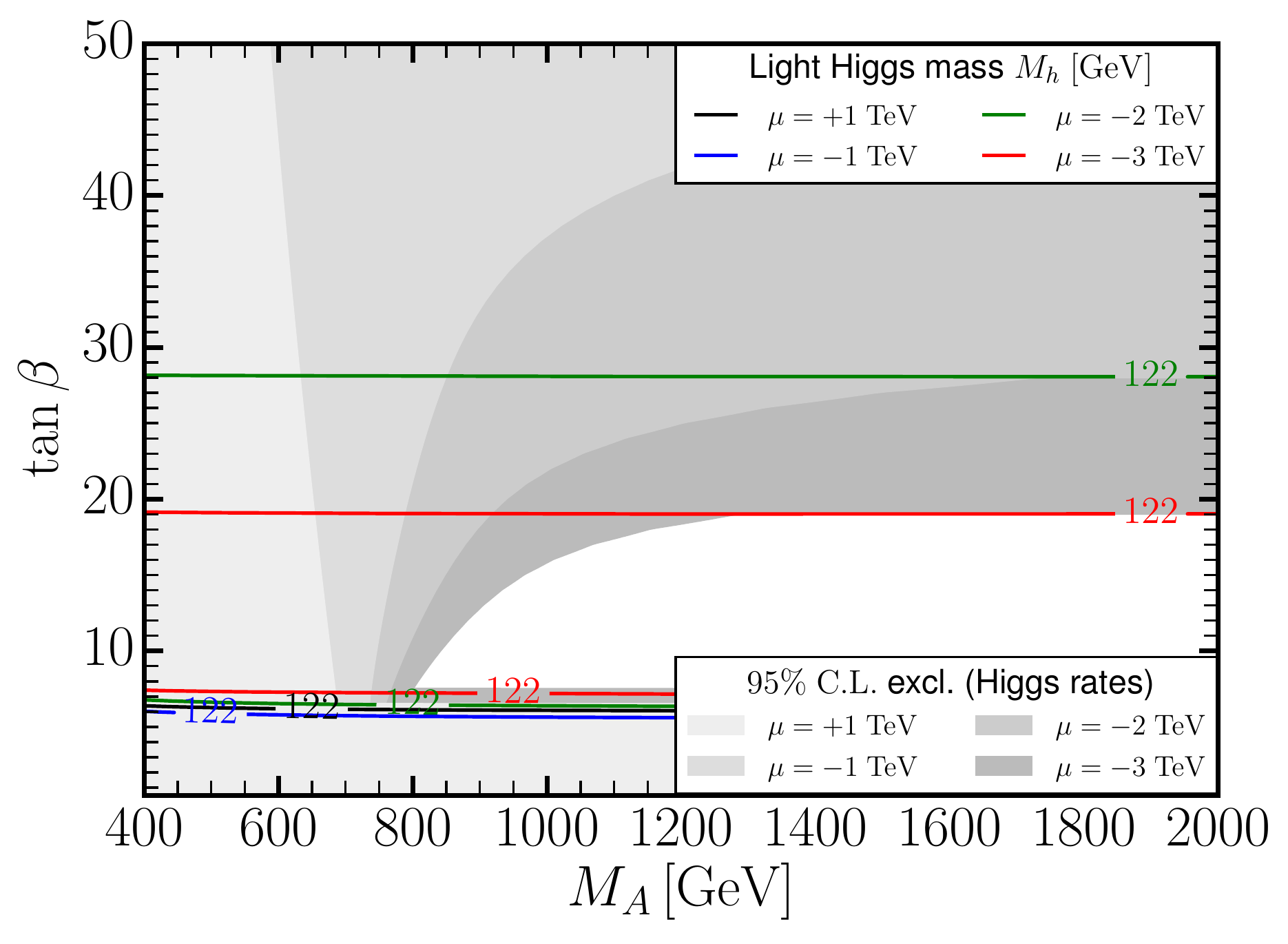}
\end{center}
\caption{$M_h^{125}$ scenario in comparison with the proposed new
    $M_h^{125}(\mu = -2\,\TeV)$ scenario and the other choices of
    $\mu = -1\,\TeV, -3\,\TeV$,
  shown in the ($M_A, \tan\beta$) parameter
  plane. \emph{Left panel}: current experimental constraints from heavy
  Higgs searches in the $b\bar{b}$~\cite{Aad:2019zwb} (dashed lines) and
  $\tau^+\tau^-$~\cite{Sirunyan:2018zut} (solid lines) final state and
  $125\,\GeV$ Higgs rate measurements (gray filled regions); \emph{Right
    panel}: Contours for the lowest acceptable value of the light-Higgs
  boson mass, $M_h = 122\,\GeV$ (taking into account a theory
    uncertainty of $\pm 3\,\GeV$), with the Higgs rate constraints (from
  the left panel) superimposed.}
\label{fig:Mh125negmu}
\end{figure}

In Fig.~\ref{fig:Mh125negmu} (\emph{left}) we show the current LHC
constraints at $95\%$ confidence level (C.L.) from heavy Higgs searches
in the $b\bar{b}$ and $\tau^+\tau^-$ final states as well as from Higgs
signal rate measurements for the standard $M_h^{125}$ scenario
(\emph{black lines} and \emph{very-light-gray shading}) and the three
variants of the $M_h^{125}$ scenario with different choices of the $\mu$
parameter: $\mu = -1~\TeV$ (\emph{blue lines} and \emph{light-gray
  shading}), $\mu = -2~\TeV$ (i.e.\ $M_h^{125,\mu-}$,
\emph{green line} and \emph{medium-gray
  shading}), $\mu = -3~\TeV$ (\emph{red line} and \emph{dark-gray
  shading}).
The heavy Higgs boson search limits are derived from the ATLAS $pp\to
H/A \to b\bar{b}$ search~\cite{Aad:2019zwb} with $27.8~\ifb$ of data,
and from the CMS $pp\to H/A \to \tau^+\tau^-$
search~\cite{Sirunyan:2018zut} with $35.9~\ifb$ of data, both at a
center-of-mass energy of $13~\TeV$,
using \texttt{HiggsBounds}~\cite{Bechtle:2008jh,Bechtle:2011sb,
Bechtle:2013gu,Bechtle:2013wla,Bechtle:2015pma,HiggsBounds5}.
The indirect constraints from Higgs
rate measurements are evaluated with \HS\ (version
\texttt{2.3.0})~\cite{Bechtle:2013xfa,HiggsSignals2} by means of a
negative log-likelihood ratio (LLR) test with the SM as alternative
hypothesis, and approximating the likelihood with a $\chi^2$
function. This test uses Run-1~\cite{Khachatryan:2016vau} and recent
Run-2 results up to around $80~\ifb$ from ATLAS~\cite{Aad:2019mbh} and
CMS~\cite{Sirunyan:2018egh,CMS:2019chr,CMS:1900lgv,Sirunyan:2018hbu,CMS:2019pyn,Sirunyan:2017elk,Sirunyan:2017dgc,CMS:2019lcn,Sirunyan:2018shy,CMS:2018dmv}.
Fig.~\ref{fig:Mh125negmu}
(\emph{left}) clearly illustrates that $pp\to H/A \to b\bar{b}$ searches
become more sensitive for scenarios with large negative $\mu$ values due
to the enhancement of the bottom-quark Yukawa coupling, as the excluded
regions probe lower values of $\tan\beta$ for larger negative $\mu$
values.\footnote{The exclusion lines for $\mu = -2\,\TeV$ and $-3\,\TeV$
  terminate, as for larger \tb\ values the light Higgs boson mass
  quickly decreases (see also the right panel of \fig{fig:Mh125negmu})
  and the prediction of Higgs-boson masses is affected by large
  uncertainties.} It is noteworthy that the exclusion limit from $pp\to
H/A\to \tau^+\tau^-$ searches does not vary significantly with
$\mu$.\footnote{The exclusion contours derived from the CMS $pp\to H/A
  \to \tau^+\tau^-$ search are practically identical for the choices
  $\mu = -1$, $-2$ and $-3\,\TeV$, and therefore plotted as a single
  contour.} This is because the $pp\to H/A\to b\bar{b}$ signal rate
profits from an enhancement in the production (in the $gg\to b\bar{b}
H/A$ production mode) \emph{and} in the decay branching ratio
($\mathrm{BR}$) of the $H/A\to b\bar{b}$ decay, while the $pp\to H/A \to
\tau^+\tau^-$ signal rate only gains from the enhancement in the
production rate whereas in combination with
the decay rate $\mathrm{BR}(H/A\to
\tau^+\tau^-)$ a large compensation of $\Delta_b$ effects
occurs~\cite{Carena:2005ek}.
Still, we observe that heavy Higgs-boson searches in the
$b\bar{b}$ final state cover significantly less parameter space
in the benchmark plane than searches in the
$\tau^+\tau^-$ final state, regardless of the choice of $\mu$.

We can furthermore see in Fig.~\ref{fig:Mh125negmu}(\emph{left})
that, currently, the indirect
exclusion derived from $125\,\GeV$ Higgs signal rate measurements is
stronger than the $H/A \to b\bar{b}$ (but not $H/A\to \tau^+\tau^-$)
search limits for all choices of $\mu$.
At larger \tb\ values these indirect constraints dominantly originate
from deviations of the light Higgs-boson Yukawa coupling to bottom
quarks. By approaching the decoupling limit, i.e.~with increasing
$M_A$,  the allowed parameter space opens towards larger
\tb\ values.  However, at around $(M_A, \tan\beta) \sim (1.75~\TeV,
28)$ for $\mu=-2\,\TeV$, and at around $(M_A, \tan\beta) \sim
(1.3\,\TeV, 19)$ for $\mu=-3\,\TeV$, and extending to larger $M_A$
values, the exclusion from the $125\;\GeV$ Higgs measurements
becomes a constant upper limit on $\tan\beta$ in this scenario.
This is caused by the
light Higgs-boson mass dropping below $122$\,GeV
(see the discussion above)
for larger values of
$\tan\beta$ for these two choices of $\mu$. Similarly, the light
Higgs-boson mass drops below $122$\,GeV for $\tan\beta\leq 7$. This is
illustrated in \fig{fig:Mh125negmu} (\emph{right}), which
shows contours for the lowest acceptable value (taking into account the
theoretical uncertainty) of the light-Higgs boson mass, $M_h =
122\;\GeV$, for the different choices of $\mu$.


\section{Experimental input for HL-LHC and ILC projections}
\label{sec:expinput}

We assess the reach of direct LHC searches in the  $\tau^+\tau^-$
final state by applying the model-independent $95\%~\mathrm{CL}$
limit projections for $6~\mathrm{ab}^{-1}$ by the CMS
experiment~\cite{Cepeda:2019klc,CMS:2018oxh},
serving as a proxy for a future ATLAS and CMS search combination using
the full HL-LHC data.
We implemented these limits --- presented as one-dimensional
(marginalized) cross section limits on either the
gluon fusion or $b\bar{b}$-associated production mode --- into
\texttt{HiggsBounds}~\cite{Bechtle:2008jh,Bechtle:2011sb,
Bechtle:2013gu,Bechtle:2013wla,Bechtle:2015pma,HiggsBounds5}
to obtain the projected $95\%~\mathrm{C.L.}$ exclusion in our scenarios.

For LHC searches for heavy Higgs bosons decaying into $b\bar{b}$ or di-Higgs ($hh$) final states, unfortunately, no experimental HL-LHC projection has been performed by ATLAS and CMS.\footnote{See Ref.~\cite{Adhikary:2018ise} for a theorists' analysis of the HL-LHC prospects of $H\to hh$ searches for various final states.} In order to estimate the future
sensitivity of these searches, we approximate
the future $95\%~\mathrm{C.L.}$ limit, $\sigma^{95\%\mathrm{CL}}_\mathrm{HL-LHC}$,
by rescaling the current expected limit, $\sigma^{95\%\mathrm{CL}}_\mathrm{current}$,
based on the current integrated luminosity, $\mathcal{L}_\mathrm{current}$,
at $\sqrt{s} = 13\,\TeV$, by the expected increase of statistics to the future combined ATLAS
and CMS integrated luminosity, $\mathcal{L}_\mathrm{HL-LHC} =
6~\mathrm{ab}^{-1}$, at a slightly higher center-of-mass energy of
$14\;\TeV$,%
\footnote{In our naive projection of the heavy Higgs-boson
searches in $b\bar{b}$ and $hh$ final states we neglect effects
related to the slightly higher center-of-mass energy $\sqrt{s} =
14\,\TeV$ in the HL-LHC runs. As the cross sections for the
background processes increase with respect to the current run, the
resulting upper cross section limit at $14\,\TeV$ will be larger
than our estimated $\sigma^{95\%\mathrm{CL}}_\mathrm{HL-LHC}$. This
degradation however depends  strongly on the experimental background
estimation methods and the specific process. Furthermore, it will
most likely be overcome by the simultaneous increase in the signal
cross section for $14\,\TeV$, thus leading overall to a net gain in
sensitivity in particular at large Higgs boson masses. In that
sense, our projection may be regarded as a conservative estimate.}
\begin{align}
\sigma^{95\%\mathrm{CL}}_\mathrm{HL-LHC} = \sigma^{95\%\mathrm{CL}}_\mathrm{current} \cdot \sqrt{\frac{\mathcal{L}_\mathrm{current}}{\mathcal{L}_\mathrm{HL-LHC}}}.
\end{align}
Obviously, this estimate has to be taken with a grain of salt, as we assumed the full uncertainty to improve like a statistical uncertainty.
Our projection of heavy Higgs boson searches in the $b\bar{b}$ final
state is  based on the current ATLAS search with
$\mathcal{L}_\mathrm{current} = 28.7\,\ifb$~\cite{Aad:2019zwb}, while we
use the current CMS $H\to hh$ analysis~\cite{Sirunyan:2018two} based on
$\mathcal{L}_\mathrm{current} = 36~\mathrm{fb}^{-1}$ (which combined
several final states) for the projection of searches in the di-Higgs
final state. The current limit of these searches is strongly dominated
by statistical uncertainties, which motivates our naive projection
approach. Only more sophisticated projection studies by
ATLAS and/or CMS would give a more reliable picture.

We do not take into account direct searches in the $A\to Zh$ final
states, as in the MSSM they are expected to be less relevant than the
$H\to hh$ searches~\cite{Bahl:2019ago}. We also do not assess the HL-LHC
reach of searches for heavy Higgs bosons in the $t\bar{t}$ final state,
as there are no official projections available and since
the limit is strongly influenced by interference effects between
signal and the SM background, as discussed in
Refs.~\cite{Djouadi:2016ack,Carena:2016npr,Hespel:2016qaf,BuarqueFranzosi:2017jrj,Bernreuther:2015fts,Bernreuther:2017yhg,Djouadi:2019cbm},
and is therefore rather model-dependent. It is clear that these
searches will be able to constrain low values of
$\tb$~\cite{Aaboud:2017hnm,CMS:2019lei},
where the heavy Higgs bosons predominantly decay into di-top final
states. Thus, in particular in the EFT-scenarios they will have a
significant impact at low values of $\tb$, see for comparison Fig.~6 in
Ref.~\cite{CMS:2019lei} for the current hMSSM~\cite{Djouadi:2013uqa,
Djouadi:2013vqa,Maiani:2013hud,Djouadi:2015jea,Liebler:2018zul}
limits obtained with an integrated luminosity of $36~\ifb$.
Similarly, also for the charged Higgs-boson searches at the HL-LHC no
official projections in benchmark planes are available.%
\footnote{In Refs.~\cite{Cepeda:2019klc,Aboubrahim:2018tpf} the
  sensitivity of charged Higgs boson searches was estimated for a set of
  minimal supergravity (mSUGRA) benchmark points. This model-dependent
  study indicates that the charged Higgs-boson search is less sensitive
  than neutral Higgs-boson searches.}%
~Consequently, we also leave charged Higgs sensitivity studies for future work.

We estimate the indirect reach through Higgs rate measurements by
using the detailed HL-LHC signal strength projections for the individual
Higgs production times decay modes, including the corresponding
correlation matrix, as evaluated by the ATLAS and CMS collaborations
(see Tab.~35 in Ref.~\cite{Cepeda:2019klc}). These projections
assume the evolution of systematic uncertainties as estimated by the
HL-LHC Working Group 2 (see Section 1.1.3 in Ref.~\cite{Cepeda:2019klc}
and references therein for details) and are referred to by future scenario
S2 or `YR18' systematic uncertainties.
We furthermore take cross-correlations of theoretical
rate uncertainties between future ATLAS and CMS measurements into
account, assuming that theoretical and parametric uncertainties are halved with
respect to current estimates~\cite{deFlorian:2016spz} (as prescribed in S2).
All projections of Higgs-boson rate measurements are implemented into
\HS, which we use to evaluate the projected reach in the MSSM parameter space.

For the indirect reach of Higgs signal rate measurements at the ILC we consider two future scenarios:
First, the inital stage scenario at $\sqrt{s} = 250\,\GeV$ with
$2~\mathrm{ab}^{-1}$
of data (denoted ILC250), and second, the
ILC program including a second run at $350\,\GeV$ with $0.2~\mathrm{ab}^{-1}$ of data, and a third run at $\sqrt{s}=500\,\GeV$ with $4~\mathrm{ab}^{-1}$ of data (for brevity we denote this future scenario by ILC500).
All ILC runs assume $-80\%$ polarization of electrons and $+ 30\%$ polarization of positrons.
The projected precisions of various Higgs-boson signal channels are listed in Tab.~21 and~22 of~Ref.~\cite{deBlas:2019rxi}.  We implemented these future anticipated measurements in \HS. For the
Higgs branching ratios, we assume that theoretical and parametric uncertainties are halved (as done for the HL-LHC, see above), while we
assume a theoretical uncertainty on the $e^+e^- \to Z h$ ($\nu\bar{\nu} h$) cross section of $0.5\%$ ($1\%$)~\cite{Freitas:2019bre}.
The projections for both future scenarios ILC250 and ILC500
are combined in our analysis with the projected HL-LHC Higgs rate measurements.

For illustration, we shall furthermore compare for specific MSSM parameter points the predictions of Higgs coupling
scale factors~\cite{deFlorian:2016spz}, $\kappa_i$, with the anticipated precision of the future $\kappa$ determination at the HL-LHC and ILC, taken from Tabs.~4 and~5 of Ref.~\cite{deBlas:2019rxi}.
Note that these projections of the future $\kappa_i$ determination are based on identical experimental input as in our study,
but have been evaluated in a Bayesian statistical analysis instead of a log-likelihood ratio (LLR) test as employed here.


\section{Sensitivity of the HL-LHC}
\label{sec:reach}

We start with a discussion of the future HL-LHC sensitivity via rate
measurements and direct searches for heavy MSSM Higgs bosons.
In the subsequent figures we present in the ($\mA, \tb$) planes of the considered MSSM benchmark scenarios the projected direct and indirect $95\%~\mathrm{C.L.}$ sensitivity of a future combination of ATLAS and CMS data
at the HL-LHC. For comparison, we also include the current sensitivity of the corresponding searches and measurements, as presented in part in Refs.~\cite{Bahl:2018zmf,Bahl:2019ago}. In the next section we shall address the improvements of the indirect reach obtained by ILC measurements.

\begin{figure}[tb!]
\begin{center}
\includegraphics[width=0.6\textwidth]{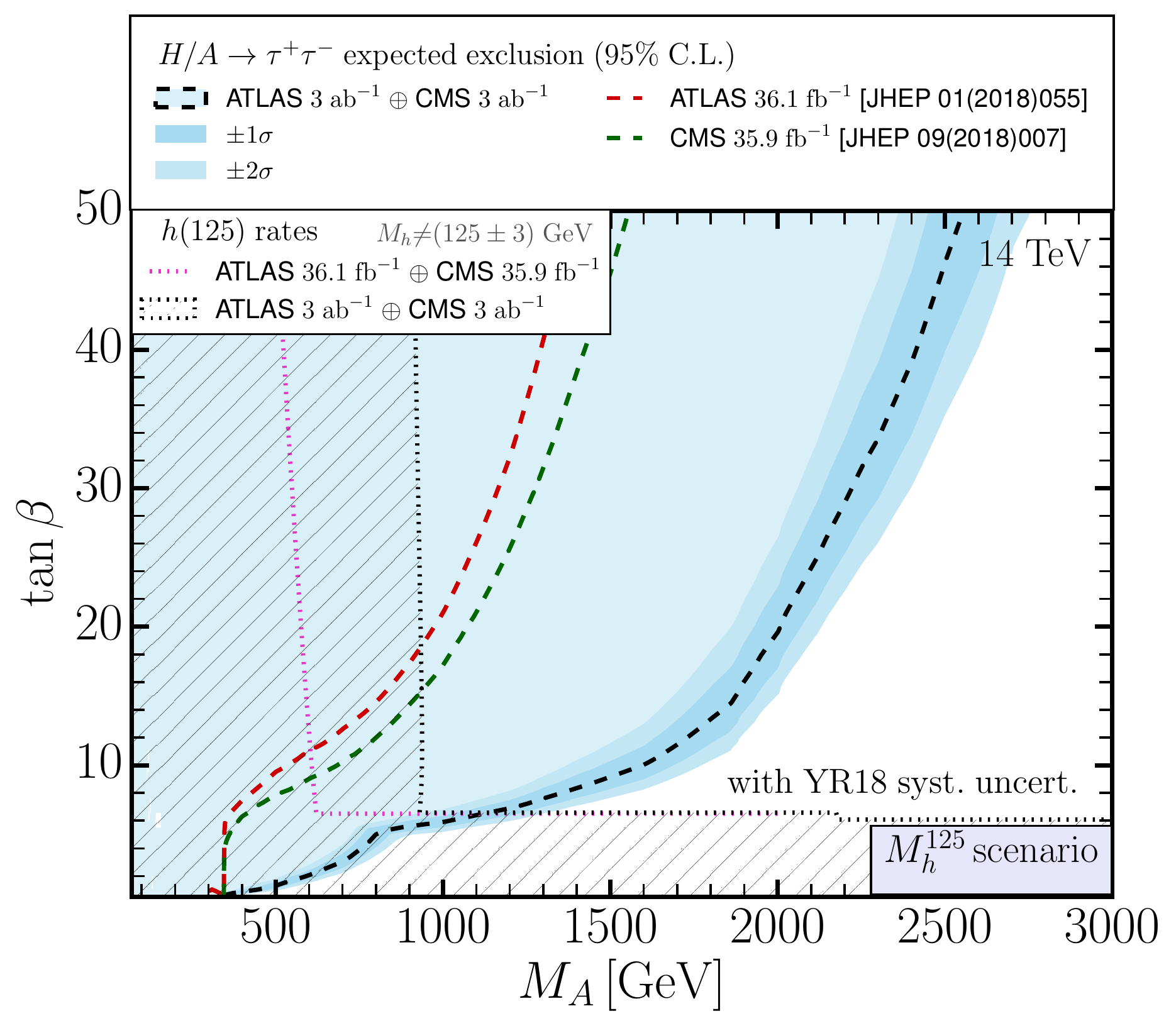}
\end{center}
\caption{HL-LHC projections in the $M_h^{125}$ scenario,
  assuming YR18 systematic uncertainties (scenario S2
  in Ref.~\cite{Cepeda:2019klc}). The dashed black curve
  and blue filled region indicate the expected HL-LHC reach via
  direct heavy
  Higgs searches in the $\tau^+\tau^-$ channel with $6~\mathrm{ab}^{-1}$
  of data (with the dark blue regions indicating the $1$ and $2\sigma$
  uncertainty), whereas
  the red and green dashed lines show the expected limit from current
  searches in this channel by ATLAS~\cite{Aaboud:2017sjh} and
  CMS~\cite{Sirunyan:2018zut}, respectively. The current and future
  HL-LHC sensitivity via  combined ATLAS and CMS Higgs rate measurements
  is shown as magenta and black dotted contours, respectively (the
  latter being accompanied with a hatching of the prospectively excluded
  region).}
\label{fig:bench1}
\end{figure}

Our projections in the $M_h^{125}$ scenario are presented in \fig{fig:bench1}.
Corresponding planes for the $M_h^{125}(\tilde\chi)$ and $M_{h, {\rm EFT}}^{125}$
scenario are given in Figs.~\ref{fig:bench2} and \ref{fig:bench4},
respectively.
In the figures we show the current limit (magenta dotted line) for the
indirect reach of the LHC in the benchmark scenarios, as evaluated
in Ref.~\cite{Bahl:2018zmf,Bahl:2019ago}, as well as the expected limit from current
direct BSM Higgs searches by ATLAS~\cite{Aaboud:2017sjh} (red dashed
line)%
\footnote{
Very recently the corresponding result based on $139\,\ifb$ has become
available~\cite{Aad:2020zxo}. Including this would not affect our
conclusions.}%
~and CMS~\cite{Sirunyan:2018zut} (green dashed line) in the
$\tau^+\tau^-$ final state, using $\sim 36~\mathrm{fb}^{-1}$ of data
from Run~2 at $13~\mathrm{TeV}$. The dashed black curve
and blue filled region indicate the HL-LHC reach via direct heavy
Higgs searches in the $\tau^+\tau^-$ channel with $6~\mathrm{ab}^{-1}$
of data (with the dark blue regions indicating the $1$ and $2\sigma$ experimental
uncertainty). The future HL-LHC sensitivity via combined ATLAS and CMS
Higgs rate measurements is shown as black dotted contours, accompanied
with a hatching of the prospectively excluded region.

Within the $M_h^{125}$ scenario, displayed in \fig{fig:bench1},
the indirect constraints at the HL-LHC have the sensitivity for a
prospective exclusion limit that is given by a nearly vertical band extending
to $M_A$ values of around $900$\,GeV and by a nearly horizontal band with
$\tan\beta$ values of up to $6$. The nearly vertical band arises from the
measurements of the Higgs signal strengths, while the nearly
horizontal band is due to the prediction for the mass of the SM-like Higgs
boson. In the nearly horizontal band this prediction is below
$122$\,GeV for the parameters of the $M_h^{125}$ scenario, such that the
interpretation of the observed Higgs signal in terms of the light \cp-even MSSM
Higgs boson $h$ is incompatible in this specific benchmark scenario
with the experimental mass value for the adopted
theory uncertainty of $3$\,GeV. \fig{fig:bench1} shows that in this scenario
the indirect sensitivity of the Higgs rate measurements at the HL-LHC is not
sufficient to probe parameter regions that are not covered by the direct
Higgs searches (black dashed line). Those direct searches in the
$\tau^+\tau^-$ final state will probe the parameter space up to
$M_A \lesssim 2.5$\,TeV for the highest displayed $\tan\beta$ values of
$\tan\beta \sim 50$. At $\tan\beta = 20$ the reach extends up to
$M_A \lesssim 2000$\,GeV.
The change in the curvature of the black dashed line around $\mA\sim
1.9$\,TeV can be understood from the fact that for larger values of $\mA$
decays of $H$ and $A$
into electroweakinos open, thus diminishing the event yield of the
$\tau^+\tau^-$ final state.
The kink in the exclusion boundary at
$M_A \sim 800$ GeV is caused by a transition of the main production
channel from gluon fusion (low \tb\ values) to bottom quark associated
production (high
\tb\ values).\footnote{It should be kept in mind that for the projected
$H/A\to \tau^+\tau^-$ search sensitivity
we used the one-dimensional profiled cross section limits for the two
relevant production modes.}
~In this scenario the prospective combined sensitivity
from direct and indirect searches in the absence of a signal would yield a
lower bound on $M_A$ of
about $M_A \gtrsim 1200\,\GeV$. In order to correctly interpret this result,
the following should be taken into account. As explained above, this
bound is {\em not\/} a consequence of prospective Higgs signal strength
measurements at the HL-LHC, but it is rather driven by the direct Higgs
search reach in combination with the Higgs-mass prediction. Since by
definition for this benchmark scenario all
parameters except $M_A$ and $\tan\beta$ are set to fixed values, the
adopted theoretical uncertainty of the Higgs-mass prediction has a major
impact on the resulting bound. For a
smaller theoretical uncertainty the allowed region in this scenario would be
shifted to larger $\tan\beta$ values, so that the lower
bound on $M_A$ would rise to values above $2$\,TeV.
On the other hand, in scenarios where the prediction for the mass of the
light Higgs boson is compatible with the measured Higgs-boson mass also for low
$\tan\beta$ values, the indirect constraints on $M_A$ from the rate
measurements can exceed the sensitivity from the direct searches (see the
discussion below).

\medskip

\begin{figure}[tb!]
\begin{center}
\includegraphics[width=0.6\textwidth]{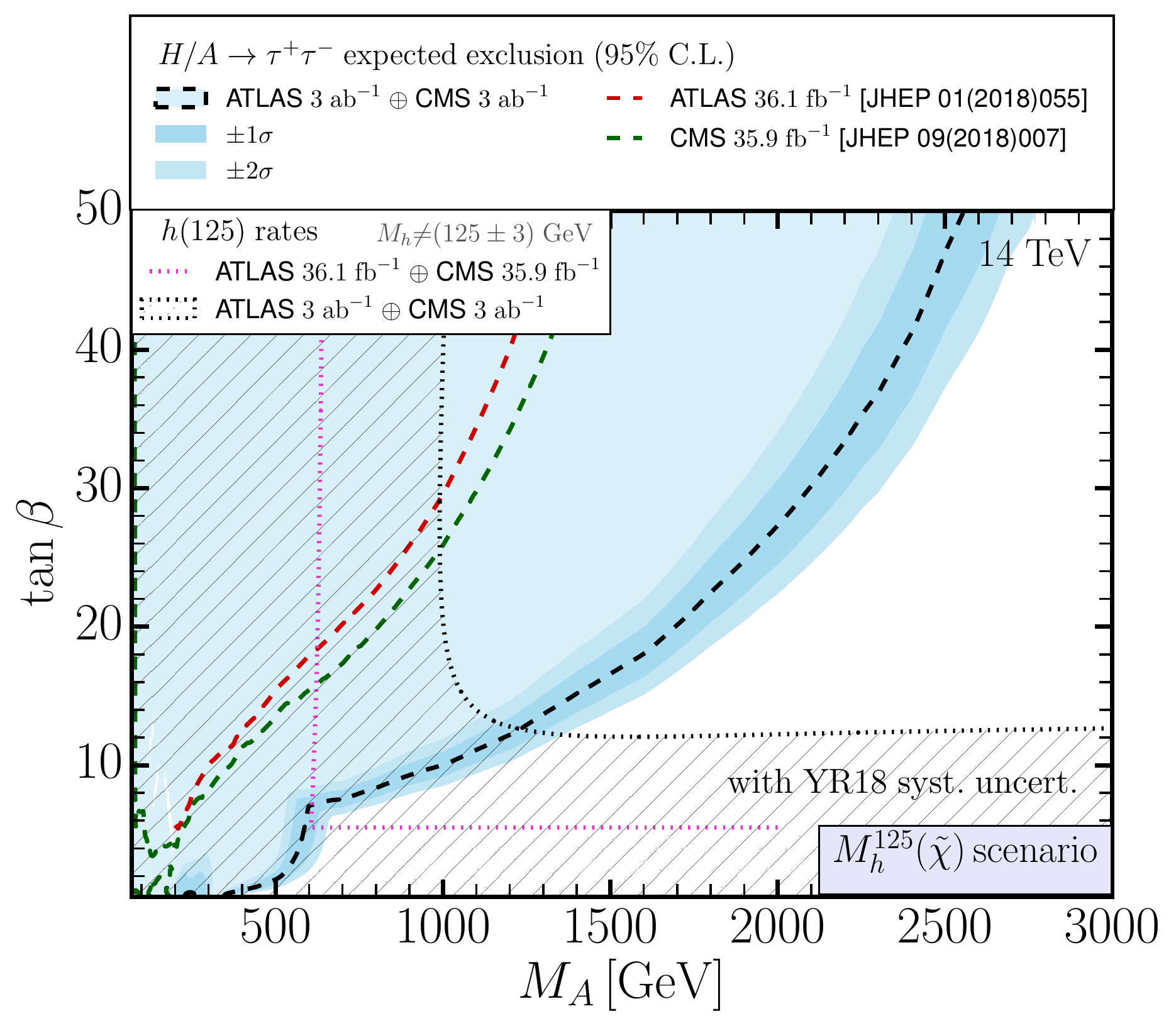}
\end{center}
\caption{HL-LHC projections in the
  $M_h^{125}(\tilde{\chi})$ scenario with the same color coding as in
  \fig{fig:bench1}.}
\label{fig:bench2}
\end{figure}

The picture is somewhat different in the $M_h^{125}(\tilde{\chi})$
scenario. Here the large branching ratio of the heavy neutral Higgs
boson decaying to charginos and neutralinos already at lower values of
$M_A$ leads to a strongly reduced
direct reach of  $H/A\to \tau^+\tau^-$ searches.
The kink in the exclusion boundary at
$M_A \sim 600$ GeV is as in \fig{fig:bench1} caused by a
transition of the most sensitive production channel from gluon fusion (at low \tb\ values)
to bottom quark associated production
(at high \tb\ values).
At $\tan\beta = 20$ the reach in the $M_h^{125}(\tilde{\chi})$ scenario is significantly
reduced to $M_A \lesssim 1700$\,GeV compared to the $M_h^{125}$ scenario with
$M_A \lesssim 2000$\,GeV.
On the other hand, at large
values of $\tan\beta \sim 50$ and thus large $\mA$ the reach is only slightly weaker than in
the $M_h^{125}$ scenario, as for those $M_A$ values
in both scenarios decays into electroweakinos
are kinematically open. In order to further strengthen the impact of direct searches
it would be useful to supplement the searches in the $\tau^+\tau^-$ and
$b \bar b$ final states with dedicated
searches for the decays of $H$ and $A$ to charginos, neutralinos and in
general also to sleptons (see Refs.~\cite{ATLAS:2009zmv,Bisset:2007mi,Charlot:2006se,Arbey:2013jla,Ball:2007zza,Craig:2015jba,Belanger:2015vwa,Barman:2016kgt,Baum:2017gbj,Profumo:2017ntc,Kulkarni:2017xtf,Arganda:2018hdn,Gori:2018pmk,Adhikary:2020ujn,Liu:2020muv,Heinemeyer:2014yya,Heinemeyer:2015pfa} for related experimental and phenomenological studies).

In this scenario the Higgs rate measurements are an
important complementary probe.
In the $M_h^{125}(\tilde{\chi})$ scenario the Higgs rate measurements at
the HL-LHC have the sensitivity to exclude values of
$M_A \le 950$\,GeV and $\tan\beta \le 12.5$.
The bound on $M_A$ arises in a similar way as for the
$M_h^{125}$ scenario discussed above from the slight deviations in the Higgs
couplings compared to the SM values that are caused by $M_A$ values below
the asymptotic region of decoupling. The displayed
bound on $\tan\beta$, on the other
hand, is not related to the Higgs-mass prediction as in \fig{fig:bench1} but
is a consequence of the loop contribution of light charginos to
the $h\to \gamma\gamma$ partial width. For values
of $\tan\beta\le 12.5$ the
enhancement of the $h\to \gamma\gamma$ partial width
is so large that the HL-LHC has the potential to exclude this parameter
region via the precise measurements of the $h\to \gamma\gamma$ rates,
which have
an anticipated precision of $2.6\%$~\cite{Cepeda:2019klc}.
The combination of
direct and indirect bounds yields
a prospective lower limit of $M_A \gtrsim 1250$\,GeV in the $M_h^{125}(\tilde{\chi})$
scenario. This bound is less sensitive to the theoretical uncertainties of the
Higgs-mass prediction than for the $M_h^{125}$ scenario but instead depends
on the mass scale of the charginos that has been chosen in the
$M_h^{125}(\tilde{\chi})$ scenario.

In fact, the ongoing searches for electroweakinos in the mass-compressed
region put
the $M_h^{125}(\tilde{\chi})$ scenario under some tension as the parameters of the
electroweakino sector are fixed to rather low values of $M_1=160\,\GeV$
and $M_2=\mu=180\,\GeV$. However, one can
increase the mass parameters of the electroweakino sector by, for instance, $+100\,\GeV$
without a major impact on the sensitivity reach of the $\tau^+\tau^-$
searches.
This can be understood from the fact that the heavy Higgs
bosons, as long as the decay modes are kinematically open, still show
dominant branching ratios into electroweakinos at low and moderate
values of $\tb$. On the other hand, increasing the electroweakino masses
significantly lowers the effect of the chargino loop
contributions on the $h\to \gamma\gamma$ partial
decay width.\footnote{For even lighter electroweakinos than in this scenario, with masses below $M_h/2$, the possibility of invisible decays of the observed Higgs boson $h$ arises. This decay leads to additional modifications of the Higgs rates and can also be searched for directly (see Sec.~6 of Ref.~\cite{Cepeda:2019klc} for the HL-LHC prospects, and Refs.~\cite{Profumo:2017ntc,Barman:2017swy,Pozzo:2018anw,Wang:2020dtb,Barman:2020vzm} for recent studies of SUSY models in the context of dark matter).} We will investigate this effect in more detail below in the
context of the $M_{h, {\rm EFT}}^{125}(\tilde\chi)$ scenario (see \fig{fig:M2muplane}).

\medskip

\begin{figure}[tb!]
\begin{center}
\includegraphics[width=0.6\textwidth]{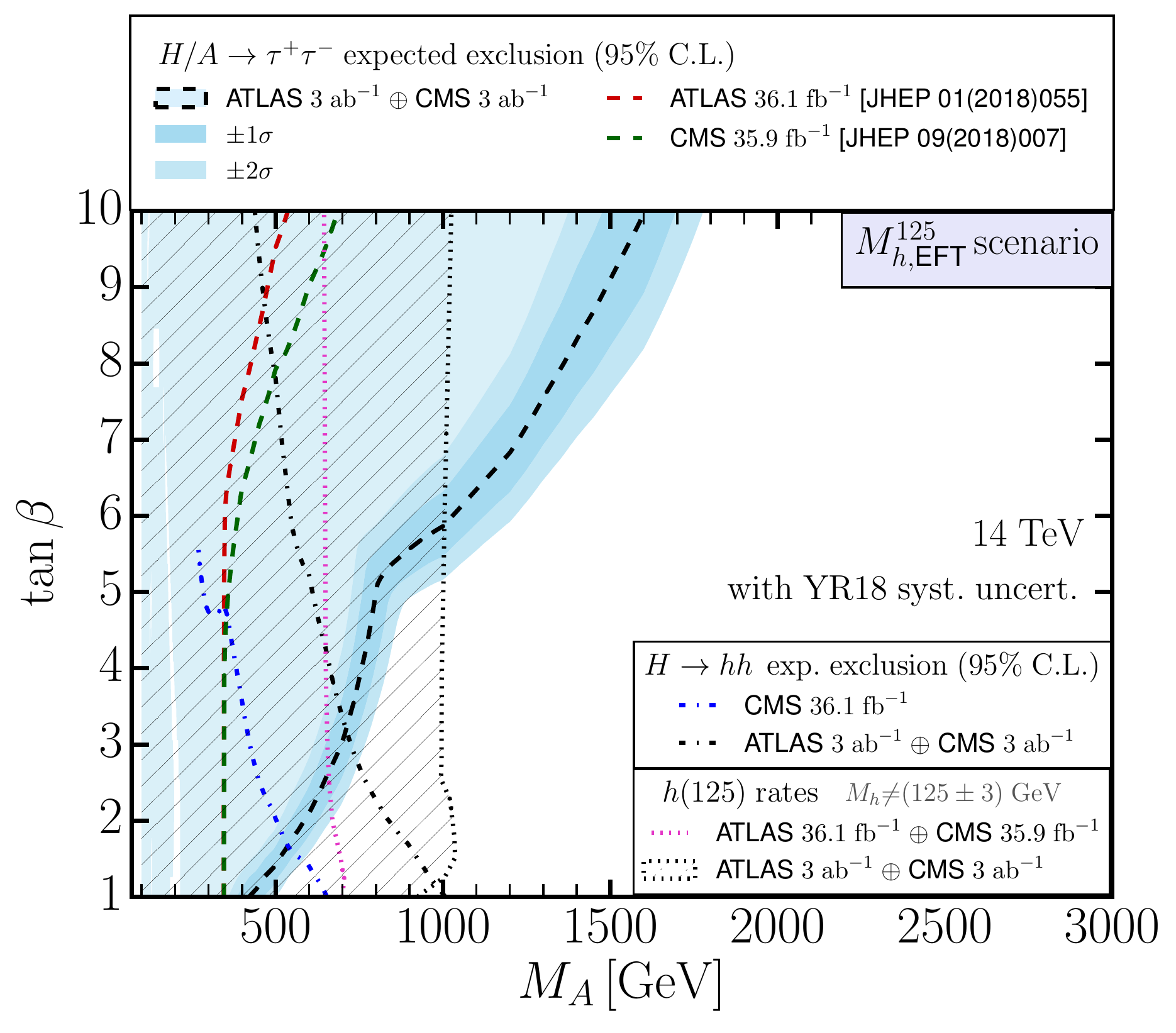}
\end{center}
\caption{HL-LHC projections in the
  $M_{h, {\rm EFT}}^{125}$ scenario with the same color coding as in
  \fig{fig:bench1}.
  The blue and black dash-dotted lines show the current CMS~\cite{Sirunyan:2018two} and future HL-LHC
  expected $95\%~\mathrm{C.L.}$ limit from a combination of $H\to hh$
  searches (see Sec.~\ref{sec:expinput} for details).}
\label{fig:bench4}
\end{figure}

The $M_{h, {\rm EFT}}^{125}$ scenario in \fig{fig:bench4} is only
shown up to $\tb = 10$, as this reflects the main feature of this
scenario, which is to provide access to the low~\tb\ region.
Since this scenario allows a light Higgs-boson mass of $125\,\GeV$
even for $\tb$ values as low as $\tb\sim 1$, the indirect reach
through Higgs rate measurements is now almost vertical.
The horizontal band yielding a lower bound on $\tan\beta$
in the $M_h^{125}$ scenario from the compatibility of the Higgs-boson mass
prediction with $125\,\GeV$ is not present
in the $M_{h, {\rm EFT}}^{125}$ scenario. Since the $M_{h, {\rm
EFT}}^{125}$ scenario does not feature light charginos, also the lower bound
on $\tan\beta$ arising
from precise measurements of the $h\to \gamma\gamma$ rates
in the $M_h^{125}(\tilde{\chi})$ scenario
does not occur in \fig{fig:bench4}.

Again the indirect reach in $M_A$ is similar to the scenarios discussed
beforehand, driven by the behavior of the couplings of the light \cp-even
Higgs boson when approaching the decoupling limit, which is mostly a
function of $\mA$ and hardly dependent
on $\tb$.
The projected sensitivity at the HL-LHC in this scenario
therefore corresponds to
an almost vertical expected exclusion region
for \cp-odd Higgs boson masses $M_A \lesssim 1000 \,\GeV$.
The direct search reach for
heavy neutral Higgs bosons is similar to the $M_h^{125}$
scenario for the displayed $\tan\beta$ region. However, due to the fact that the low
$\tb$ region is not excluded by the light Higgs-boson mass predictions,
the decay $H \to hh$ can cover
some additional parameter space, up to $M_A \sim 1000 \,\GeV$ for
$\tb = 1$. At low $\tb$ searches for heavy neutral Higgs bosons in the di-top final state would
be of relevance as well, but are not further discussed here  (see
Sec.~\ref{sec:expinput}).
As a result, in the $M_{h, {\rm EFT}}^{125}$ scenario we find that the
indirect sensitivity from the Higgs rate measurements has the largest
coverage for $\tan\beta$ values up to $\tan\beta \sim 5.5$, while for higher
values of $\tan\beta$ the direct searches for heavy Higgs bosons in the
$\tau^+\tau^-$ final state have the best prospects.

In order to cover the low-$\tb$ region,
further experimental sensitivity studies for
direct searches for ${H/A\to t\bar t}$, ${H\to hh}$ and ${A\to Zh}$
decays as well as heavy Higgs boson decays into electroweakinos are of
interest
(see Refs.~\cite{Djouadi:2019cbm, Adhikary:2018ise} for recent theorists'
projections of $H/A\to t\bar{t}$ and $H\to hh$, and
Ref.~\cite{Cepeda:2019klc} for experimental projections in different
scenarios). The searches for decays to electroweakinos are
of particular importance in both the $M_h^{125}(\tilde\chi)$ and the
$M_{h, {\rm EFT}}^{125}(\tilde\chi)$ scenario, see also
Ref.~\cite{Gori:2018pmk,Bahl:2018zmf, Bahl:2019ago}.

\medskip

\begin{figure}[tb!]
\begin{center}
\includegraphics[width=0.6\textwidth]{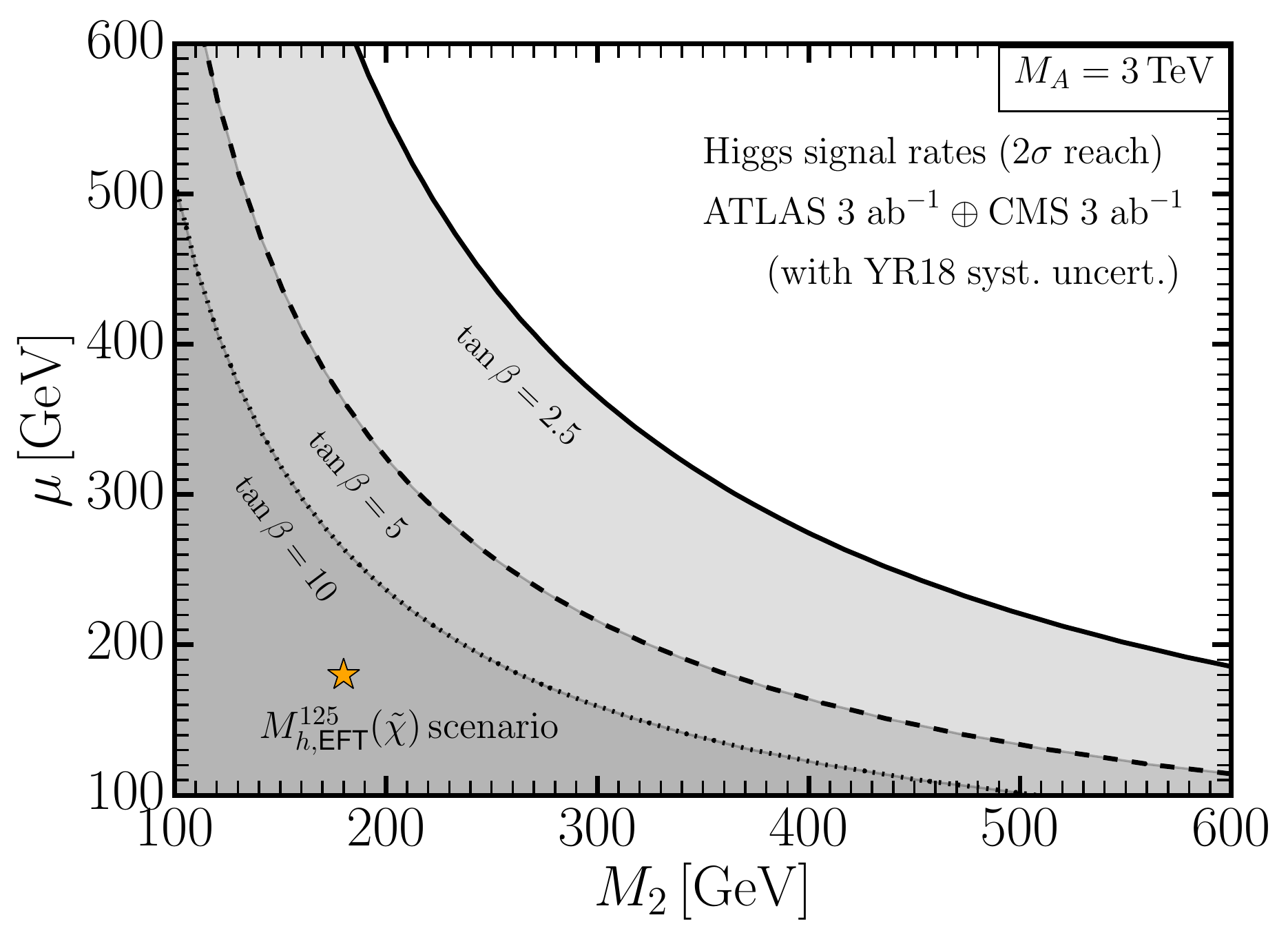}
\end{center}
\caption{Projected reach of future Higgs signal rate measurements at
  ATLAS and CMS with $3~\mathrm{ab}^{-1}$ assuming YR18 systematic
    uncertainties in the ($M_2$, $\mu$) parameter plane around the
  $M_{h,\text{EFT}}^{125}(\tilde{\chi})$ scenario (denoted by the orange
  star), for fixed $M_A = 3\;\TeV$. The solid, dashed and dotted contour
  lines and the corresponding gray areas indicate the $2\sigma$ reach
  for $\tan\beta$ values of $2.5$, $5$ and $10$, respectively.
}
\label{fig:M2muplane}
\end{figure}

We now turn to the second EFT scenario, $M_{h, {\rm EFT}}^{125}(\tilde\chi)$,
with a light EWino spectrum.
As for the case of the $M_h^{125}(\tilde{\chi})$ scenario discussed
above, the HL-LHC measurement of the di-photon Higgs-boson signal rate
has the potential to set a lower bound on $\tan\beta$ for the chosen values
of the chargino masses. In fact, restricting ourselves to the
$\tan\beta$ range between 1 and 10 that was originally proposed for this
scenario, the entire ($\mA$,~$\tb$) plane of the
$M_{h, {\rm EFT}}^{125}(\tilde\chi)$ scenario can be probed by the HL-LHC
measurement of the di-photon Higgs-boson signal rate. Accordingly, this
parameter plane could be
excluded at the HL-LHC if no deviation from the SM prediction is observed.
Therefore, instead of displaying the ($\mA$, $\tb$) plane, we instead
investigate the reach of the HL-LHC
in the ($M_2$, $\mu$) parameter plane, where $M_2$ is the soft-breaking wino mass parameter
and $\mu$ the Higgs mixing parameter. This is shown in \fig{fig:M2muplane}, where we highlight
the prospective $2\sigma$ excluded region, assuming HL-LHC Higgs signal
rate measurements that agree with the SM expectation.
The results are shown for three different values of $\tb = 2.5, 5, 10$
and fixed $M_A = 3\,\TeV$.
As can be seen in \fig{fig:M2muplane}, the reach in the chargino
mass parameters $M_2$ and $\mu$ increases with decreasing $\tb$,
caused by a larger mixing of the charginos with decreasing $\tb$,
which directly impacts the $h \to \gamma\gamma$ partial decay width. Similarly,
the largest values of the light chargino mass, $M_{\tilde{\chi}_1^\pm}$, can
be probed if $M_2 \approx \mu$, as in this case the chargino mixing is large,
and in turn, the Higgs boson coupling to charginos is maximized.
For instance, for $\tb=2.5~(5)$ and $M_2 \approx \mu$, light chargino masses
up to $\sim 255~(190)\,\GeV$  can be probed at the 2\,$\sigma$ level
(in this case, the heavier chargino
mass is $\sim 410~(320)\,\GeV$). In contrast, in case of a larger hierarchy,
$M_2 \gg \mu$ or $M_2 \ll \mu$, the smaller of the two mass parameters has to be
rather low in order to be able to probe the electroweakino sector via the di-photon signal strength measurements.
The nominal values of $M_2$ and $\mu$ that were chosen in the definition of
the $M_{h, {\rm EFT}}^{125}(\tilde\chi)$ scenario,
marked by an orange star in \fig{fig:M2muplane},
could be probed for $\tb \lesssim 12.5$, which is in agreement with the findings in the $M_h^{125}(\tilde{\chi})$ scenario, see \fig{fig:bench2}.
We emphasize that this indirect
probe for electroweakinos via their loop contributions to the
$h \to \gamma\gamma$ partial decay width
is complementary to the
direct searches for electroweakinos at the HL-LHC~\cite{CidVidal:2018eel}.

Finally, in \fig{fig:bench5} we show the HL-LHC
sensitivity for the proposed new $M_h^{125}(\mu = -2\,\TeV)$ scenario
in comparison with the $M_h^{125}$ scenario and the other choices of
$\mu = -1\,\TeV, -3\,\TeV$, as introduced in
Section~\ref{sec:models}. The exclusion lines and filled regions are
analogous to those in \fig{fig:Mh125negmu} (\emph{left}), but
are now determined using the HL-LHC prospective searches and
measurements, instead of the current experimental results.
The main qualitative features observed in \fig{fig:Mh125negmu}
(\emph{left}) can be
found here for the HL-LHC projections as well: Searches for heavy Higgs
bosons in the $\tau^+\tau^-$ final state cover a
larger area in the ($M_A, \tan\beta$) parameter plane
than those in the $b \bar b$ final state, and
the $H/A\to b\bar{b}$ search sensitivity shows a strong dependence
on the size and sign of $\mu$ while there is only a moderate impact on
the searches in the $\tau^+\tau^-$ final state. On the other hand,
\fig{fig:bench5} shows that
the anticipated reach of  heavy Higgs boson searches in
the $b\bar b$ final state is competitive with the indirect reach of the
anticipated Higgs-boson rate measurements. Except for $\mu=-3\,\TeV$
the direct searches in the $b\bar b$ final state
yield a stronger expected exclusion in the high-$M_A$ region than the
Higgs-boson rate measurements.
The flat regions towards large values of $M_A$
in the upper bounds on $\tan\beta$
for $\mu=-2$\,TeV and $\mu=-3$\,TeV are again caused by the
fact that the prediction for the light Higgs-boson mass is below $122$\,GeV
in this region (see~\fig{fig:Mh125negmu} (\emph{right})),
and the same applies to the lower limit in $\tan\beta$
(which is almost identical for all values of $\mu$).
However, for $M_A \lesssim 2$\,TeV in the scenario with $\mu=-2$\,TeV
and for $M_A \lesssim 1.5$\,TeV in the scenario with $\mu=-3$\,TeV
the Higgs rate measurements provide sensitivity for a non-trivial upper bound
on $\tan\beta$.

\begin{figure}[tb!]
\begin{center}
\includegraphics[width=0.6\textwidth]{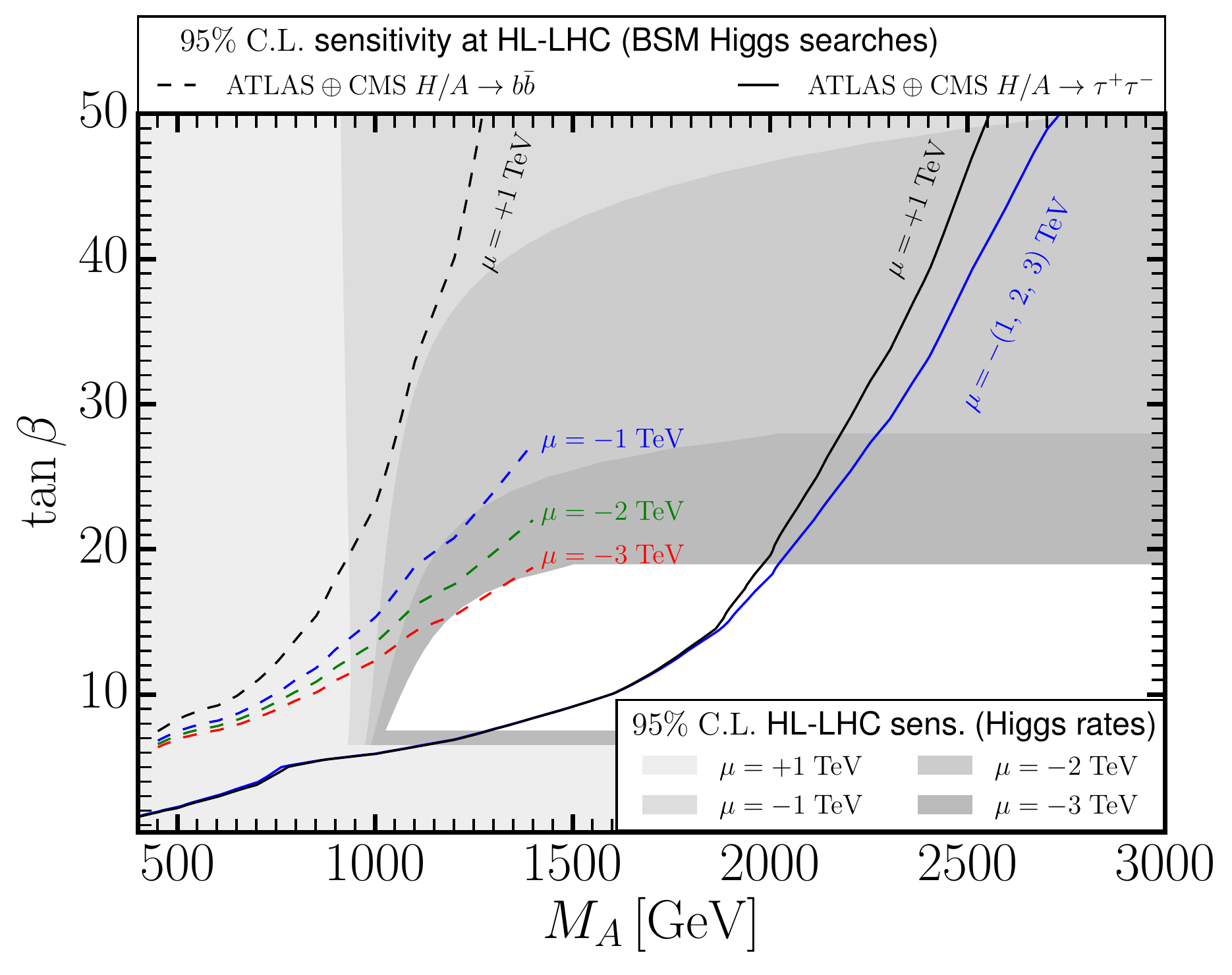}
\end{center}
\caption{HL-LHC projections for the
proposed new $M_h^{125}(\mu = -2\,\TeV)$ scenario
in comparison with the $M_h^{125}$ scenario and the other choices of
$\mu = -1\,\TeV, -3\,\TeV$,
  shown in the ($M_A, \tan\beta$)
  parameter plane; the dashed and solid lines show the expected
  exclusion from heavy Higgs boson searches in the $b\bar{b}$ and
  $\tau^+\tau^-$ final state, respectively, and the gray filled regions
  indicate the indirect reach of HL-LHC Higgs rate measurements.}
\label{fig:bench5}
\end{figure}


\section{Constraints on BSM physics from the
prospective rate measurements at the HL-LHC and the ILC}
\label{sec:reachILC}

We now extend our investigations to the situation where the results from
the HL-LHC are combined with prospective high-precision measurements of the
Higgs signal rates at a future linear $e^+e^-$ collider (for definiteness,
we focus on the ILC as
the currently most advanced project for which the
anticipated precision levels are based on full detector simulations).
We do not take into account in this context the capabilities of
an $e^+e^-$ linear collider for detecting new light states, like the light
electroweakinos occurring in the benchmark scenarios discussed above and the
possible production of additional light Higgs
bosons~\cite{Biekotter:2017xmf,Drechsel:2018mgd,Wang:2018awp,Wang:2019mzd,Biekotter:2019gtq}.
The latter
possibility is of less relevance in the benchmark scenarios that we discuss
in the present paper, but within the MSSM context can occur in scenarios with
(approximate) alignment without decoupling~\cite{Bahl:2018zmf}. Light
additional Higgs bosons that are compatible with present experimental
constraints can occur as a generic
feature in extended Higgs sectors with an additional singlet, see
e.g. Refs.~\cite{Domingo:2015eea,Drechsel:2016jdg,Domingo:2018uim,Biekotter:2019kde}.

In the following we first discuss the indirect constraints on $M_A$ that
could be inferred in the absence of a deviation from the SM predictions,
i.e.\ under the assumption that the measured rates exactly agree with the SM
prediction. It should be obvious from the discussion of the HL-LHC
sensitivities in the previous section, where in the considered benchmark
scenarios the prospective constraints on $M_A$ are already quite far in the
decoupling region of the MSSM, that an improved precision of the detected
Higgs-boson rates will only have a moderate effect in such a scenario. This
is due to the fact that the dependence of the Higgs-boson rates on $M_A$ is
essentially flat in this region, with only very small deviations from the SM
value. It should be noted that a precision measurement of the Higgs-boson
rates would have a much higher impact for the case, for instance, where the
Higgs boson at $125~\GeV$ would have a non-vanishing decay mode into BSM
particles.\footnote{As mentioned above, if the mass of the lightest neutralino is below $M_h/2$ the possibility of invisible Higgs decays can also be searched for directly, with an anticipated $95\%~\text{C.L.}$ upper limit on the branching ratio of $2.6\%$ ($0.3\%$) at the HL-LHC (for the combination of HL-LHC and ILC)~\cite{Cepeda:2019klc, deBlas:2019rxi}.}

In a second line of analysis we investigate scenarios that would correspond
to the situation where a particular MSSM parameter point was
realized in
nature. We show in this context how on the basis of the Higgs rate
measurements alone such a scenario could be distinguished from the SM case
and how well the parameters $M_A$ and \tb\ could be indirectly constrained.


\subsection{Impact of the rate measurements for the case where
the SM is realized}
\label{sec:nosignal}

\begin{figure}[tb!]
\begin{center}
\includegraphics[width=0.5\textwidth]{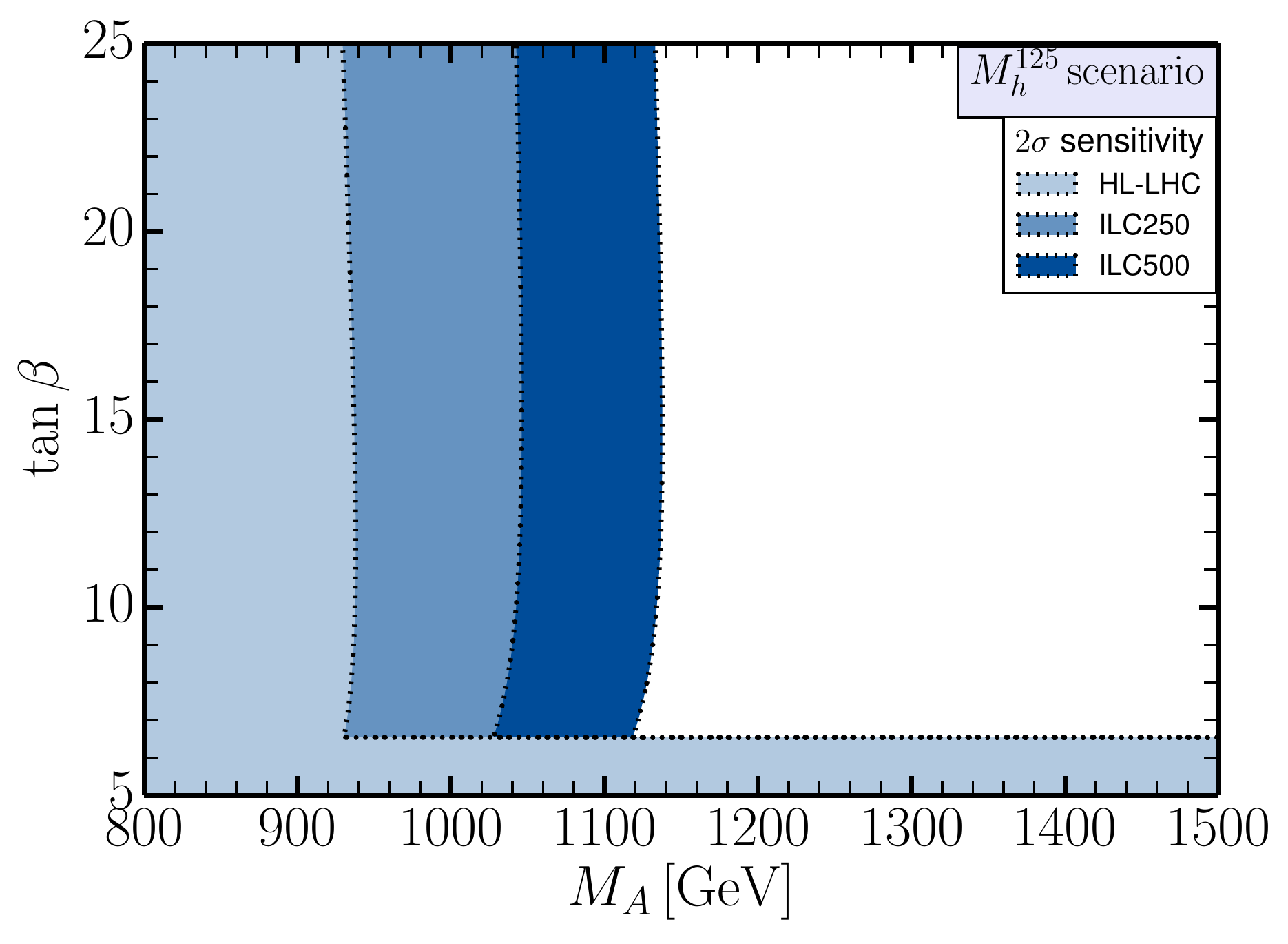}\hfill
\includegraphics[width=0.5\textwidth]{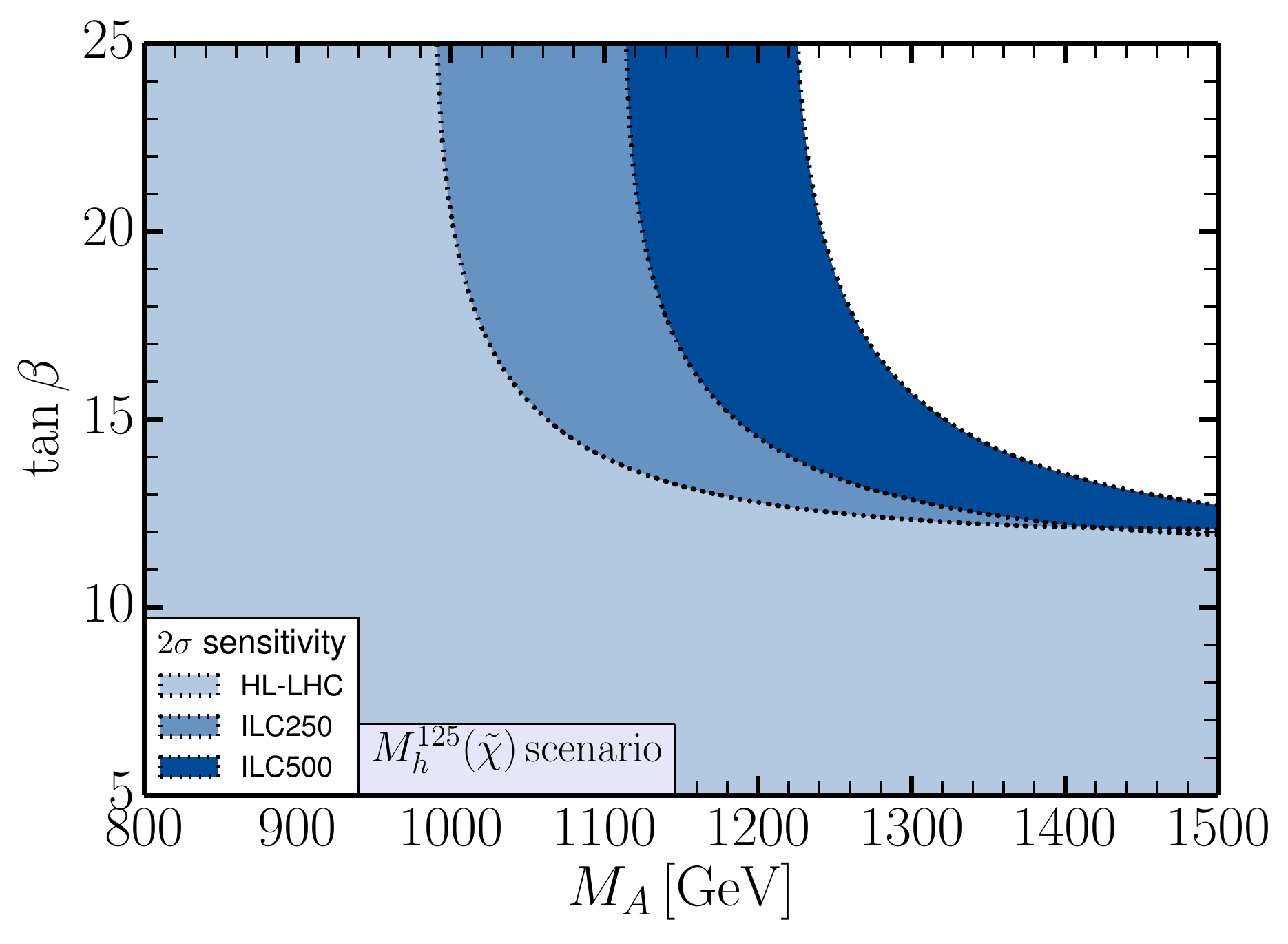}\\
\includegraphics[width=0.5\textwidth]{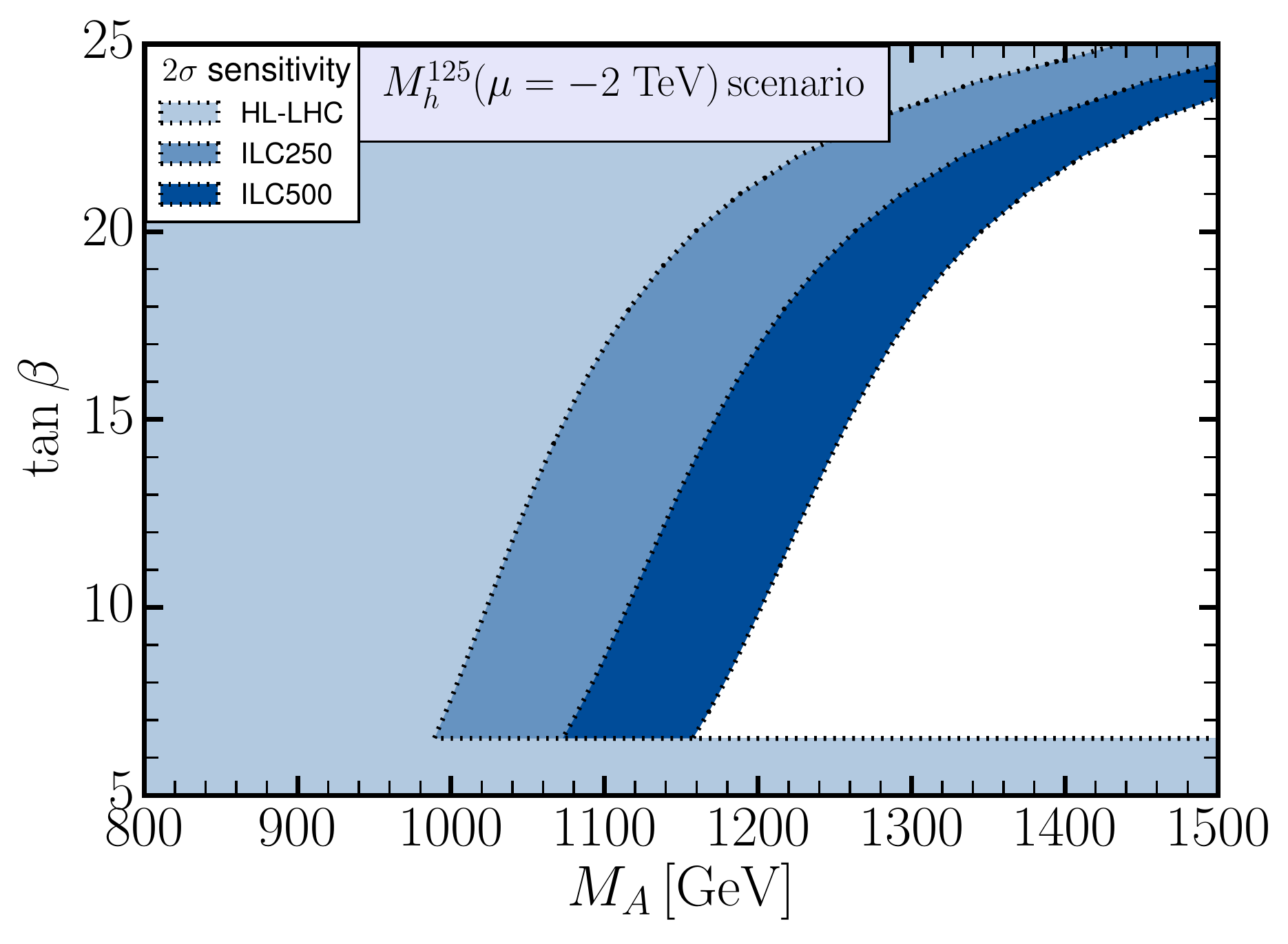}
\end{center}
\caption{Prospective $2\sigma$ indirect exclusion region
from Higgs rate
  measurements at the HL-LHC and the ILC, assuming agreement with the SM
  predictions, in the $M_h^{125}$ scenario \how{upper left panel},
the $M_h^{125}(\tilde\chi)$ scenario \how{upper right panel}, and the
$M_h^{125,\mu-}$ scenario \how{bottom panel}.}
\label{fig:HLLHCILCexclusion}
\end{figure}

In \fig{fig:HLLHCILCexclusion} we present the indirect $2\sigma$
expected exclusion regions
through precision Higgs-boson rate measurements that are obtained
under the assumption that the measured rates exactly agree with the SM
predictions. The results are given
in the $M_h^{125}$ scenario \how{upper left panel}, the
$M_h^{125}(\tilde\chi)$ scenario \how{upper right panel} and the
$M_h^{125, \mu-}$ scenario \how{lower panel}. We show the results for the three considered future scenarios: In light blue the HL-LHC projection is displayed,
corresponding to the results already shown in
Figs.~\ref{fig:bench1}, \ref{fig:bench2} and \ref{fig:bench5}.
In medium blue the
projection corresponding to the combination of the HL-LHC and
ILC250 measurements is shown, while the dark blue
shading indicates the case where in addition also
ILC500 measurements are included
(see Sec.~\ref{sec:expinput}).

In the first two scenarios, $M_h^{125}$ and $M_h^{125}(\tilde\chi)$, the
sensitivity at large \tb\ is determined by the decoupling behavior
with $M_A$, resulting in roughly
vertical exclusion lines for $\tan\beta \gtrsim 25$
(not explicitly shown in \fig{fig:HLLHCILCexclusion}). In this
large-\tb\ region the HL-LHC will already probe masses of the heavy Higgs
bosons far in the decoupling regime, $\mA\gtrsim 920\,\GeV$ and
$1000\,\GeV$ for the $M_h^{125}$ and $M_h^{125}(\tilde\chi)$ scenario,
respectively. The ILC measurements at ILC250 and ILC500 will be able to extend
the HL-LHC reach in $\mA$ by around $+(110$--$125)\,\GeV$ and
$+(200$--$235)\,\GeV$, respectively.
In the $M_h^{125}$ scenario, due to the absence of light SUSY particles,
this lower bound on $M_A$ roughly remains the same for lower
\tb\ values.
These indirect constraints on $M_A$ are complementary to the sensitivity
of the direct searches discussed in \fig{fig:bench1}, which depend on the
details of the decay patterns of the heavy Higgs bosons. The indirect
constraints from the rate measurements can potentially exceed the direct
search sensitivity for heavy Higgs bosons in the lower-\tb\ region.
In the $M_h^{125}$ scenario
this parameter space is largely covered by the indirect
constraint from confronting the prediction for the mass of the light Higgs
boson with the measured value
(see also \fig{fig:bench1}), but as discussed above this kind of constraint is
highly sensitive to the details of the considered benchmark scenario.

The result for the $M_h^{125}(\tilde\chi)$ scenario \how{upper right panel
of \fig{fig:HLLHCILCexclusion}} shows that there is additional sensitivity in the
low-\tb\ region in comparison to the case of the $M_h^{125}$ scenario.
As already noted in the discussion of \fig{fig:bench2}, the presence of light
charginos in the $M_h^{125}(\tilde\chi)$ scenario induces a shift in the
$h \to \gamma\gamma$ partial width. If the HL-LHC measurements of the
$\gamma\gamma$ rate of the Higgs boson at $125~\GeV$ yield exactly the SM
value, a lower limit of $\tb\gtrsim 12$ could be obtained almost independently of
$M_A$. Since the $h \to \gamma\gamma$ channel is not expected to be
significantly improved at the ILC, the measurements at ILC250 and ILC500
would only have a minor impact on this lower bound on $\tb$ in the
$M_h^{125}(\tilde\chi)$ scenario.
For higher values of $\tb$ the ILC measurements would improve the indirect
sensitivity on $M_A$ by about $200~\GeV$, similarly to the case of the
$M_h^{125}$ scenario.

In the $M_h^{125, \mu-}$ scenario the situation is qualitatively different
than in the other two displayed benchmark scenarios, since in this case
the projected exclusion becomes stronger for increasing \tb\ values.
Below $\tb \sim 20$ the improvement of the expected lower $M_A$ limit
achieved by the ILC precision is similar as
in the other two scenarios. For $\tan\beta \gtrsim 20$
the indirect sensitivity up to rather high values of $M_A$
arises from the strong enhancement of the
bottom-quark Yukawa coupling, which in turn influences all Higgs decay
rates as the $h\to b\bar{b}$ partial width dominates the total decay
width.
In this region the main impact on constraining the parameter space at the
HL-LHC arises from the $h\to \gamma\gamma$ rate measurements, since the
enhancement of the bottom-quark Yukawa coupling and the corresponding
increase in the total width diminishes
the branching ratio into $\gamma\gamma$. The ILC measurements would add
complementary information from the other branching ratios and from the
production cross sections of the Higgs signal at $125~\GeV$.

It should be noted that the displayed results in the scenarios with negative
$\mu$, see Figs.~\ref{fig:bench5} and~\ref{fig:HLLHCILCexclusion},
can also be interpreted in the following way:
the Higgs-rate measurements in the scenarios with negative $\mu$,
as a consequence of their dependence on potentially large $\Delta_b$
corrections,
provide sensitivity to set an {\em upper\/} bound on $|\mu|$ (assuming that
$\mu < 0$ holds) depending on $\mA$ and $\tan\beta$.
As an example, measuring SM Higgs rates at the HL-LHC would exclude
the point $M_A=1300~\GeV$ and $\tan\beta=20$ for $\mu = -3~\TeV$ but
not for $\mu = -2~\TeV$.
While this sensitivity is restricted to a certain range of $\mA$ (see the
discussion of Fig.~\ref{fig:bench5}) and relies on the assumption
$\mu < 0$, it could nevertheless be of interest in the context of future
direct searches for supersymmetric particles.
We come back to the discussion of the influence of the bottom-quark
Yukawa coupling enhancement on the Higgs rates in the next subsection,
however at a lower value of $\tan\beta$, where the effect is not
as pronounced, see \fig{fig:Mh125m_fit} below.

As stressed above, the analysis within specific benchmark scenarios,
where besides $M_A$ and $\tb$ all other SUSY parameters are fixed to specific
values by definition, and the assumption used in this section
that the detected rates exactly agree with
the SM predictions, cannot demonstrate the full potential of the precision
measurements of the Higgs boson rates at the HL-LHC and the ILC.
In particular, assuming that the underlying nature of the probed physics
scenario is the MSSM implies very important correlations between the couplings,
the production and decay rates of the Higgs boson at $125~\GeV$. Thus, a
precise measurement of only few observables is already sufficient to
determine the decoupling behavior if the underlying structure of the physics
scenario is assumed to be known. Instead, if no such assumption is made the
full breadth of precision measurements at the HL-LHC and the ILC will be
crucial in order to either determine the nature of observed patterns of
deviations from the SM or to set constraints on wide classes of possible
extensions or alternatives to the SM.
Specifically, the model-independent measurement of the total Higgs
production cross section of $e^+e^- \to Zh$ at the ILC (which allows a
model-independent determination of the Higgs branching ratios and
provides a robust method for obtaining the total width with high precision)
has no direct impact in our benchmark scenarios, while it is crucial for
probing models that both allow for additional Higgs boson decay mode(s)
as well as a compensating enhancement of the production
rates~\cite{Bechtle:2014ewa}.

Furthermore, the assumption that all future HL-LHC and ILC
measurements of Higgs production and decay modes will yield \emph{exact
agreement} with the SM predictions is of course not realistic. Even in the
absence of any contribution of new physics one would still expect that the
measured values are scattered around the SM values according to their
statistical
uncertainty. In order to answer the question whether the actually observed
pattern of measurements with a certain amount of data
hints at a non-zero deviation from the SM or not, the
full set of observables accessible at the HL-LHC and the ILC measured with
the highest possible accuracy will be instrumental.


\subsection{Impact of the rate measurements for the case where a particular
MSSM scenario is realized}
\label{sec:MSSMrealization}

We now take a different perspective and assume that future
precision Higgs-boson rate measurements reveal deviations from the SM
prediction, which are compatible with MSSM predictions for the
Higgs-boson rates.
Specifically, we assume that a certain parameter point within the
considered MSSM benchmark scenario is realized and set the central values of all projected
HL-LHC and ILC measurements to the predicted rates at this parameter
point (while keeping the same relative uncertainty as for the SM case). We
then again analyze how well the MSSM parameter space can be indirectly
constrained from precision measurements of the rates of the
Higgs boson at $125\,\GeV$.\footnote{An earlier analysis of this type can
be found in Ref.~\cite{Desch:2004cu}, see also Ref.~\cite{Weiglein:2004hn}.}
We first assume the realization of MSSM parameter points at a moderate
value of $\tb$ that are expected to
elude a direct $5\sigma$ discovery in heavy Higgs boson
searches. The chosen points lie close to the border of the $2\sigma$
sensitivity of HL-LHC direct searches in the $\tau^+\tau^-$ channel.
Furthermore we briefly discuss the case that a relatively large value of
$\tb$ could be realized in nature. In light of the existing search
limits this would mean that the associated value of $M_A$ has to be
significantly higher, and correspondingly the impact of such a scenario on
the rates of the Higgs boson at $125\,\GeV$ is expected to be rather small.

\medskip

We start with the $M_h^{125}$~scenario in \fig{fig:Mh125_fit},
where we assume in the left panels $(M_A,\tb) = (700\,\GeV,8)$ and in the
right panels $(M_A,\tb) = (1000\,\GeV, 8)$ being realized in nature.
In fact, both points will be probed at the $\gtrsim 2\sigma$ level by
heavy Higgs searches in the $\tau^+\tau^-$
final state at the HL-LHC, see \fig{fig:bench1}, which would thus
give complementary information.
Here we want to focus on the indirect constraints on the parameters
that can be obtained from the precision
rate measurements of the Higgs boson at $125\,\GeV$.
In the upper panels of \fig{fig:Mh125_fit} we show  the parameter regions which are
preferred at the $2\sigma$ level by the HL-LHC alone \how{faint red}, and in
combination with the ILC250 measurements \how{medium red}
and the ILC500 measurements \how{dark red}.
The lower panels of \fig{fig:Mh125_fit} show contours
for $R^{Vh}_{bb}$, i.e.~the Higgs boson signal rate for $pp\to Vh$ ($V=W^\pm,Z$)
production, followed by the decay $h\to b\bar{b}$, and normalized to
its SM prediction (at the same Higgs boson mass). In the considered
parameter space $R^{Vh}_{bb}$ is very similar to the SM-normalized rates
of the ILC processes $e^+e^- \to Zh, h\to b\bar{b}$ and $e^+e^- \to
\nu\bar{\nu} h, h\to b\bar{b}$,
as the $gg\to Zh$ contribution to $R^{Vh}_{bb}$ is
subdominant, and the rates of the other production processes
contributing to $R^{Vh}_{bb}$ scale
uniformly with the squared $hVV$ coupling.
This signal process is the most sensitive search channel for the
$b\bar{b}$ final state, with an anticipated HL-LHC (ATLAS and CMS
combined) precision
for the signal strength of around $4.6\%$ in the $pp\to Zh$ production
mode~\cite{Cepeda:2019klc}.

\begin{figure}[tb!]
\begin{center}
\includegraphics[width=0.5\textwidth]{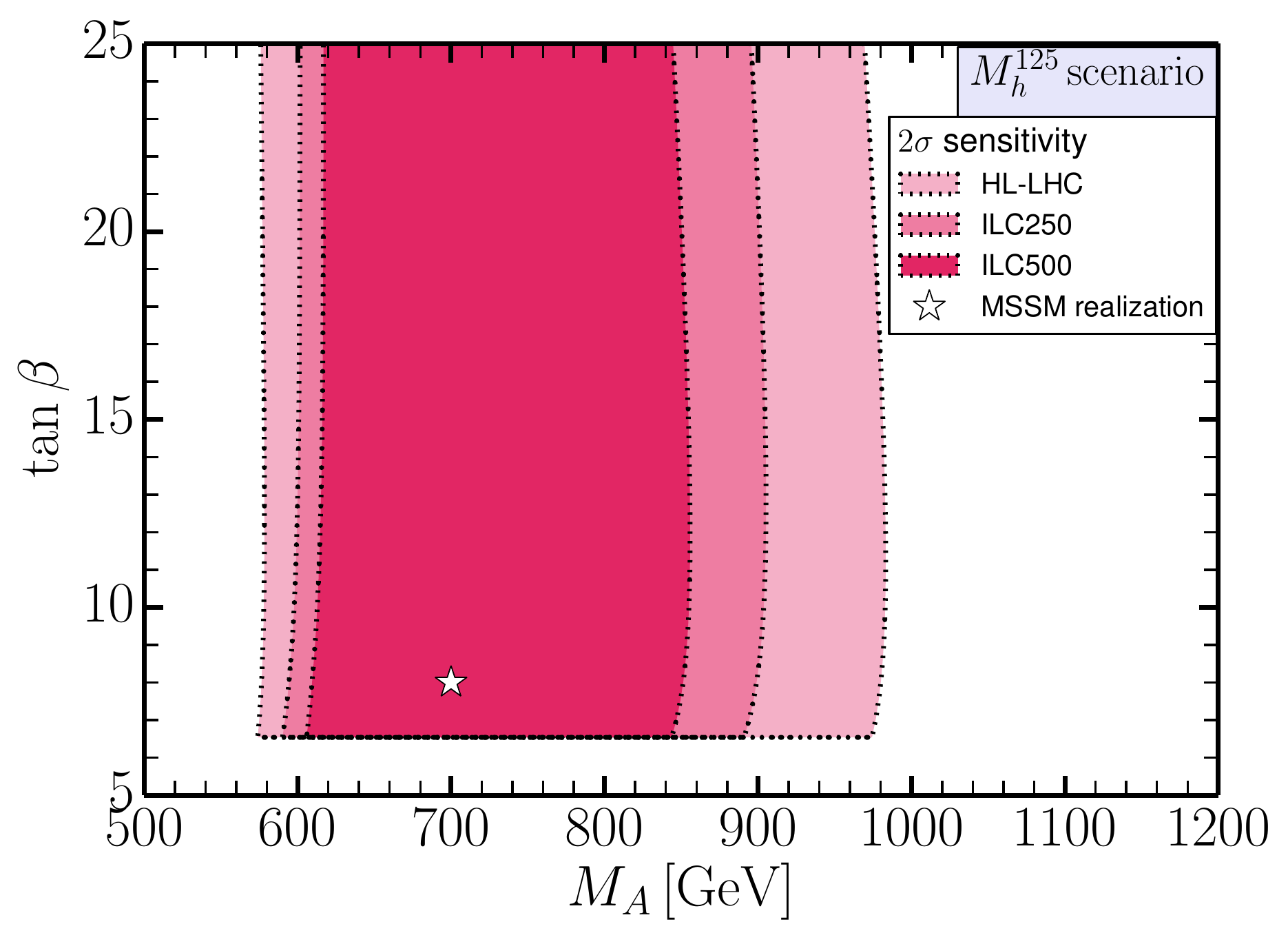}\hfill
\includegraphics[width=0.5\textwidth]{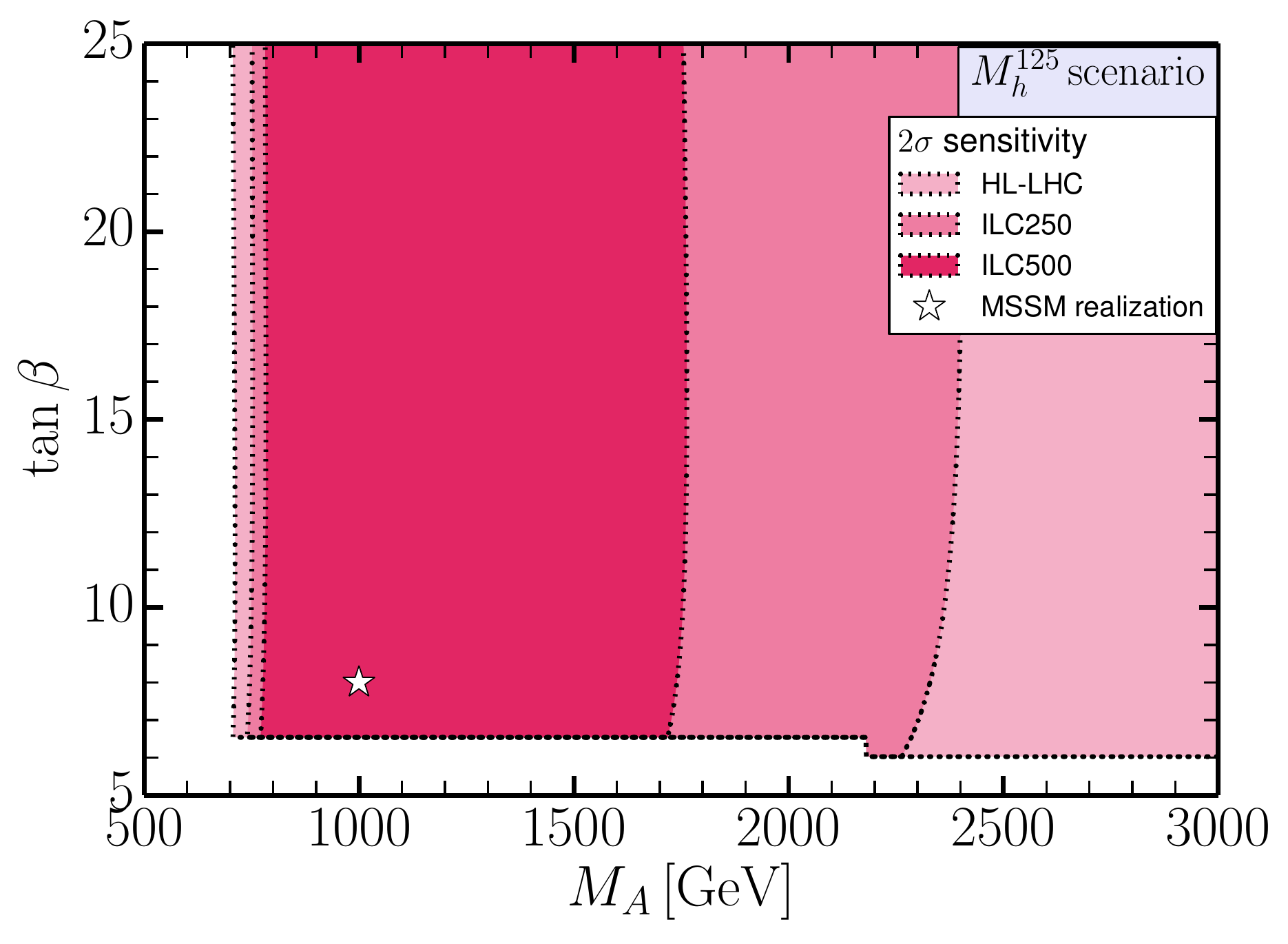}\\
\includegraphics[width=0.5\textwidth]{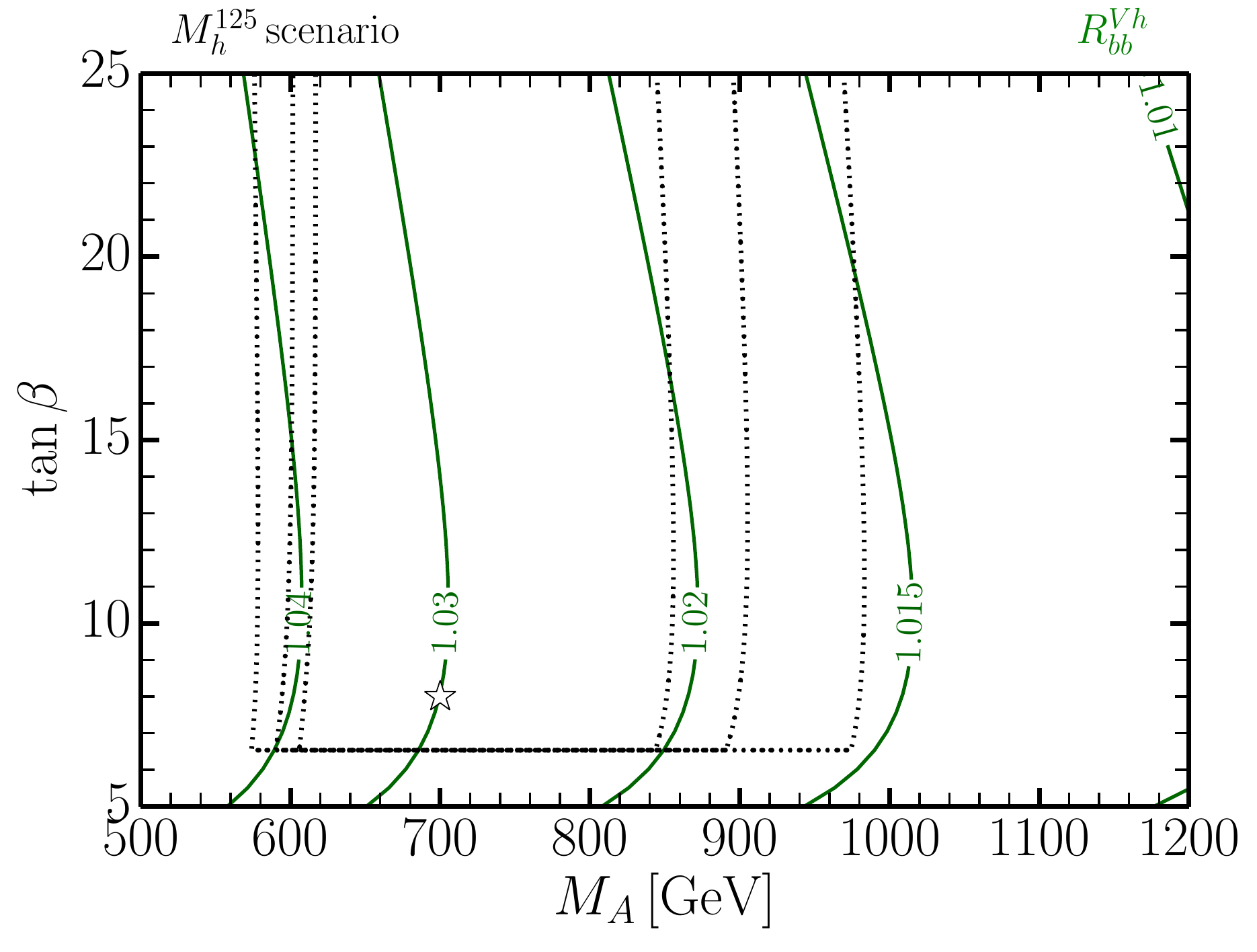}\hfill
\includegraphics[width=0.5\textwidth]{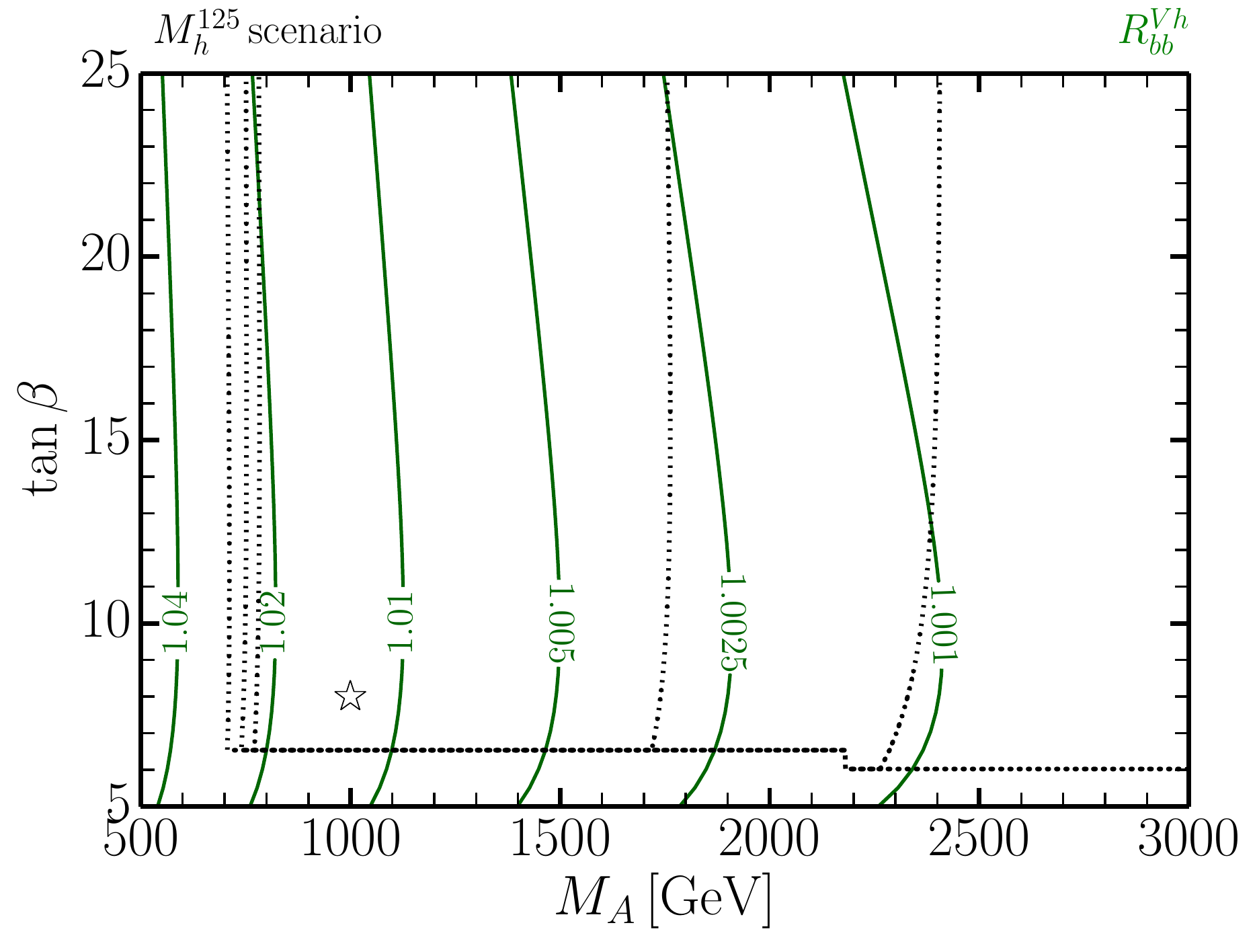}
\end{center}
\caption{\emph{Upper panels}: Indirect $2\sigma$ constraints
in the ($M_A$, $\tan\beta$) parameter plane of the $M_{h}^{125}$ scenario
  from prospective Higgs-boson signal-rate measurements at the HL-LHC
  \how{faint red} and in combination with ILC250
  \how{medium red} and ILC500 \how{bright red} measurements,
  assuming that the point, indicated by a star,
 $(M_A, \tan\beta) = (700\,\GeV, 8)$
  \how{left panel} or $(M_A, \tan\beta) =(1\,\TeV, 8)$  \how{right
      panel} is realized in nature.
\emph{Lower panels}: SM-normalized Higgs rate in the $pp\to Vh, h\to b\bar{b}$ channel, $R_{bb}^{Vh}$ (green contours), with the $2\sigma$ parameter ranges from the upper panels superimposed.}
\label{fig:Mh125_fit}
\end{figure}

As already discussed above, the Higgs rate measurements have almost no
sensitivity on \tb\ in the $M_h^{125}$~scenario (in the range $\tan\beta \gtrsim
6$ {where in this benchmark scenario the predicted value of the mass of
the light Higgs boson is compatible with the experimental value of the
detected Higgs signal). For the assumed point $(M_A,\tb) = (700\,\GeV,8)$
the heavy Higgs mass scale, $M_A$, could be constrained
by prospective Higgs rate measurements to
\begin{align*}
575\,\GeV \lesssim M_A  \lesssim 980 \,\GeV	&\qquad (\text{HL-LHC}),\\
600\,\GeV \lesssim M_A  \lesssim 900 \,\GeV	&\qquad (\text{$+$ ILC250}),\\
615\,\GeV \lesssim M_A  \lesssim 850 \,\GeV	&\qquad (\text{$+$ ILC500}),
\end{align*}
at the $2\sigma$ level. In contrast, assuming the point closer to the decoupling limit, $(M_A,\tb) = (1000\,\GeV,8)$, the expected $2\sigma$ ranges are
\begin{align*}
700\,\GeV \lesssim M_A \qquad\qquad\quad\;\;	&\qquad (\text{HL-LHC}),\\
750\,\GeV \lesssim M_A  \lesssim 2400 \,\GeV	&\qquad (\text{$+$ ILC250}),\\
780\,\GeV \lesssim M_A  \lesssim 1750 \,\GeV	&\qquad (\text{$+$ ILC500}) .
\end{align*}
In this case the prospective accuracy of the signal strength measurements
at the HL-LHC is not sufficient to place an indirect upper bound on $M_A$ at
the $2\sigma$ level. The incorporation of the precision measurements at the ILC
would significantly improve the sensitivity of the Higgs rate measurements
for a distinction between the SM and the MSSM.
The possibility to establish an upper bound on $M_A$ via the rate
measurements of the Higgs state at $125\,\GeV$ at the HL-LHC and the ILC would
provide a clear target for the direct searches for additional Higgs bosons.

The bottom panels in \fig{fig:Mh125_fit} show that in the $2\sigma$-allowed region,
the deviations of $R^{Vh}_{bb}$ are $\lesssim 2\%$
for the case where $M_A = 1000\,\GeV$ has been assumed and
$\lesssim 4\%$ for the case with $M_A = 700\,\GeV$. Accordingly,
the future HL-LHC measurements of the $pp\to Zh$, $h\to b\bar{b}$
rate have only a limited sensitivity for constraining the parameter space in
the considered scenarios. On the other hand, the
inclusive rate measurements in the $h\to \gamma\gamma$, $h\to VV$ ($V=W^\pm,
Z$) and $h\to \tau^+\tau^-$ channels at the HL-LHC
have sensitivity to modifications of the
$hb\bar{b}$ coupling through the dependence of the involved branching ratio
on the total width of the Higgs boson at $125\,\GeV$.
For instance, near the assumed points
$(M_A,\tb) = (700\,\GeV,8)$ and $(1000\,\GeV, 8)$,
the Higgs-to-diphoton rate is suppressed by $7\%$ and $3\%$ with respect
to the SM prediction, respectively,
as a result of a slightly enhanced bottom-quark Yukawa coupling and
its impact on the total Higgs width.
The combination of the
measurements of the Higgs signal rates at the HL-LHC in various channels
involving the
product of the production cross sections and decay branching ratios will
therefore provide sensitive information on possible deviations from the SM,
while it will be non-trivial to disentangle the source of the deviations.
Concerning the prospective rate measurements at the ILC,
the most precise Higgs rate measurements will be performed in the
$e^+e^-\to Zh$, $h\to b\bar{b}$ channel during the run at $250\,\GeV$ and in the
$e^+e^-\to \nu\bar{\nu}h$, $h\to b\bar{b}$ channel in the $500\,\GeV$ run~\cite{deBlas:2019rxi}, each with a precision at the sub-percent
level.
The ILC measurements will therefore complement the information obtainable
at the HL-LHC with high-precision input on observables that cannot be well
exploited at the LHC. The ILC will furthermore provide model-independent
measurements of individual branching ratios. This kind of information will
be crucial in order to determine the source of
possible deviations without invoking model assumptions.
In order to investigate the underlying nature of detected deviations
from the SM, the indirect constraints that we have discussed here should of
course be applied in the context of the information that is available
from the direct searches for additional
Higgs bosons (see in particular \fig{fig:bench1}) and other states of new
physics. The limits from these searches may in fact exclude large
regions of
the parameter space that is favored by the indirect constraints. Naturally,
in case of a significant excess (or more than one) in the direct searches
the prospects for narrowing down the possible nature of new physics with
the combination of direct and indirect information would of course much
improved.

\begin{figure}[t!]
\begin{center}
\includegraphics[width=0.5\textwidth]{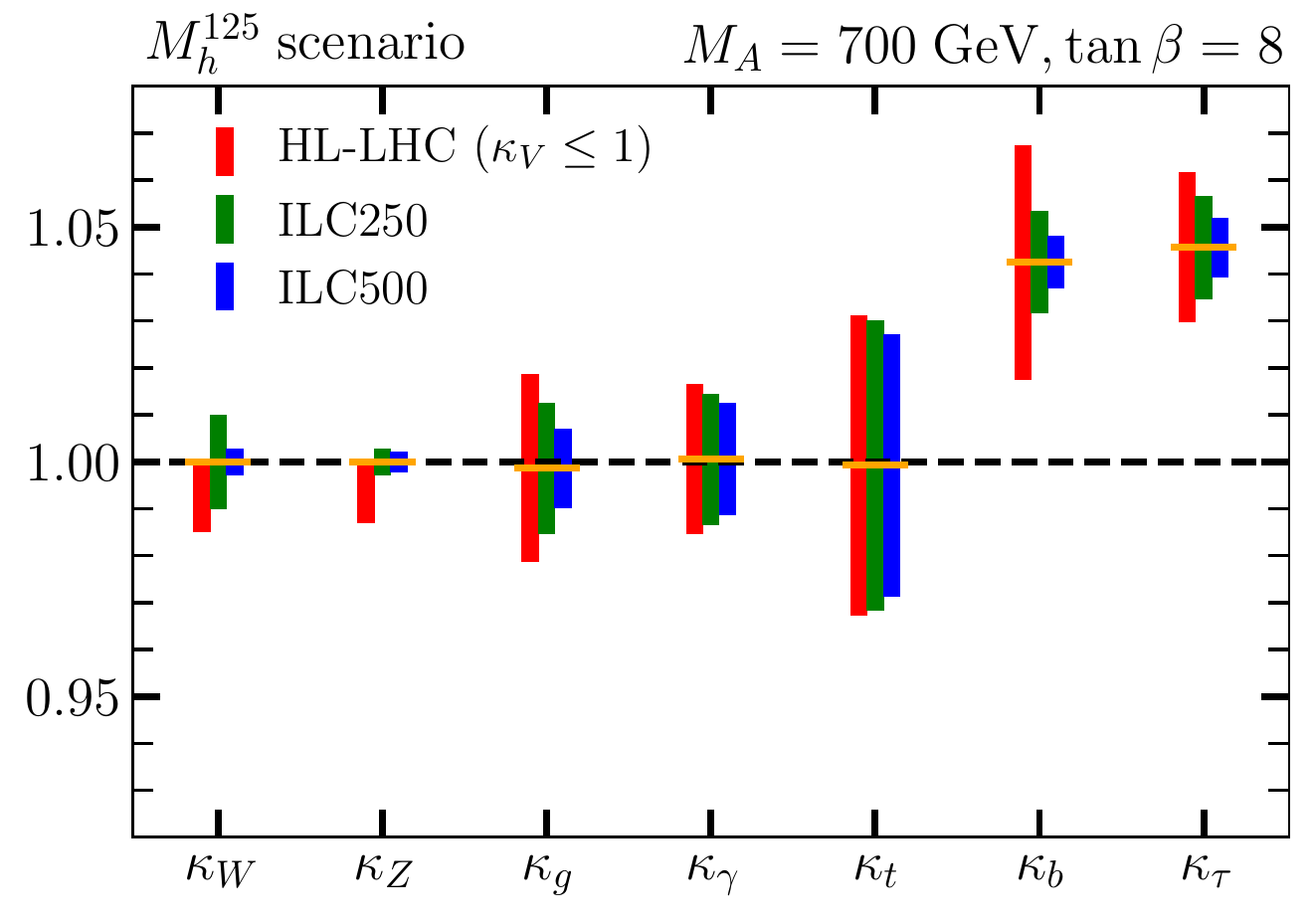}\hfill
\includegraphics[width=0.5\textwidth]{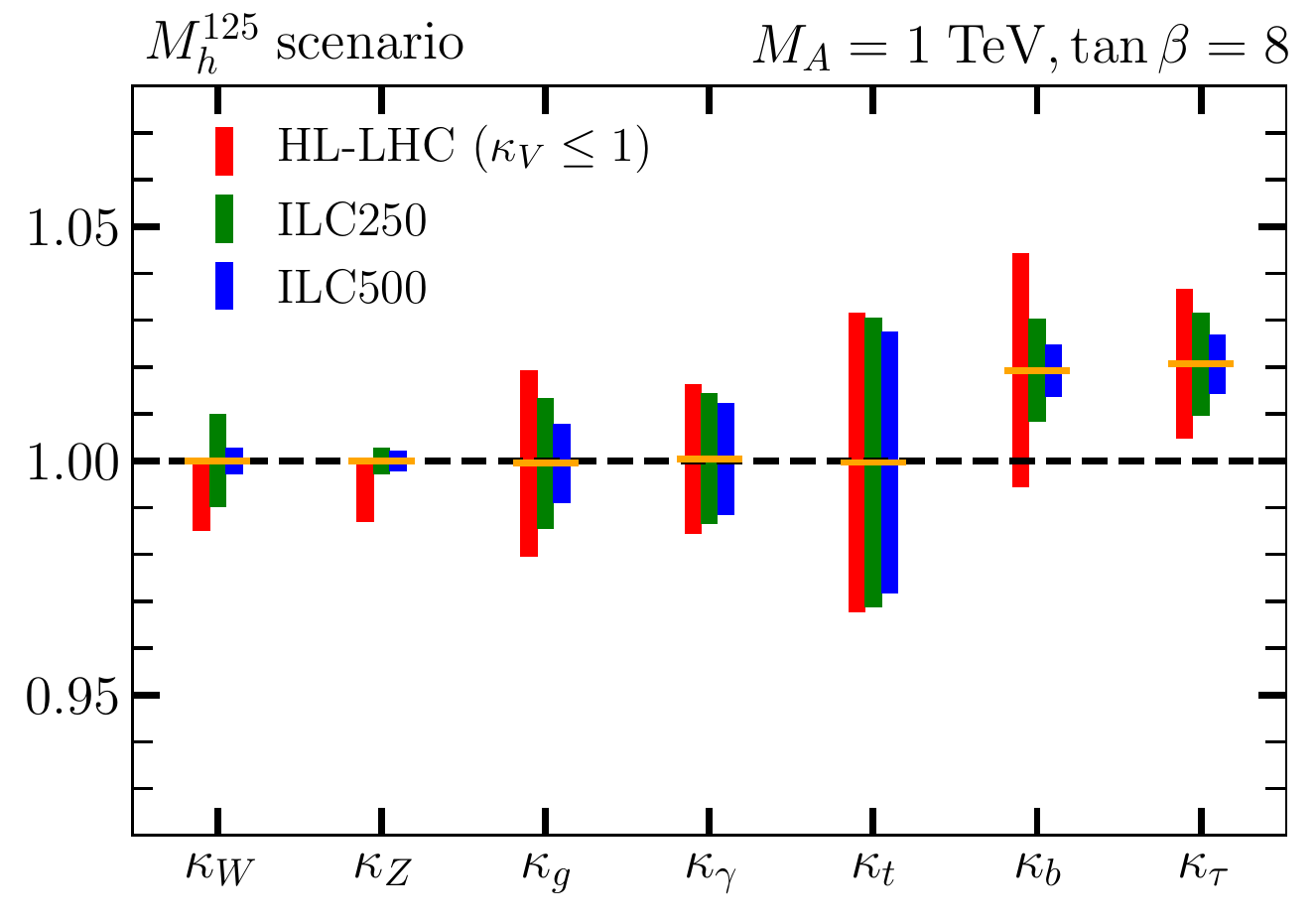}
\end{center}
\caption{\emph{W\"ascheleinen-plots} for the two assumed MSSM parameter
  points $(M_A, \tan\beta) = (700\,\GeV, 8)$ \how{left} and $(M_A,
  \tan\beta) =(1\,\TeV, 8)$  \how{right} in the $M_h^{125}$ scenario:
  Predicted Higgs couplings in the $\kappa$ framework
  \how{orange horizontal bars} along with the anticipated $1\sigma$
  precision from a global fit~\cite{deBlas:2019rxi} to Higgs rate
  measurements at the HL-LHC, where the theoretical assumption
  $\kappa_V \leq 1$ is employed, and including prospective measurements at
ILC250 and ILC500 (but without imposing an assumption on $\kappa_V$).
}
\label{fig:WaescheleineMh125}
\end{figure}

The pattern of the deviations from the SM predictions corresponding to
the situation where the parameter point $(M_A,\tb) = (700\,\GeV,8)$
or the point $(M_A,\tb) = (1000\,\GeV, 8)$
of the $M_h^{125}$~scenario is realized in nature is shown in
\fig{fig:WaescheleineMh125}. The displayed plots, which we denote as
\emph{``W\"ascheleinen-plots''} (washing line plots)
in the following, show
the predicted light Higgs couplings (normalized to the SM prediction) at
the assumed MSSM points in the $\kappa$
framework~\cite{deFlorian:2016spz}, along with the anticipated $1\sigma$
precision of a rather general $\kappa$
determination~\cite{deBlas:2019rxi} from prospective Higgs rate
measurements at the HL-LHC and the ILC.\footnote{More specifically, we
  show the expected precision of the $\kappa$ parameters in the
  kappa-3 scenario~\cite{deBlas:2019rxi}.}
It should be noted that the coupling determination at the HL-LHC is based
on the theoretical assumption $\kappa_V \leq 1$, while no such assumption is
needed for the coupling determinations at the ILC.
The plots in \fig{fig:WaescheleineMh125} show that for the assumed parameter
points sizable deviations from the SM only occur for
the bottom-quark and
tau-lepton Yukawa couplings, represented by $\kappa_b$ and
$\kappa_\tau$, respectively. At the HL-LHC the precision of the
$\kappa_b$ and $\kappa_\tau$ determination is at the $2\%$ and $1.5\%$
level, respectively.
One can see that in particular for the assumed point $(M_A,\tb) =
(1000\,\GeV,8)$ the ILC accuracy will be crucial in order to experimentally
establish a significant deviation of the rates of the Higgs boson at
$125\,\GeV$ from the SM predictions.
Moreover, the ILC measurements will provide a crucial consistency test of
the correlations between couplings, for instance, between $\kappa_b$ and
$\kappa_\tau$, and will therefore help to further discriminate between
different BSM interpretations.

\begin{figure}[tb!]
\begin{center}
\includegraphics[width=0.5\textwidth]{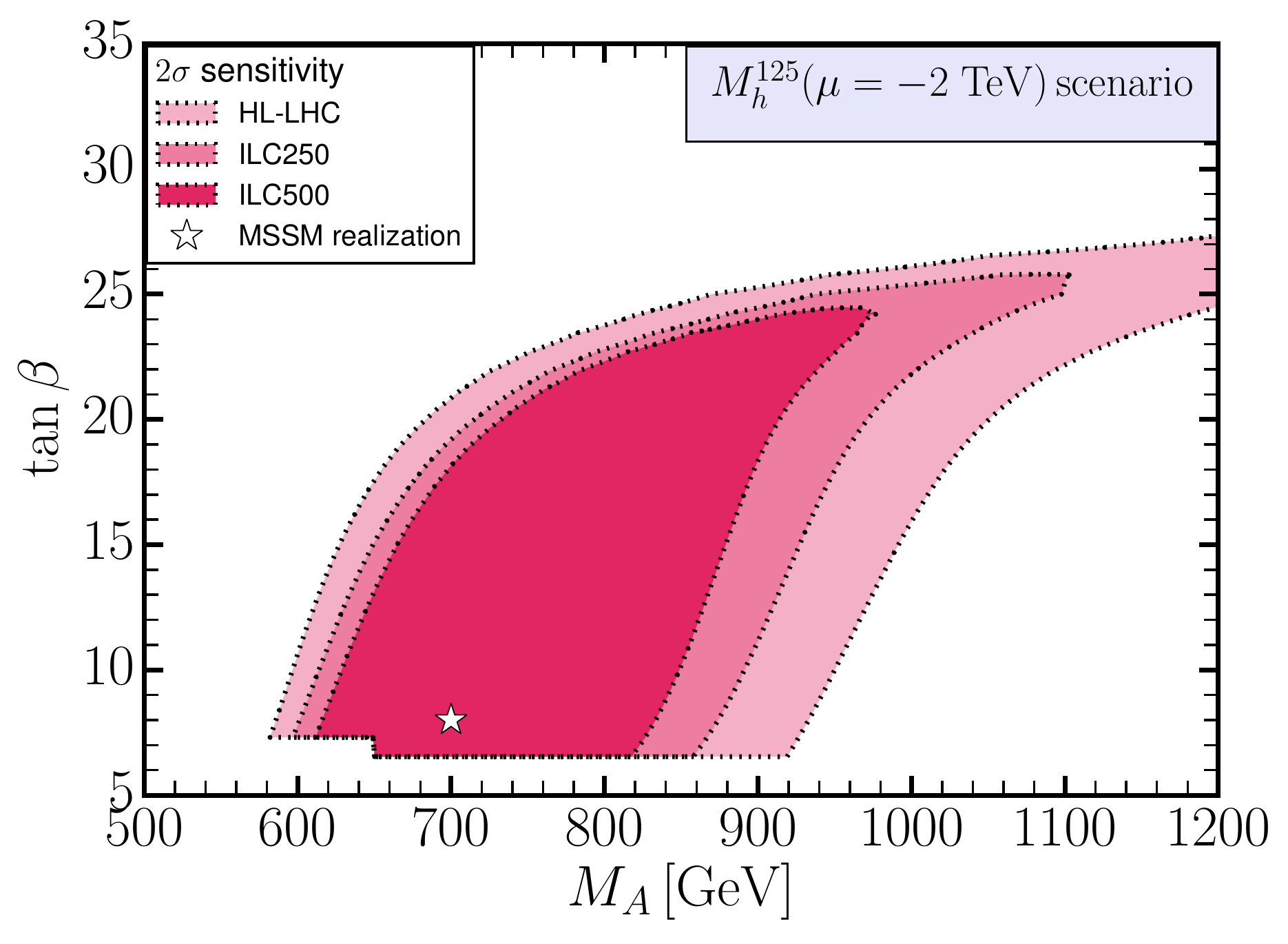}\hfill
\includegraphics[width=0.5\textwidth]{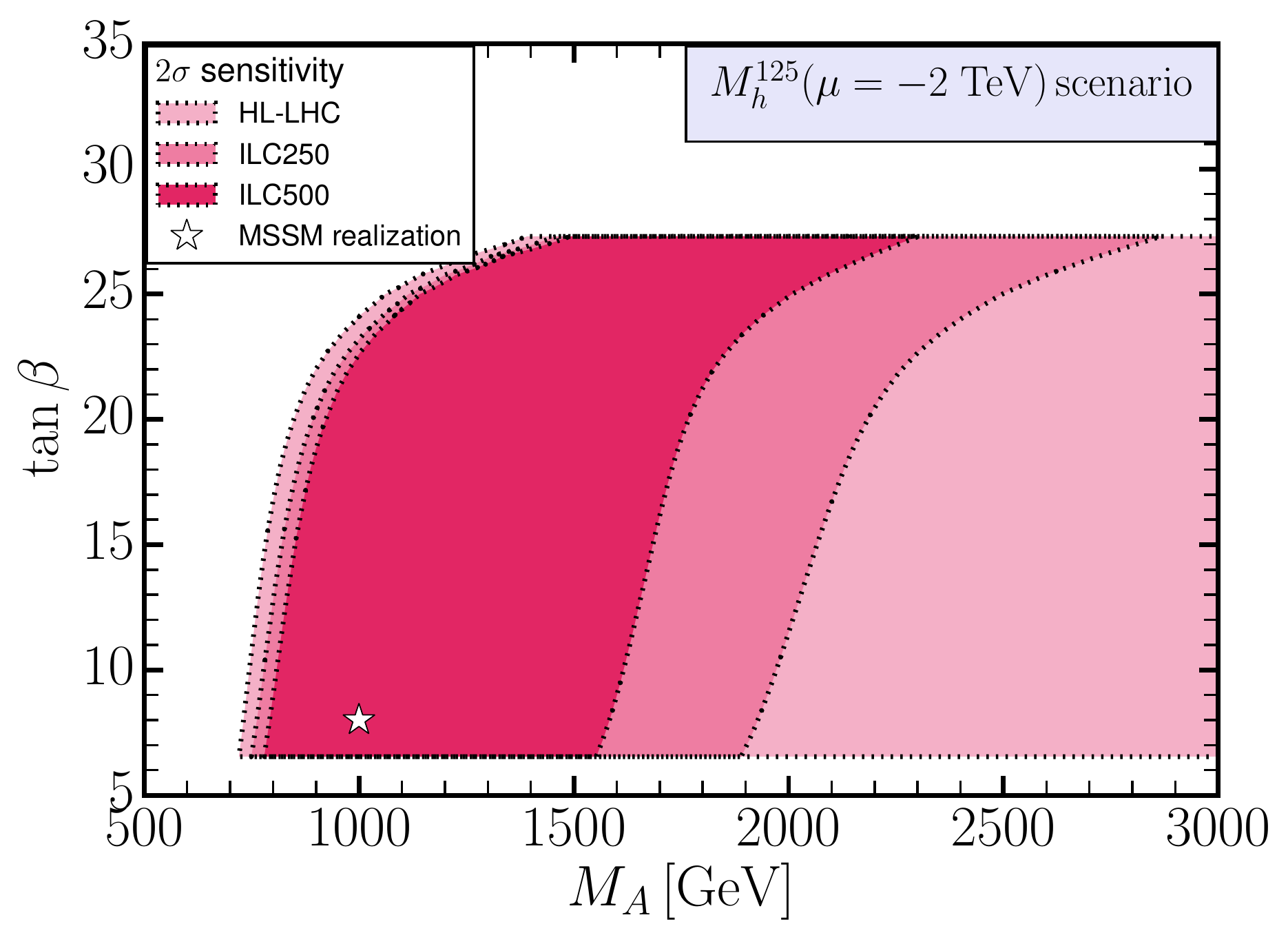}\\
\includegraphics[width=0.5\textwidth]{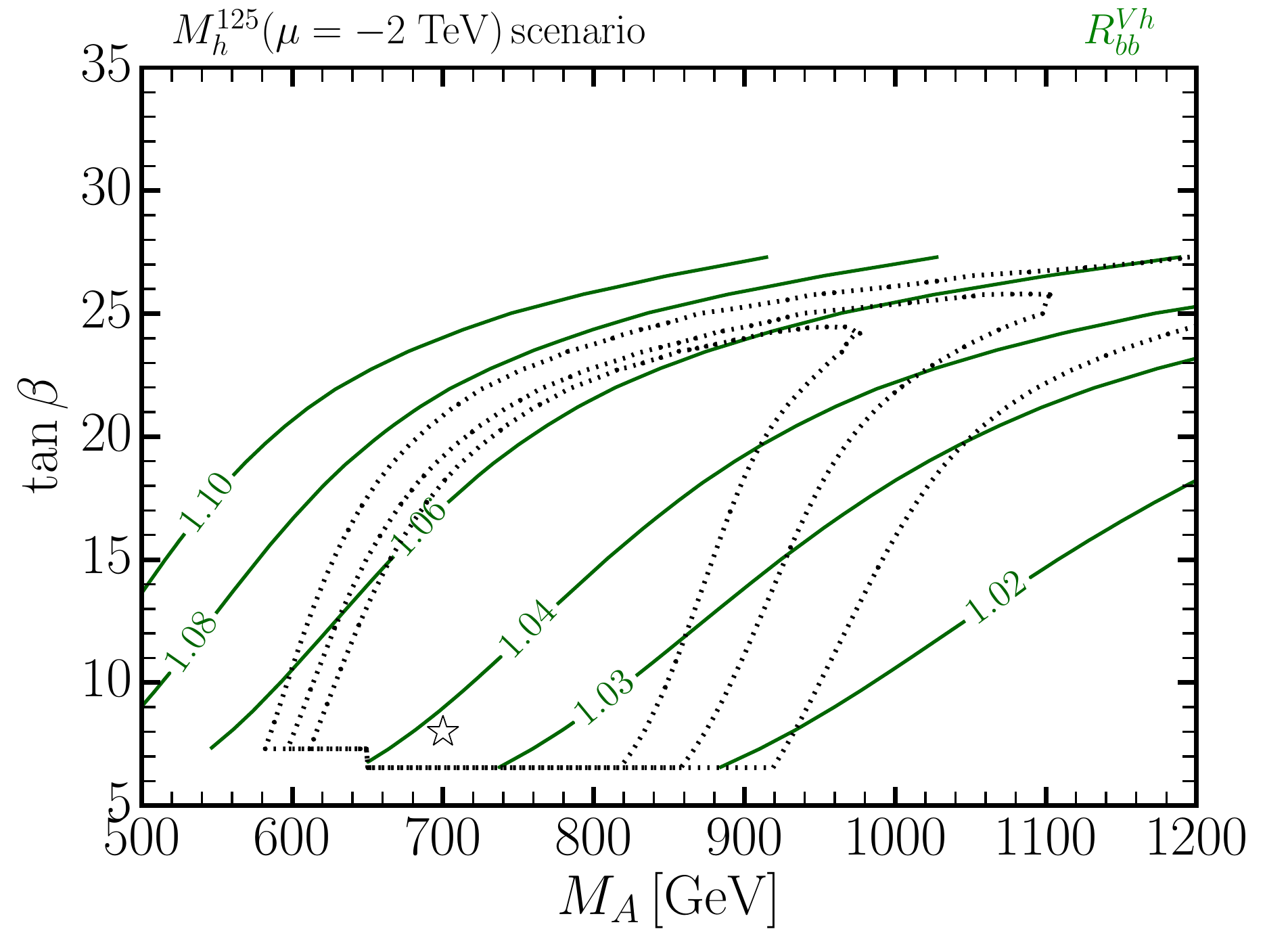}\hfill
\includegraphics[width=0.5\textwidth]{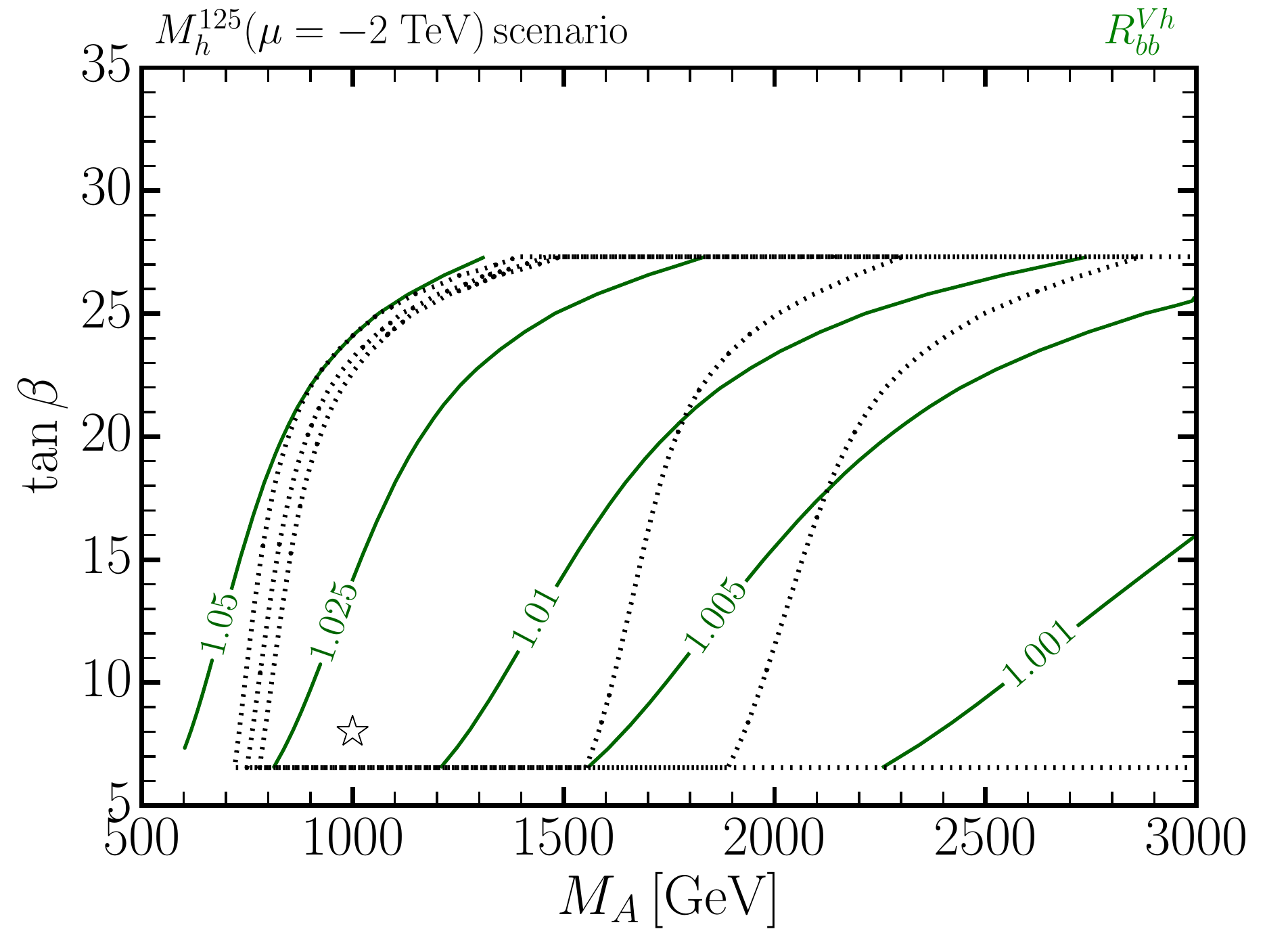}
\end{center}
\caption{Indirect $2\sigma$ constraints
in the ($M_A$, $\tan\beta$) parameter plane
from prospective Higgs-boson signal-rate measurements at the HL-LHC
and the ILC (upper row) and $R_{bb}^{Vh}$ contours
(lower row) in the $M_h^{125,\mu-}$ scenario, assuming that the point, indicated by a star, $(M_A,\tan\beta) = (700~\GeV,8)$ (left panels) or $(M_A,\tan\beta) = (1~\TeV,8)$ (right panels) is realized in nature.
The same color coding as in \fig{fig:Mh125_fit} is used.}
\label{fig:Mh125m_fit}
\end{figure}

The prospects for the indirect constraints from the Higgs rate
measurements at the HL-LHC and ILC for
the assumed parameter points $(M_A, \tb) = (700\,\GeV, 8)$ and
$(M_A, \tb) = (1000\,\GeV, 8)$ in the $M_h^{125,\mu-}$ scenario are
shown in the upper panels of \fig{fig:Mh125m_fit},
while the lower panels show
the corresponding predictions for the (SM-normalized) Higgs rate for the
$pp\to Vh$, $h\to b\bar{b}$ channel, $R_{bb}^{Vh}$.
The main effect of changing from
$\mu = +1\,\TeV$ (the value chosen in the previously discussed
$M_h^{125}$ scenario) to $\mu = -2\,\TeV$ is the non-trivial
\tb-dependence of the preferred parameter region. In the case of the
first assumed point $(M_A, \tb) = (700\,\GeV, 8)$ \how{left panels} the
Higgs rate measurements
at the HL-LHC (and including the measurements at ILC250 and ILC500)
yield an upper limit of $\tb \lesssim 27$
($\tb \lesssim 26$, $24.5$, respectively).
These upper limits become significantly stronger at smaller $M_A$
values.
On the other hand, upper bounds on $M_A$ from the Higgs rate measurements
can be inferred at the largest values of $\tb$, namely
$M_A \lesssim 1420\,\GeV$ at the HL-LHC,
$M_A \lesssim 1100\,\GeV$ including measurements at ILC250, and
$M_A \lesssim 980\,\GeV$ including measurements at ILC500.
For the plots showing the indirect constraints for the case of the
other assumed
parameter point $(M_A, \tb) = (1000\,\GeV, 8)$ \how{right panels},
there is a cutoff visible for
$M_A \gtrsim 1400\,\GeV$ that restricts the displayed values of \tb\ to
$\tb\lesssim 27$ (this feature is not visible in the left panels due to the
smaller displayed range of $M_A$). This cutoff is
not a consequence of the rate measurements but arises in this benchmark
scenario because of the
incompatibility of the light Higgs mass with the observed value, as
discussed above (see the right panel of \fig{fig:Mh125negmu}). The impact of
the Higgs rate measurements for constraining \tb\
can be seen in this case for $M_A \lesssim 1400\,\GeV$.
For the assumed parameter point
$(M_A, \tb) = (1000\,\GeV, 8)$ an indirect upper limit
on $M_A$ from the Higgs rate measurements
can only be obtained using the ILC measurements.
Here the measurements from later ILC stages (ILC500) will sharpen the
upper limits on $M_A$ obtained from using the measurements at
the initial ILC run (ILC250) by around
$400\;\GeV$, independently of the \tb\ value.
For both assumed parameter points the incorporation of the Higgs rate
measurements at the ILC leads to a large reduction of the allowed parameter
space in the ($M_A$, $\tan\beta$) plane.

The pattern of the deviations in $R_{bb}^{Vh}$ in the preferred regions of
the ($M_A$, $\tan\beta$) plane \how{lower panels of \fig{fig:Mh125m_fit}}
is similar to the case of \fig{fig:Mh125_fit}. Thus, also in this case the
incorporation of the ILC measurements would lead to a much larger set of
observables showing sizable deviations from the SM. This kind of information
will be crucial to determine the underlying nature of the detected
deviations. As discussed above, those investigations should of course be
based on both the direct information from searches and the indirect
constraints.
For the $M_h^{125,\mu-}$ scenario large parts of the parameter
region that would be preferred by the prospective Higgs rate measurements
are within the $2\sigma$ reach of heavy Higgs searches in the
$\tau^+\tau^-$ and possibly even $b\bar{b}$ final states at the HL-LHC,
see \fig{fig:bench5}. A robust excess in these searches would provide
clues for the mass scale of the heavy Higgs bosons, $M_A$. The
$125\;\GeV$ Higgs rate measurements could then, together with first
potential measurements of the strength of such a heavy Higgs boson
signal, allow one to put new physics interpretations under scrutiny and, within
the considered scenario, lead to strongly improved constraints on the model
parameters.

\begin{figure}[tb!]
\begin{center}
\includegraphics[width=0.5\textwidth]{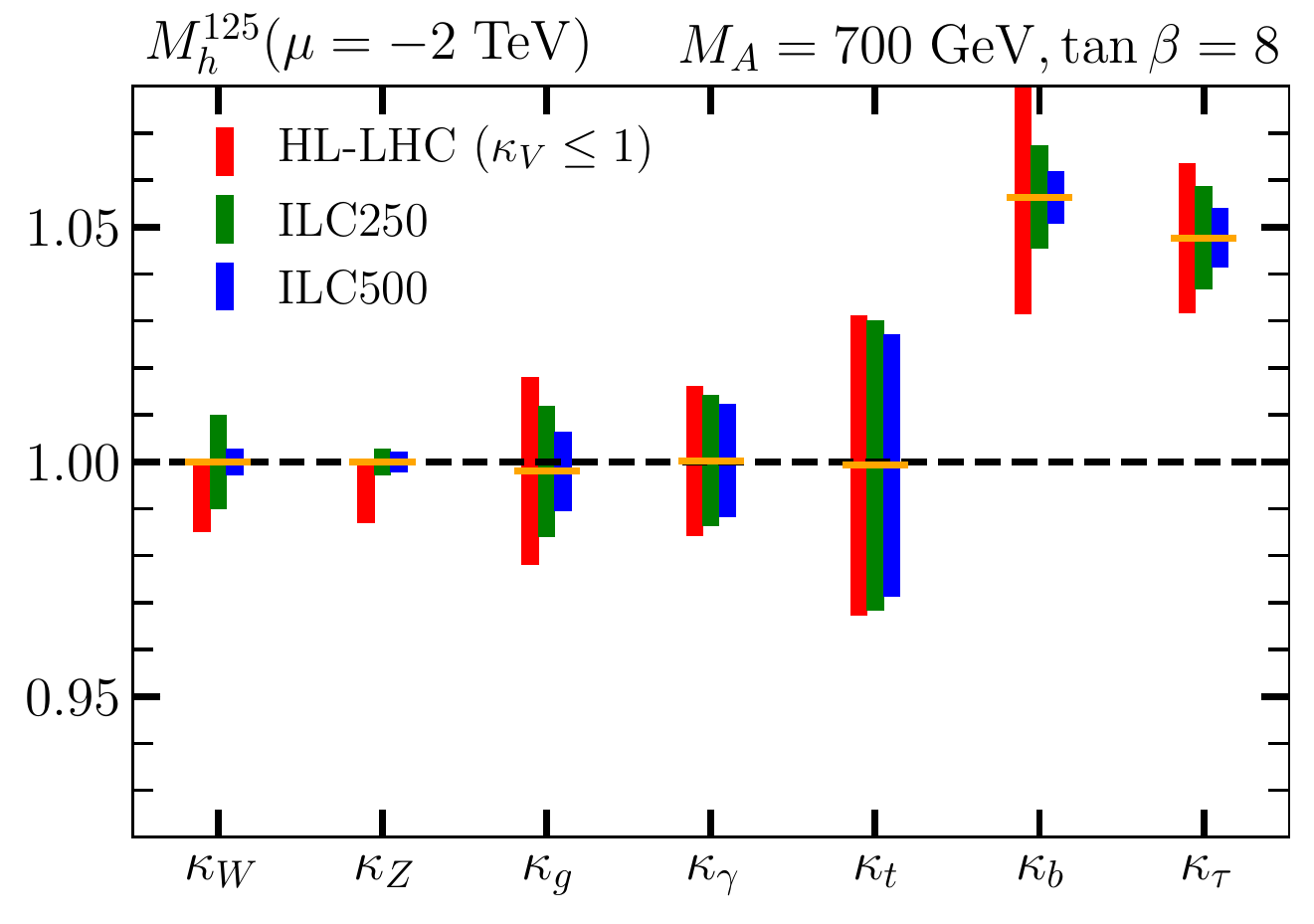}\hfill
\includegraphics[width=0.5\textwidth]{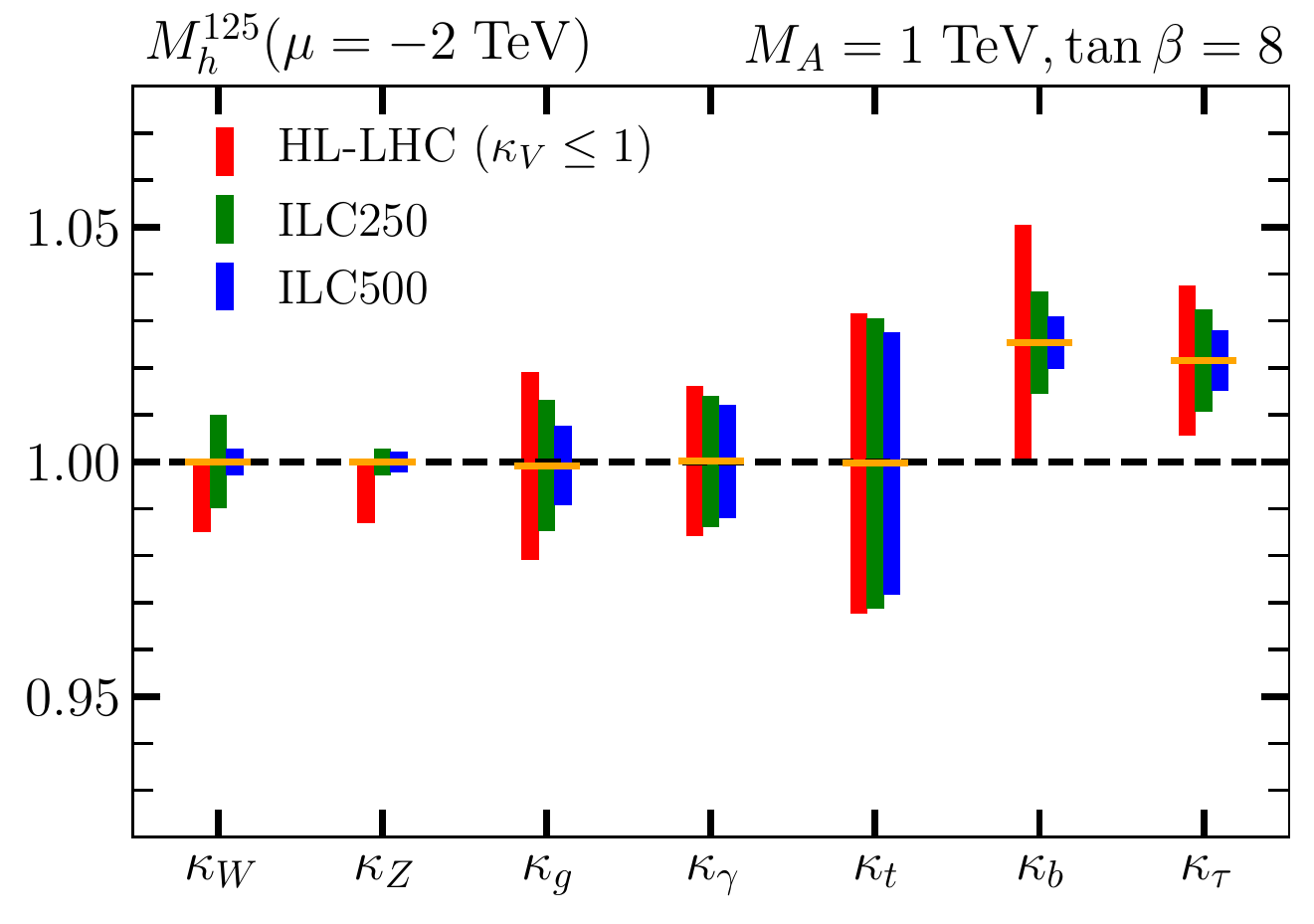} \\[.2em]
\includegraphics[width=0.65\textwidth]{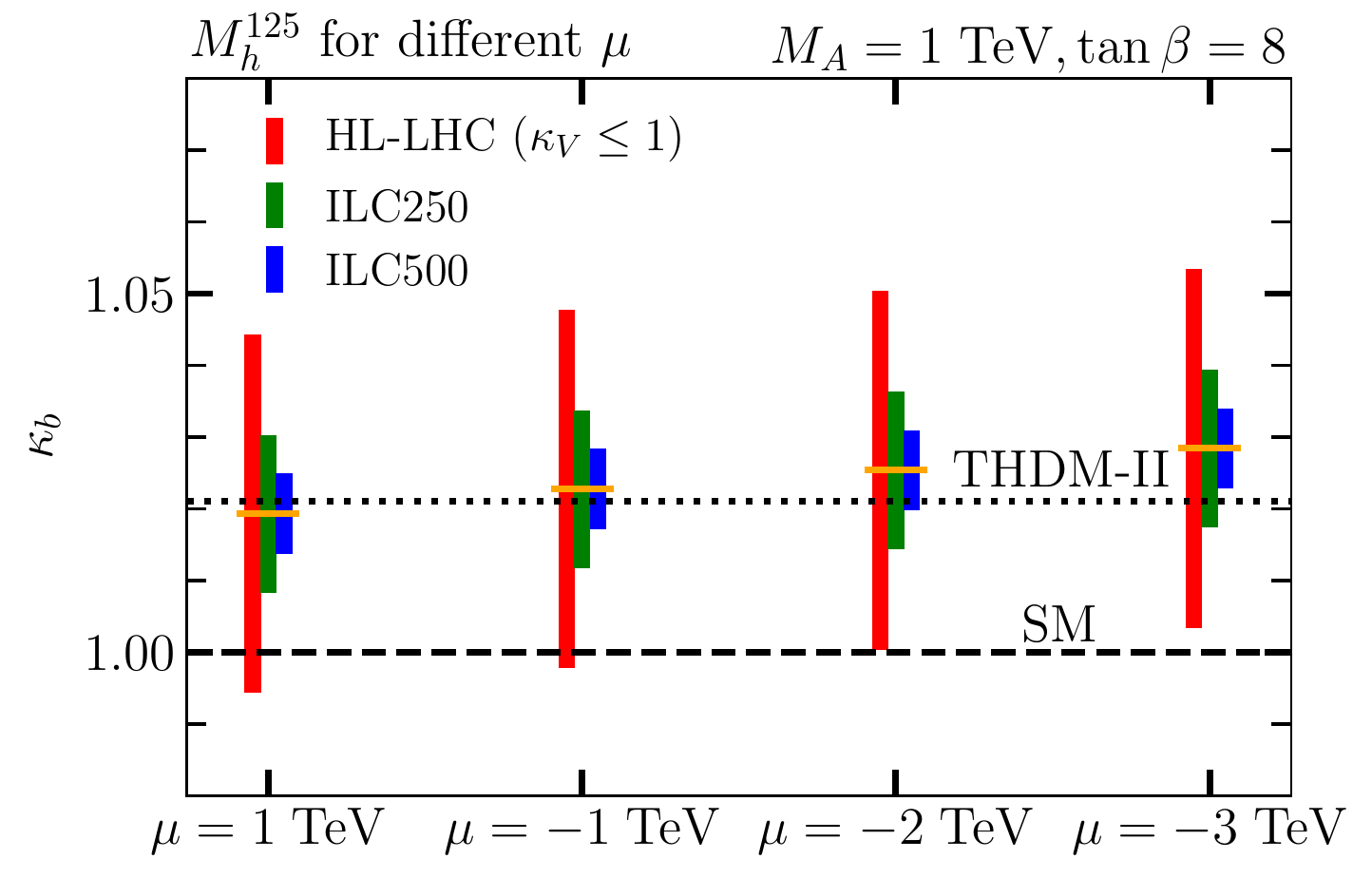}
\end{center}
\caption{\emph{W\"ascheleinen-plots},
using the the same color coding as
in \fig{fig:WaescheleineMh125},
for the two assumed MSSM parameter points $(M_A, \tan\beta) = (700\,\GeV, 8)$
\how{upper left panel} and $(M_A, \tan\beta) =(1\,\TeV, 8)$  \how{upper right
panel} in the $M_h^{125,\mu-}$ scenario.
The lower panel shows
for the assumed point $(M_A, \tb) = (1\,\TeV, 8)$ and different values of
$\mu$ the prospects for $\kappa_b$, where for comparison also the
corresponding prediction in the THDM-II (see text)
is indicated \how{dotted
line}, see text for details.
The Higgs couplings in the $\kappa$ framework predicted in the displayed
scenarios are compared with
the anticipated $1\sigma$ precision from Higgs rate
  measurements, where at the HL-LHC the theoretical assumption
  $\kappa_V \leq 1$ is employed, while for the results
including prospective measurements at
ILC250 and ILC500 no assumption on $\kappa_V$ is employed.}
\label{fig:WaescheleineMh125m}
\end{figure}

In \fig{fig:WaescheleineMh125m} we show
\emph{W\"ascheleinen-plots} for the parameter points $(M_A, \tb) =
(700\,\GeV, 8)$ \how{upper left panel} and $(M_A, \tb) = (1000\,\GeV, 8)$
\how{upper right panel} in the $M_h^{125,\mu-}$ scenario, i.e.~we
show the
predicted Higgs couplings represented by $\kappa$ scale factors in the
displayed scenarios along
with the prospective $1\sigma$ precision levels of their determination
from a global fit~\cite{deBlas:2019rxi} to Higgs rate
measurements. For the precisions from HL-LHC the theoretical assumption
$\kappa_V \leq 1$ is employed, while including prospective measurements at
ILC250 and ILC500 the assumption on $\kappa_V$ is dropped.
The lower panel shows the comparison between the predicted Higgs couplings
$\kappa_b$ for the assumed point $(M_A, \tb) = (1\,\TeV, 8)$ and
the anticipated experimental precisions for different values of
$\mu$. Besides the predictions for different MSSM parameters also the
prediction in the THDM-II for $(M_A, \tb) = (1\,\TeV, 8)$ is indicated
(obtained
approximately from the MSSM predictions as described below).
The results for the $M_h^{125,\mu-}$ scenario displayed in the upper panels
of \fig{fig:WaescheleineMh125m} are
both qualitatively and quantitatively similar to the $M_h^{125}$ scenario
as shown in \fig{fig:WaescheleineMh125}.
As expected, the value of $\mu = -2\,\TeV$ in
the $M_h^{125,\mu-}$ scenario leads to larger deviations in $\kappa_b$ and
$\kappa_\tau$ from the SM value in comparison to the case of
$\mu = +1\,\TeV$ adopted in the $M_h^{125}$ scenario.
This is most visible in the shift of $\kappa_b$
for the assumed point $(M_A, \tan\beta) = (700\,\GeV, 8)$ \how{upper left
panel}
in comparison to the left plot of \fig{fig:WaescheleineMh125}.
The deviations from the SM predictions are smaller for the assumed point
$(M_A, \tb) = (1000\,\GeV, 8)$, and also the impact of different choices for
$\mu$ is smaller in this case.
This is illustrated for $\kappa_b$ in the lower panel of
\fig{fig:WaescheleineMh125m} for $(M_A, \tb) = (1000\,\GeV, 8)$
and $\mu = +1, -1, -2, -3\,\TeV$. The variation of the $\mu$ values
via the $\Delta_b$ corrections yields upward (downward) shifts of
$\kappa_b$ for negative (positive) values of $\mu$ which in total amount to
modifications of up to $1\,\%$.
Accordingly, while the induced variation of the genuine SUSY loop
contribution $\Delta_b$ (see Eq.~\ref{Deltab}) can be relatively large,
for $M_A = 1\,\TeV$ these effects are already heavily suppressed as a
consequence of the decoupling properties of the MSSM. The behavior of
$\kappa_b$ for $M_A = 1\,\TeV$ is therefore mainly governed by the
lowest-order mixing of the two Higgs doublets. Indeed, the observed deviation
from the SM value in $\kappa_b$ of about $2\,\%$ for
$M_A = 1\,\TeV$ is in agreement with the expected behavior arising from the
tree-level prediction, $\kappa_b\approx 1+2\tfrac{M_Z^2}{M_A^2}$.
The dotted horizontal line in the lower panel of
\fig{fig:WaescheleineMh125m} indicates the
prediction for the
THDM type~II, disregarding any additional SUSY effect
beyond the mixing of the \cp{}-neutral Higgs bosons
(obtained approximatively by averaging between the MSSM
prediction for $\mu = -1\,\TeV$ ($\Delta_b$-enhanced) and the MSSM
prediction for $\mu = +1\,\TeV$ ($\Delta_b$-suppressed) in the
$M_h^{125}$~scenario).
The comparison of the $\kappa_b$ value for the THDM-II
with the variation of $\kappa_b$ and $\kappa_\tau$ in the MSSM shows that for
$M_A \gtrsim 1\,\TeV$ the prospects for distinguishing between the MSSM and a
THDM-II via the differences in the higher-order contributions to
$\kappa_b$ and $\kappa_\tau$ do not appear to be promising. We note this
point since parameter points that have been used elsewhere in the literature
seem to indicate that such a distinction could be possible.%
\footnote{The pMSSM example point ``(1)'' that was used
in Ref.~\cite{Barklow:2017suo}
has meanwhile been found to be invalid due to corrected theory calculations.}

\medskip

\begin{figure}[tb!]
\begin{center}
\includegraphics[width=0.5\textwidth]{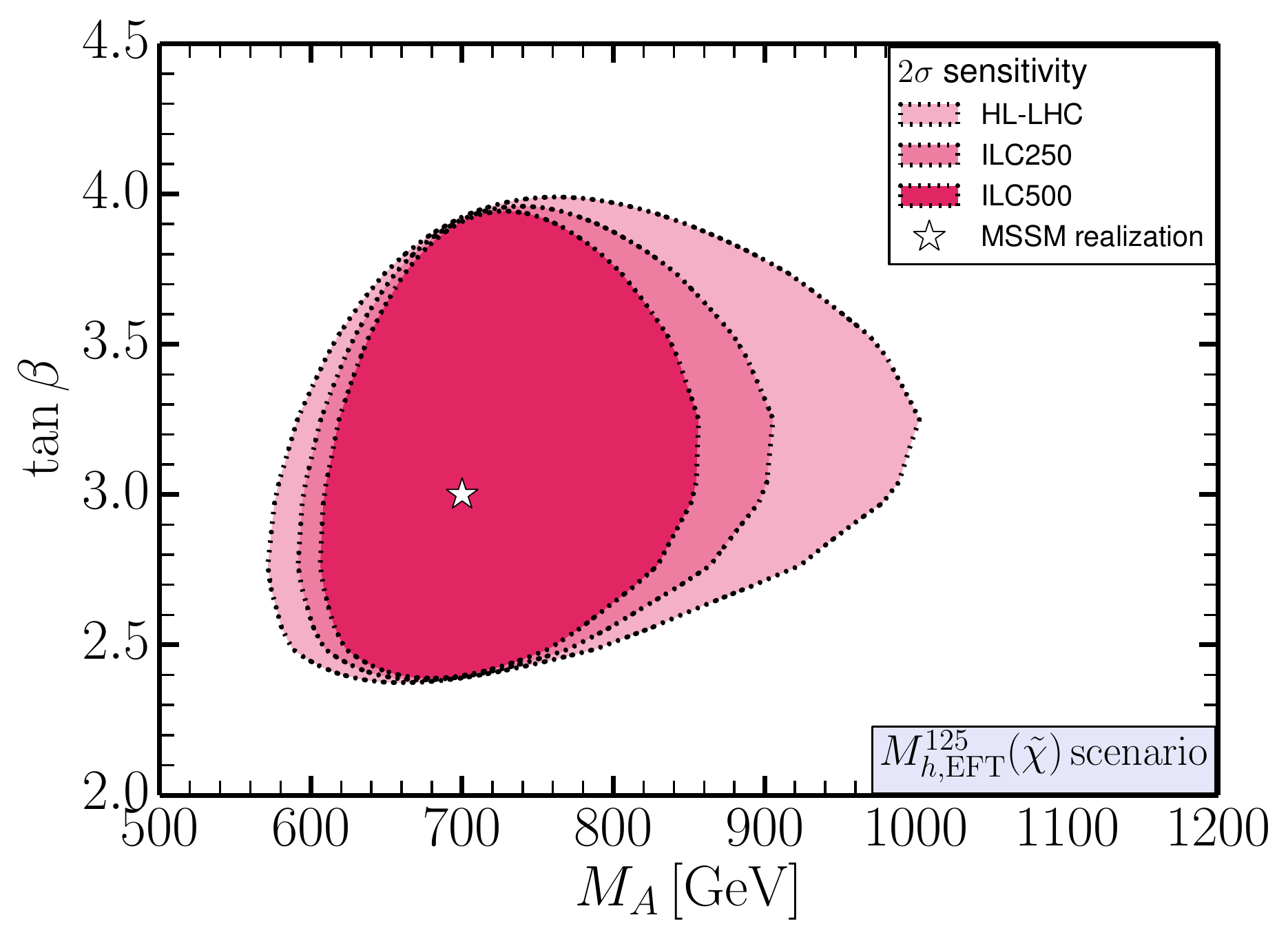}\hfill
\includegraphics[width=0.5\textwidth]{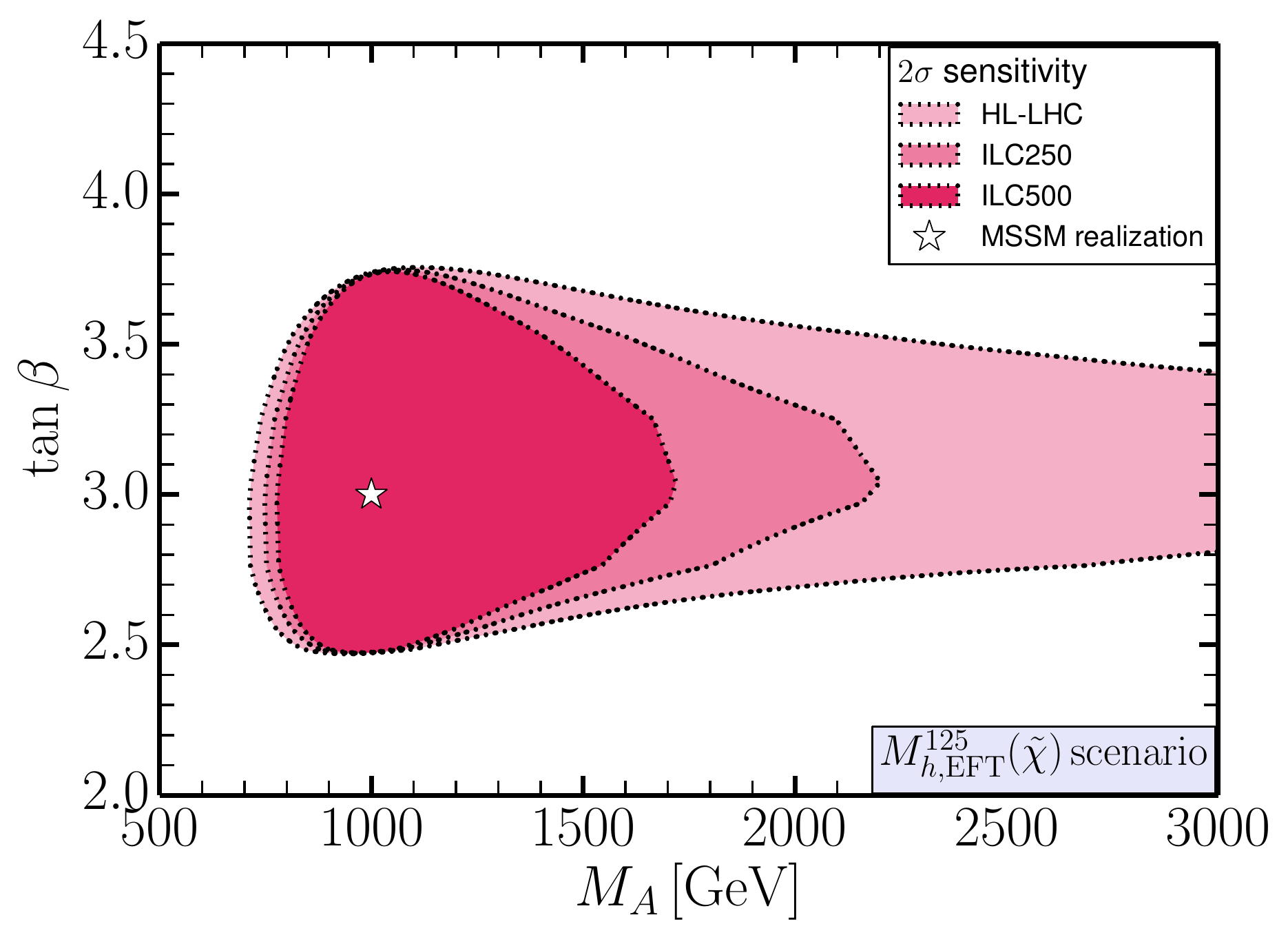}\\[-2mm]
\includegraphics[width=0.5\textwidth]{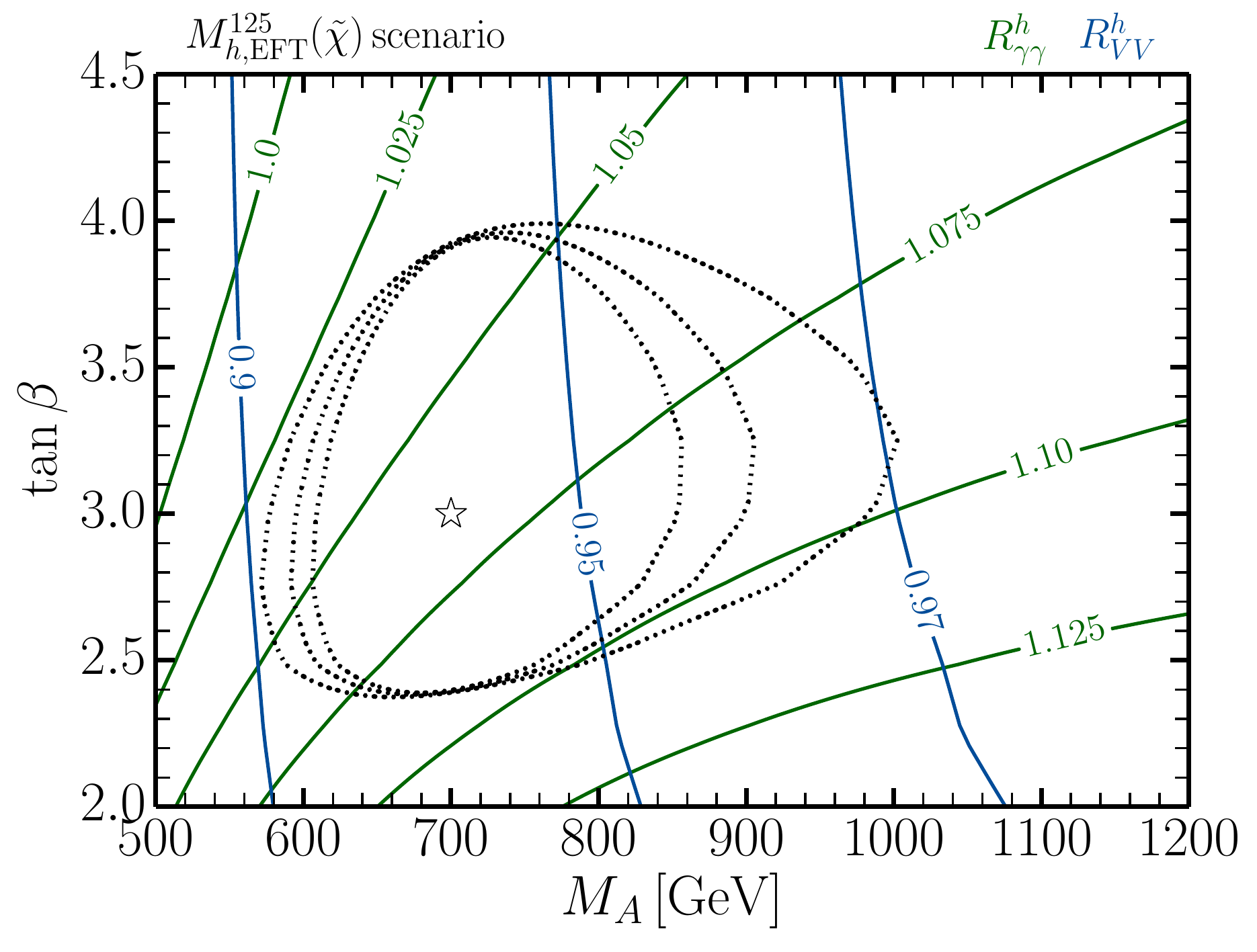}\hfill
\includegraphics[width=0.5\textwidth]{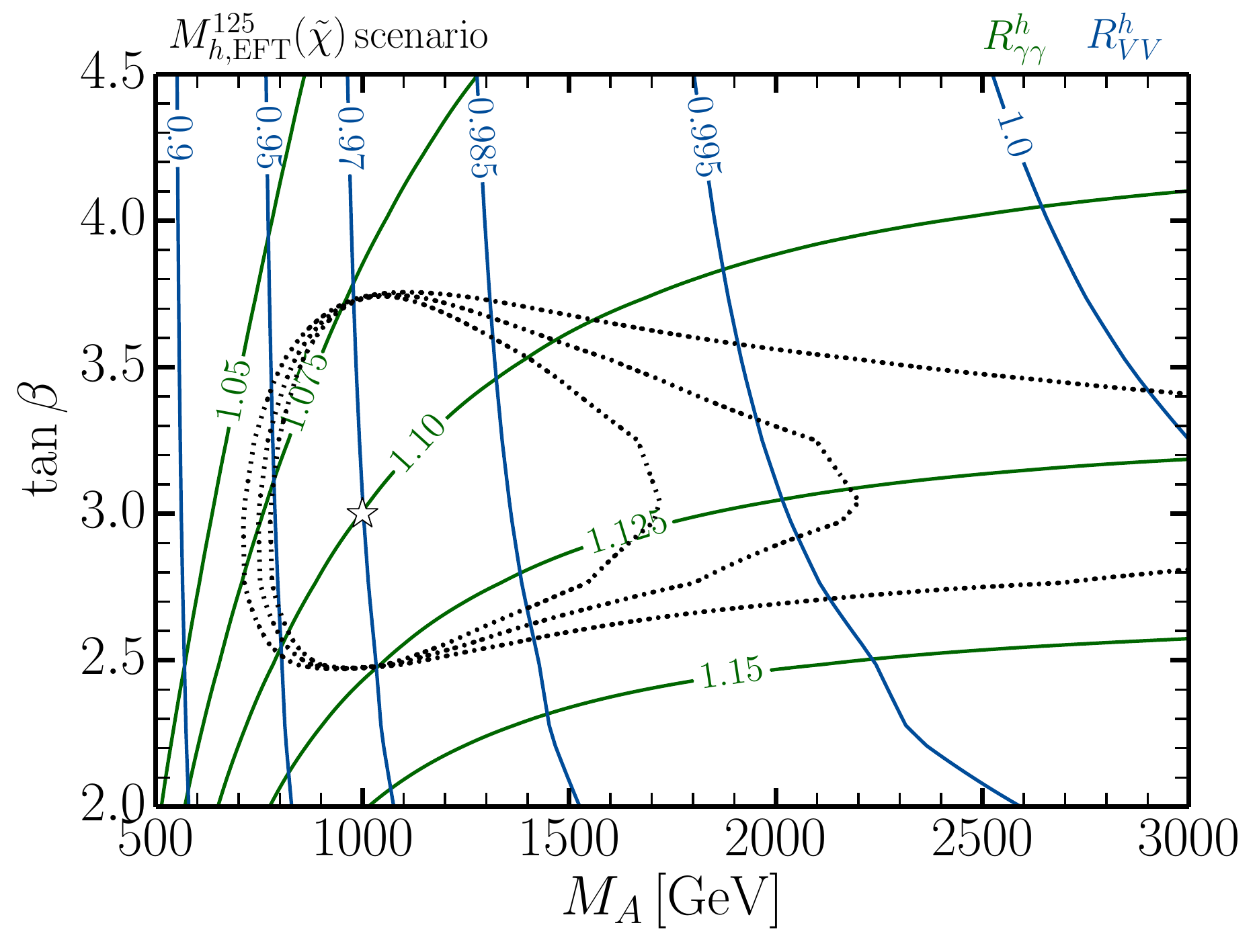}
\end{center}
\caption{Indirect $2\sigma$ constraints
in the ($M_A$, $\tan\beta$) parameter plane
of the $M_{h, {\rm EFT}}^{125}(\tilde\chi)$ scenario
from prospective Higgs-boson signal-rate measurements at the HL-LHC
and the ILC (upper row) and contours for
SM-normalized Higgs rates (lower row)
with the same color coding as in \fig{fig:Mh125_fit}, assuming that
 the point, indicated by a star,
$(M_A, \tan\beta) = (700\,\GeV, 3)$ \how{left panel}
or $(M_A, \tan\beta) =(1\,\TeV, 3)$  \how{right panel} is realized in nature.
The blue and green contours in the lower panels show
the inclusive rate for $pp\to h\to VV$ ($V = W^\pm, Z)$,
$R^h_{VV}$, and the inclusive rate for $pp \to h\to \gamma\gamma$,
$R^h_{\gamma\gamma}$, respectively,
with the $2\sigma$ parameter ranges from the upper panels superimposed.}
\label{fig:Mh125EFT_fit}
\end{figure}

We now turn to the $M_{h, {\rm EFT}}^{125}(\tilde\chi)$~scenario where we
consider as possible realizations the points $(M_A, \tb) = (700\,\GeV, 3)$
and $(M_A, \tb) = (1000\,\GeV, 3)$. In comparison to the previously discussed
points in the $M_{h}^{125}$ and $M_h^{125,\mu-}$ scenarios (as well as
its variations with different values of $\mu$), these parameter points are
more difficult to probe directly by experimental searches for heavy
Higgs bosons, since at such low
values of $\tb$ the $pp\to H/A \to \tau^+\tau^-$ rate does not contain a
large enhancement factor. Other
direct heavy Higgs searches, in particular in the di-top final state and in electroweakino final states as suggested in Refs.~\cite{Bahl:2018zmf,Bahl:2019ago,Gori:2018pmk}
could improve the sensitivity.
Moreover, as discussed above, LHC searches for direct
production of mass-compressed electroweakinos have the potential to probe a
scenario with such low values of $\tan\beta$
in the electroweakino sector
and can thus provide complementary
information~\cite{CidVidal:2018eel}.

In \fig{fig:Mh125EFT_fit} we show the results in the
$M_{h, {\rm EFT}}^{125}(\tilde\chi)$~scenario, with the assumed parameter point
$(M_A, \tb) = (700\,\GeV, 3)$ in the left panels, and $(M_A, \tb) = (1000\,\GeV, 3)$  in the right panels.
The expected $2\sigma$ allowed parameter ranges obtained by Higgs-boson signal-rate
measurements are shown in the upper panel of \fig{fig:Mh125EFT_fit}
with the same color coding as before in Figs.~\ref{fig:Mh125_fit}~and~\ref{fig:Mh125m_fit}.
The indirect bounds from the Higgs rate measurements on $\mA$
for the first assumed point at $(M_A, \tb) = (700\,\GeV, 3)$ amount to
\begin{align*}
575\,\GeV \lesssim M_A  \lesssim 1000 \,\GeV	&\qquad (\text{HL-LHC}),\\
590\,\GeV \lesssim M_A  \lesssim 900 \,\GeV	&\qquad (\text{$+$ ILC250}),\\
610\,\GeV \lesssim M_A  \lesssim 850 \,\GeV	&\qquad (\text{$+$ ILC500}),
\end{align*}
whereas for the second assumed point, $(M_A, \tb) = (1000\,\GeV, 3)$, we find
\begin{align*}
700\,\GeV \lesssim M_A \qquad\qquad\quad\;\;	&\qquad (\text{HL-LHC}),\\
750\,\GeV \lesssim M_A  \lesssim 2200 \,\GeV	&\qquad (\text{$+$ ILC250}),\\
780\,\GeV \lesssim M_A  \lesssim 1710 \,\GeV	&\qquad (\text{$+$ ILC500}).
\end{align*}
These $M_A$ ranges are very similar to those found in the $M_h^{125}$ scenario for the two assumed points at the same $M_A$ values.
In particular, for the assumed point with $M_A = 1\,\TeV$
the prospective accuracy of the signal strength measurements at the HL-LHC is
not sufficient to place an indirect upper bound on $M_A$ at the
$2\sigma$ level.
As in the previous scenarios, the
ILC measurements would be essential in this case to obtain an upper bound on
$M_A$. In contrast to the scenarios discussed above, however,
for the assumed value of $\tb = 3$ in this scenario with light charginos the
Higgs rate measurements would not only provide a high sensitivity for a
distinction from the SM but would also allow one to
constrain $\tb$ to a narrow range
\begin{align*}
2.5 \le \tan\beta \le 4.0	&\qquad \text{for $(M_A, \tb) = (700\,\GeV, 3)$,} \\
2.5 \le \tan\beta \le 3.75	&\qquad \text{for $(M_A, \tb) = (1000\,\GeV, 3)$.}
\end{align*}
This high sensitivity to $\tb$ is caused by the fact that
the chargino contributions to the $h\to\gamma\gamma$ decay
rate strongly depend on the chargino mixing, which in turn depends on
$\tb$. The Higgs rate measurements would complement the
information from the direct searches for new particles, which in such a scenario
could yield a significant excess or even a signal from the production of
light electroweakinos.

The lower panels of \fig{fig:Mh125EFT_fit} display two SM-normalized Higgs
rates that are of particular importance at the
HL-LHC in this case:
The inclusive rate for $pp\to h\to VV$ ($V = W^\pm, Z)$, denoted
$R^h_{VV}$, and the inclusive rate for $pp \to h\to \gamma\gamma$, denoted
$R^h_{\gamma\gamma}$. The displayed results for
$R^h_{\gamma\gamma}$ indicate that the
di-photon rate is strongly influenced by loop
contributions of charginos, which become large at small $\tb$ values. In
contrast, the $VV$ rate follows the basic trend that its decoupling
behavior is
mostly governed by $\mA$, see also the rate $R^{Vh}_{bb}$ in the
discussion of the $M_h^{125}$~scenario. The pattern of how the decoupling
is approached, however, shows that for low $\tan\beta$ values slightly larger
deviations from the SM value occur for the same value of $\mA$ than for
larger $\tb$.
The interplay of the constraints from in particular these two rate
measurements at the HL-LHC leads to the displayed shape of the allowed region
that extends to the highest values of $\mA$ (going beyond the plotted range
in the right plot) in the vicinity of $\tb \sim 3$.
It should be noted that for the assumed point at $M_A=1\,\TeV$ (right plot)
the enhancement of
$R^h_{\gamma\gamma}$ compared to the SM value is even somewhat larger
($10\%$) than for
the assumed point at $M_A=700\,\GeV$ ($\sim 6.5\%$, left plot). This is
because at lower $M_A$ values the enhancement of the $h\to b\bar{b}$
decay is stronger, which in turn suppresses the $h\to \gamma\gamma$
decay rate via its contribution to the total decay width.
For the considered scenario the impact of the Higgs rate measurements at the ILC
would mainly be a
significant improvement of the indirect constraints on $\mA$.

\begin{figure}[tb!]
\begin{center}
\includegraphics[width=0.5\textwidth]{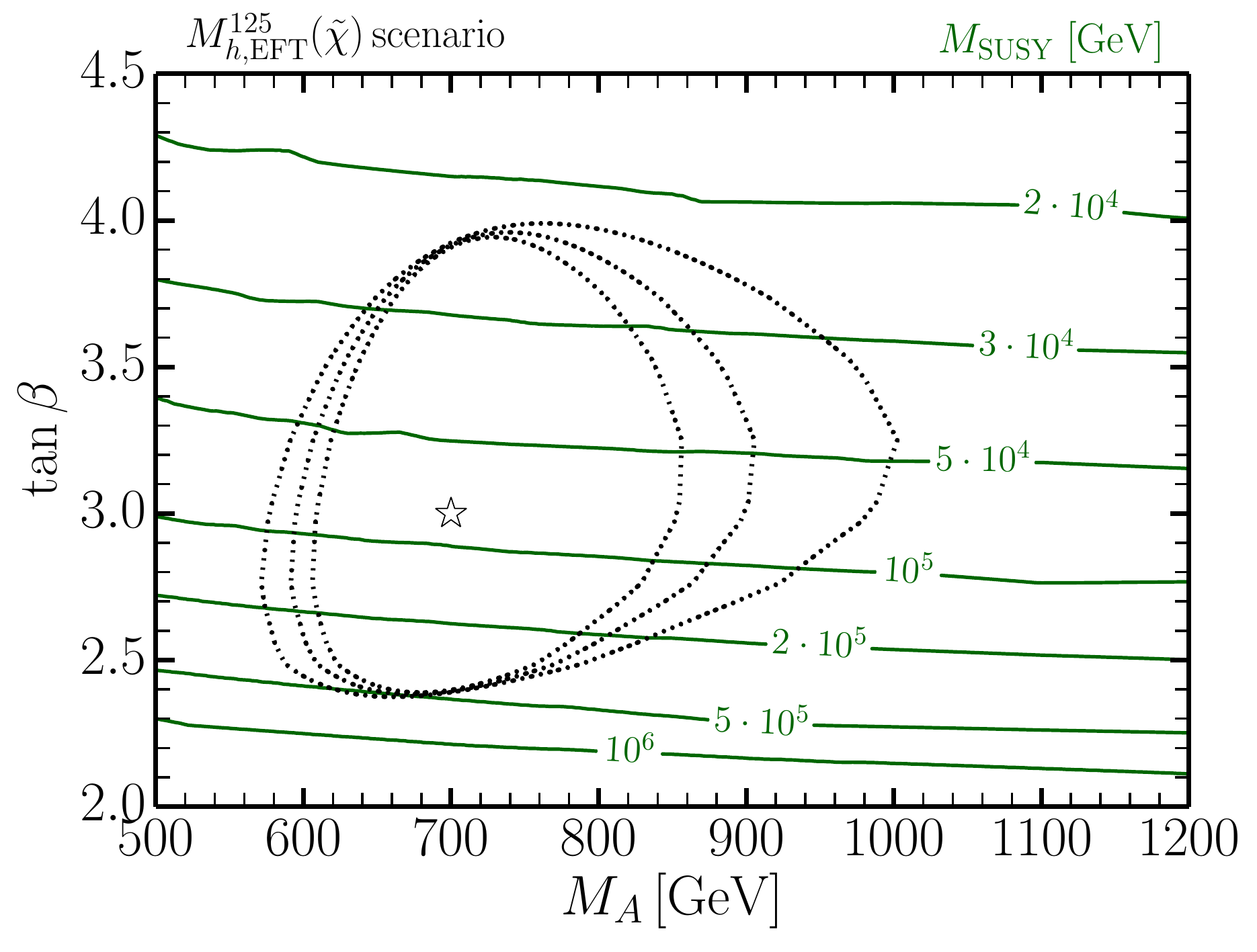}\hfill
\includegraphics[width=0.5\textwidth]{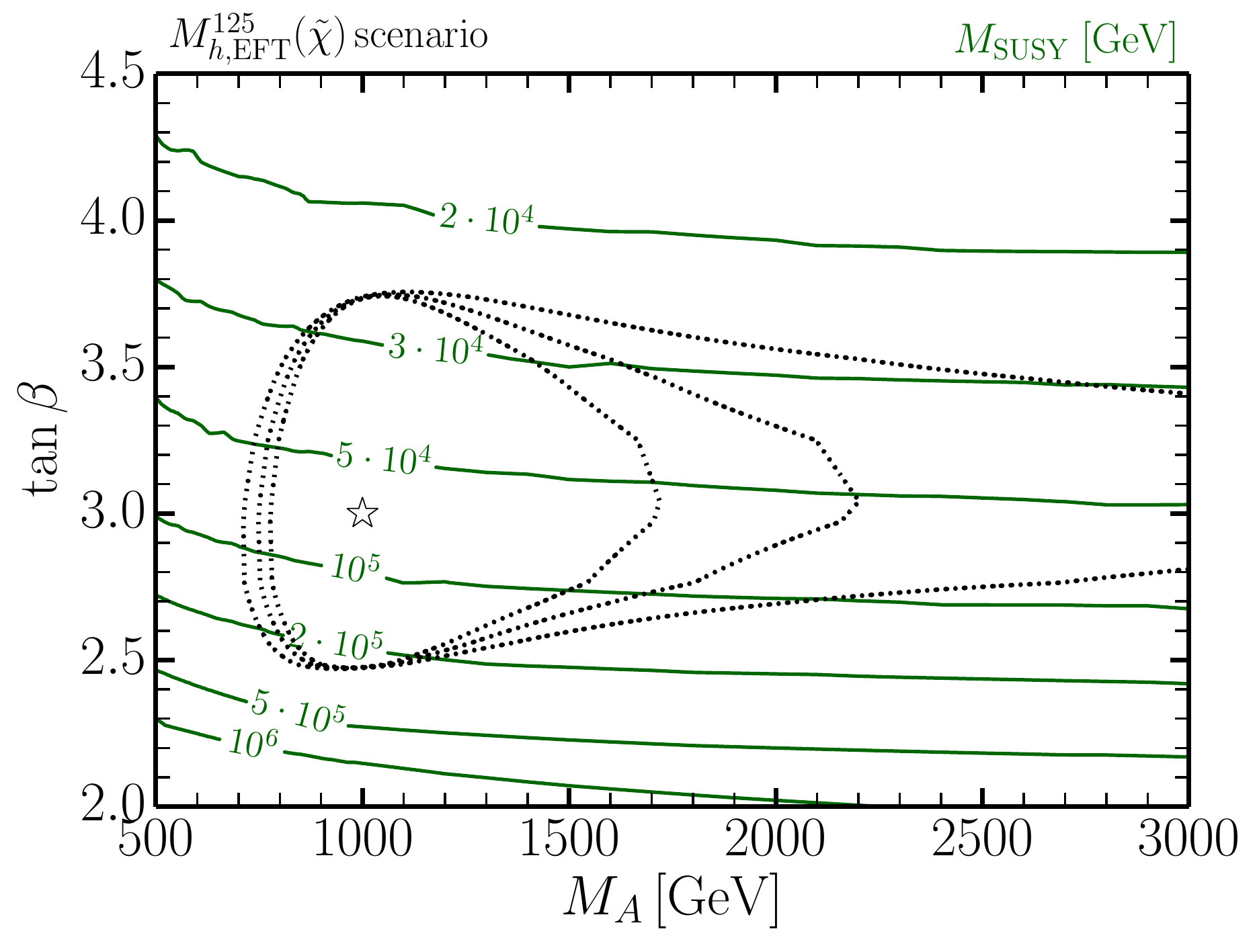}
\end{center}
\caption{
Contours of the scalar fermion
soft-SUSY breaking mass $M_{\text{SUSY}}$
in the ($M_A$, $\tan\beta$) parameter plane
of the $M_{h, {\rm EFT}}^{125}(\tilde\chi)$ scenario
discussed in \fig{fig:Mh125EFT_fit},
assuming that the point, indicated by a star,
$(M_A, \tan\beta) = (700\,\GeV, 3)$ \how{left panel}
or $(M_A, \tan\beta) =(1\,\TeV, 3)$  \how{right panel} is realized in nature.
The indirect $2\sigma$ constraints
from prospective Higgs-boson signal-rate measurements at the HL-LHC
and the ILC obtained in
the upper panels of \fig{fig:Mh125EFT_fit} are
superimposed.}
\label{fig:Mh125EFT_fit_MSUSY}
\end{figure}

In \fig{fig:Mh125EFT_fit_MSUSY} we show contour lines
of equal $M_{\text{SUSY}}$ in the same parameter space as considered in
\fig{fig:Mh125EFT_fit}. Superimposed \how{as dotted lines} are the expected
$2\sigma$-allowed parameter regions shown previously in
\fig{fig:Mh125EFT_fit} for the same MSSM points that we assume to be
realized. $M_{\text{SUSY}}$ denotes the scale of all
scalar fermion soft-SUSY breaking masses. As explained in
Sec.~\ref{sec:models}, in the $M_{h, {\rm EFT}}^{125}(\tilde\chi)$~scenario
$M_{\text{SUSY}}$ is adjusted at every point in the parameter plane
such that $M_h \simeq 125\,\GeV$. Thus the constraints in the
$(M_A, \tb)$ parameter plane for a given assumed realization of the MSSM can
be translated into a constraint on the sfermion mass scale in this scenario.
As a result, if such a scenario with light electroweakinos and a rather
low value of $\tb$ was
realized in nature, the sensitivity to $\tb$ arising
from the loop contributions of the light charginos
to the di-photon rate could be exploited to constrain $M_{\text{SUSY}}$
to the ranges
\begin{align*}
2.3\,\TeV \lesssim M_\text{SUSY} \lesssim 50\,\TeV &\qquad \text{for $(M_A, \tb) = (700\,\GeV, 3)$,} \\
2.3\,\TeV \lesssim M_\text{SUSY} \lesssim 30\,\TeV	&\qquad \text{for $(M_A, \tb) = (1000\,\GeV, 3)$.}
\end{align*}
Those indirect constraints could of course be significantly improved with
the results of the direct searches for additional Higgs bosons and
electroweakinos, which in the considered scenario would have good prospects for a
significant excess or even a discovery.

The predicted Higgs couplings
in the $M_{h, {\rm EFT}}^{125}(\tilde\chi)$~scenario,
parametrized in terms of $\kappa$ scale
factors, are shown in \fig{fig:WaescheleineMh125EFTchi} for the assumed
MSSM points $(M_A, \tan\beta) = (700\,\GeV, 3)$ \how{left panel} and
$(M_A, \tan\beta) =(1\,\TeV, 3)$  \how{right panel} in comparison to the
anticipated $1\sigma$ precision of the future $\kappa$ determination. In
contrast to the $M_h^{125}$ scenario, \fig{fig:WaescheleineMh125}, and
in line with the previous discussion,
because of the large loop contributions of the light charginos
to the di-photon rate
in the $M_{h, {\rm EFT}}^{125}(\tilde\chi)$~scenario
a sizable deviation in
$\kappa_\gamma$ is clearly visible already with the HL-LHC
precision. This precision on the effective Higgs-photon-photon coupling
can only mildly be improved by the ILC measurements. On the other hand,
$\kappa_b$ and $\kappa_\tau$ show deviations similar to the points
considered in the $M_h^{125}$ scenario, \fig{fig:WaescheleineMh125}, and
here the ILC measurements will be crucial to achieve a significant
discrimination with respect to the SM prediction.

\begin{figure}[tb!]
\begin{center}
\includegraphics[width=0.5\textwidth]{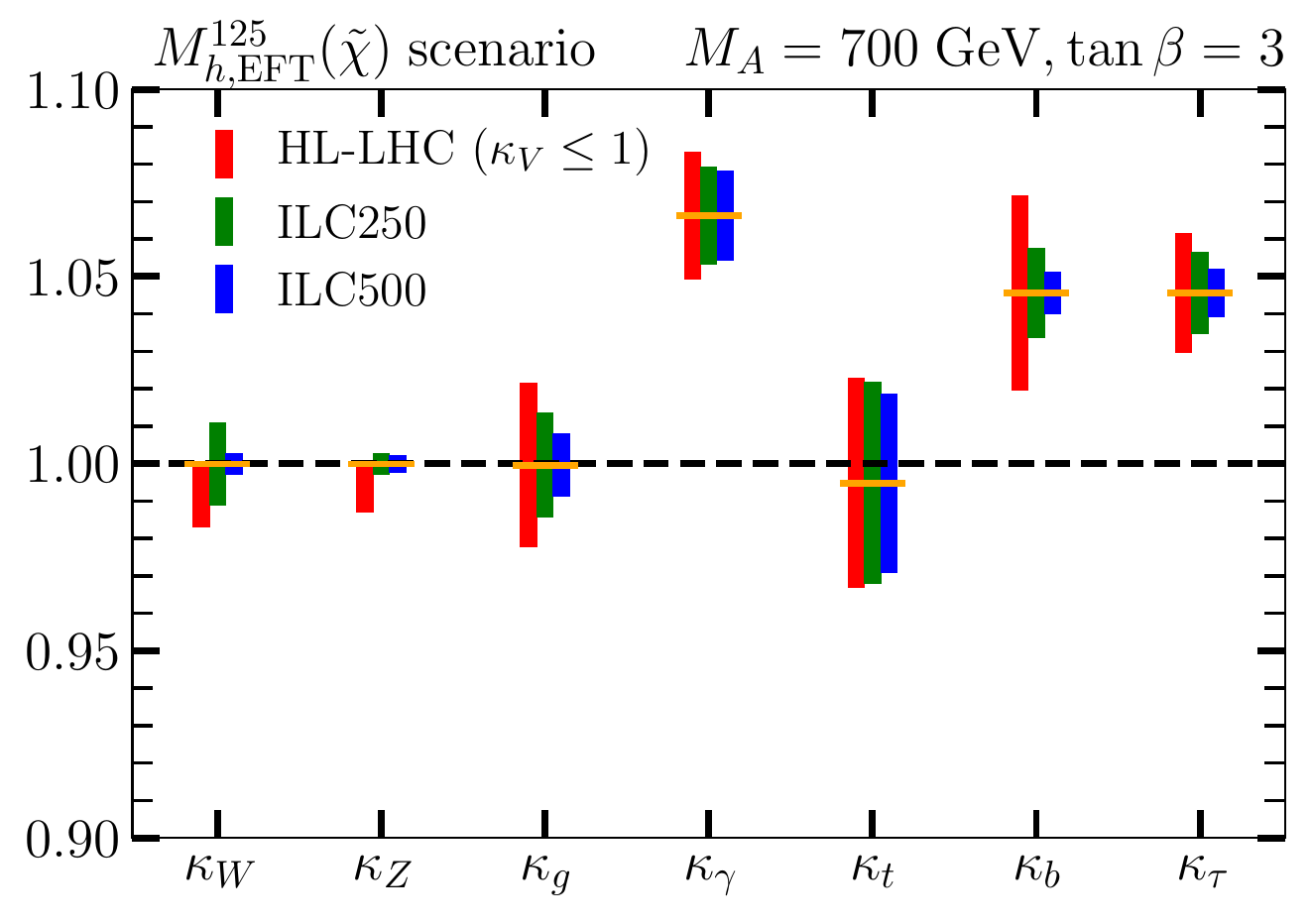}\hfill
\includegraphics[width=0.5\textwidth]{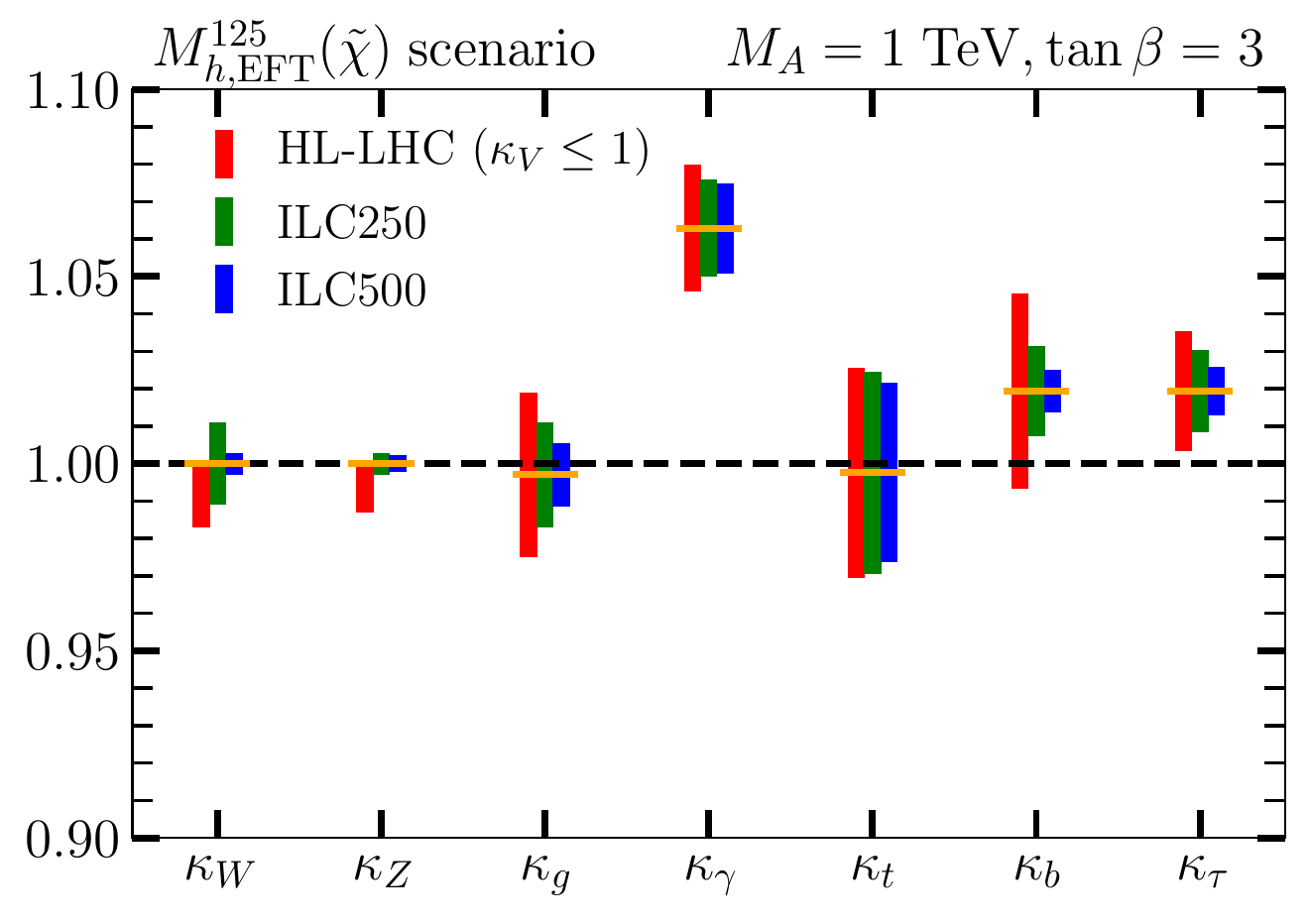}
\end{center}
\caption{\emph{W\"ascheleinen-plots},
using the the same color coding as in \fig{fig:WaescheleineMh125},
for the two assumed MSSM parameter
points $(M_A, \tan\beta) = (700\,\GeV, 3)$ \how{left panel} and $(M_A,
\tan\beta) =(1\,\TeV, 3)$ \how{right panel} in the   $M_{h, {\rm
EFT}}^{125}(\tilde\chi)$ scenario.
The predicted Higgs couplings in the $\kappa$ framework
are compared with
the anticipated $1\sigma$ precision from Higgs rate
  measurements, where at the HL-LHC the theoretical assumption
  $\kappa_V \leq 1$ is employed, while for the results
including prospective measurements at
ILC250 and ILC500 no assumption on $\kappa_V$ is employed.}
\label{fig:WaescheleineMh125EFTchi}
\end{figure}

\medskip

We now turn to the discussion of the case that a relatively large value of
$\tb$ could be realized in nature. For this purpose
we choose a heavy Higgs-boson mass of $M_A = 1.75\,\TeV$.
In the $M_h^{125}$ and $M_h^{125}(\tilde\chi)$ scenarios the
\tb\ value is chosen to be $\tb=50$,
close to the expected exclusion bound of the current $pp \to H/A\to
\tau^+\tau^-$ analysis~\cite{Sirunyan:2018zut}.
For the $M_h^{125,\mu-}$ scenario we fix $\tb=25$,
close to the current indirect exclusion from Higgs rate measurements.
The chosen value of $M_A = 1.75\,\TeV$ is a ``best-case'' scenario if the
MSSM with a large value of $\tb$ is realized, in the sense that it would
certainly lead to a discovery of heavy Higgs bosons at the HL-LHC (see our
discussion above of the projections in the different benchmark scenarios) and
possibly even already in the near future.

For definiteness, we quote here the
$13\;\TeV$ signal rates of the processes $pp\to H/A \to \tau^+\tau^-$ and $pp\to H/A \to b\bar{b}$,
whose production is completely dominated by bottom-quark associated Higgs
production at these parameter points. They are given by
\begin{align*}
\sigma (pp\to H/A \to \tau^+\tau^-) = 1.94~\mathrm{fb} & \qquad \text{in
  the $M_h^{125}$ scenario
  for } (M_A, \tb) = (1.75\,\TeV, 50) , \\
\sigma (pp\to H/A \to \tau^+\tau^-) = 1.51~\mathrm{fb} & \qquad \text{in
  the $M_h^{125}(\tilde\chi)$ scenario
  for } (M_A, \tb) = (1.75\,\TeV, 50) , \\
\sigma (pp\to H/A \to \tau^+\tau^-) = 0.49~\mathrm{fb} & \qquad \text{in
  the $M_h^{125,\mu-}$ scenario
  for } (M_A, \tb) = (1.75\,\TeV, 25) ,
\end{align*}
and
\begin{align*}
\sigma (pp\to H/A \to b\bar{b}) = 5.17~\mathrm{fb} & \qquad \text{in the
  $M_h^{125}$ scenario
  for } (M_A, \tb) = (1.75\,\TeV, 50) , \\
\sigma (pp\to H/A \to b\bar{b}) = 7.16~\mathrm{fb} & \qquad \text{in the
  $M_h^{125}(\tilde\chi)$ scenario
  for } (M_A, \tb) = (1.75\,\TeV, 50) , \\
\sigma (pp\to H/A \to b\bar{b}) = 11.42~\mathrm{fb} & \qquad \text{in
  the $M_h^{125,\mu-}$ scenario
  for } (M_A, \tb) = (1.75\,\TeV, 25) ,
\end{align*}
respectively. According to our discussion in Sec.~\ref{sec:reach} we
expect that the additional Higgs bosons would be first detectable in the
searches in the $\tau^+\tau^-$ final states, while an excess in $H/A\to
b\bar{b}$ searches would at best show up in the full HL-LHC dataset (in
the $M_h^{125,\mu-}$ scenario).
It should be noted that for additional Higgs bosons in the mass range
considered here the possibility that they could have a sizable branching
fraction into other BSM particles should be taken seriously.
For instance, for the considered parameter point in the
$M_h^{125}(\tilde\chi)$ scenario, the decay rates of $H$ and $A$ to
electroweakinos amount to around $32\%$. Therefore,
in this scenario not only the decays of the
heavy Higgs bosons to third generation leptons and
quarks but also the decays into electroweakinos would be
promising channels for their detection.

\begin{figure}[htb!]
\begin{center}
\includegraphics[width=0.5\textwidth]{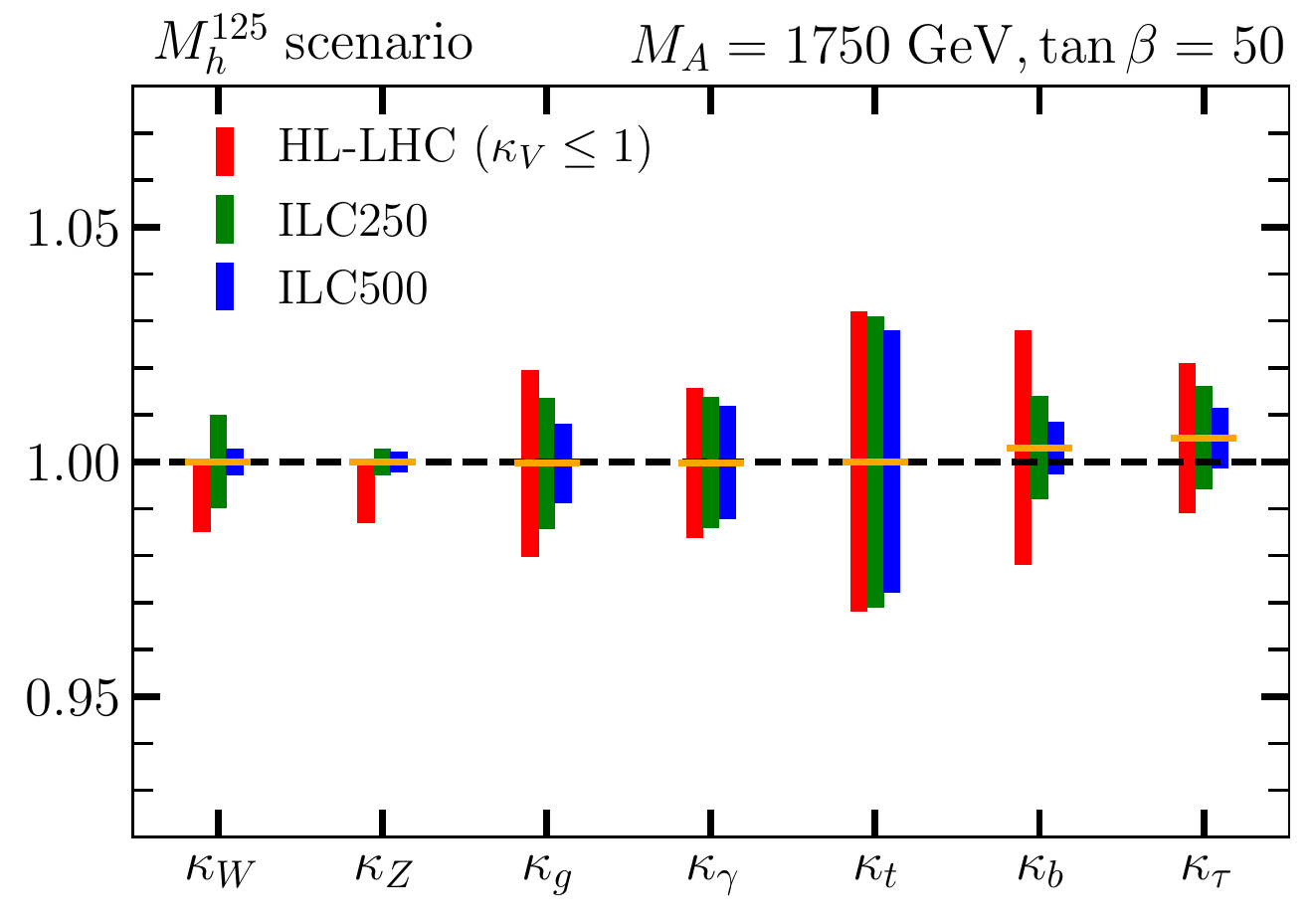}\hfill
\includegraphics[width=0.5\textwidth]{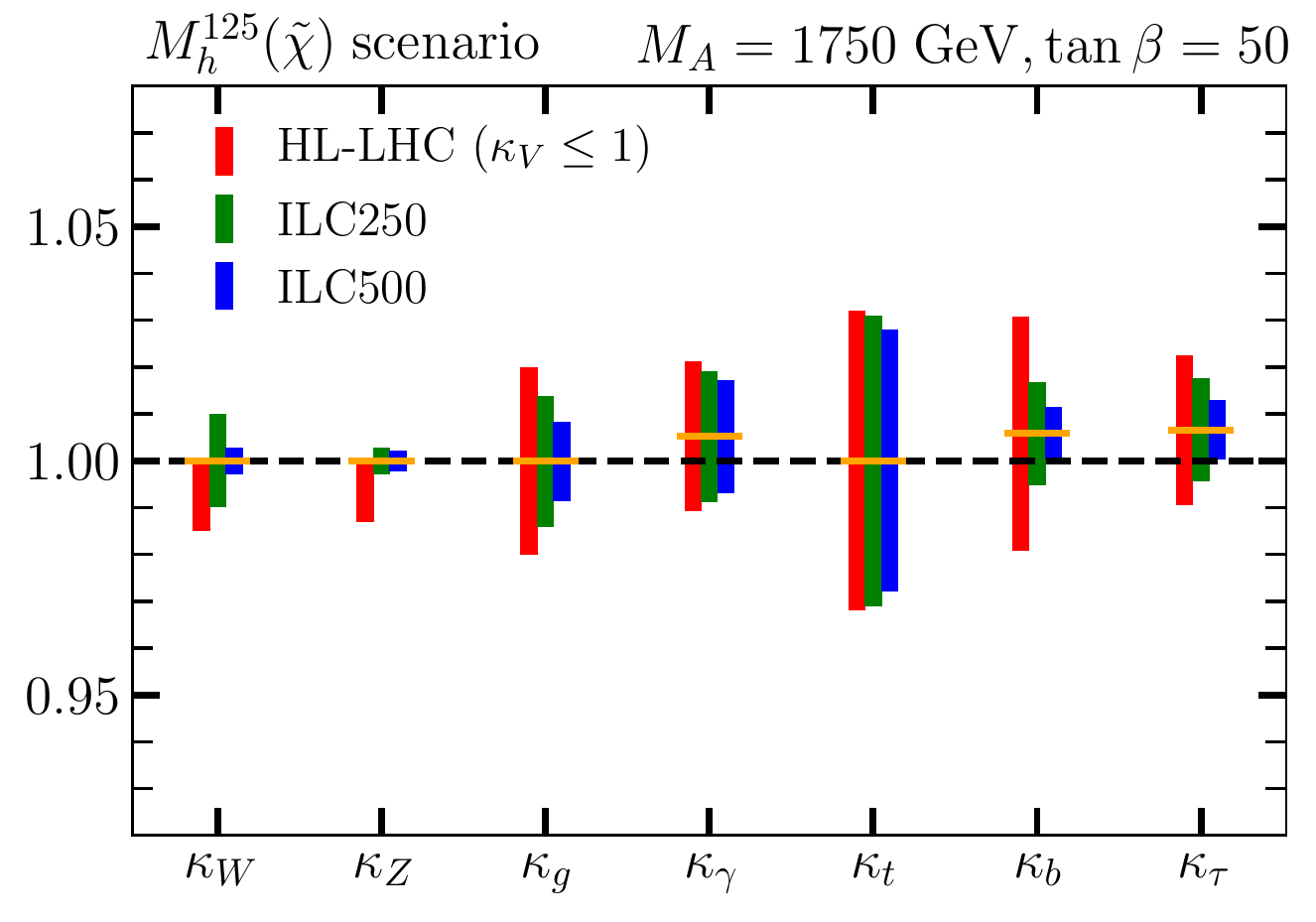}\\
\includegraphics[width=0.5\textwidth]{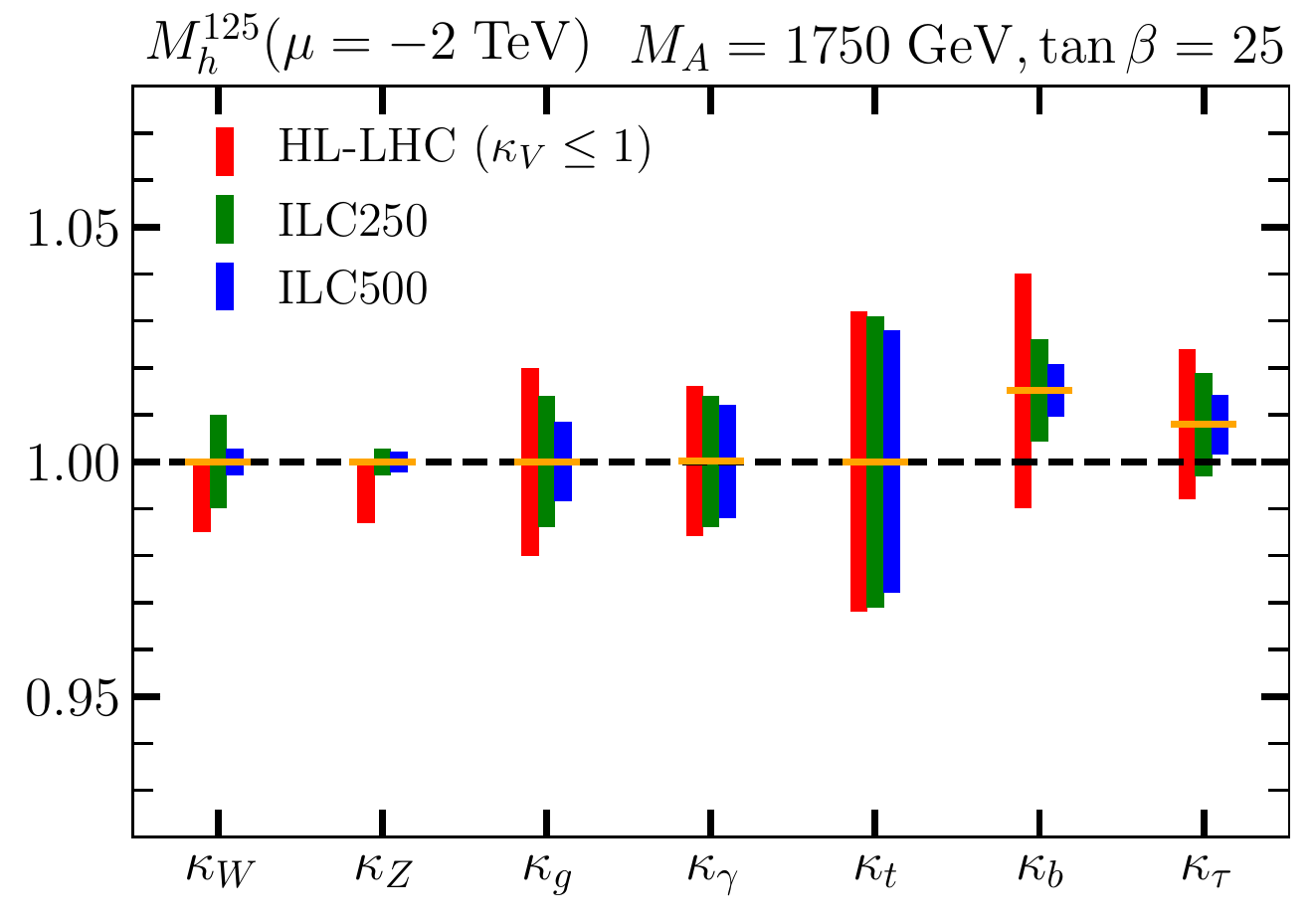}
\end{center}
\caption{\emph{W\"ascheleinen-plots},
using the the same color coding as in \fig{fig:WaescheleineMh125},
for the following three assumed MSSM scenarios: $(M_A, \tb) = (1750\,\GeV,
50)$ in the $M_h^{125}$ scenario \how{upper left panel}, $(M_A, \tb) =
(1750\,\GeV, 50)$ in the $M_h^{125}(\tilde\chi)$ scenario \how{upper right},
and $(M_A, \tb) = (1750\,\GeV, 25)$ in the $M_h^{125,\mu-}$ scenario
\how{lower panel}.
The predicted Higgs couplings in the $\kappa$ framework
are compared with
the anticipated $1\sigma$ precision from Higgs rate
  measurements, where at the HL-LHC the theoretical assumption
  $\kappa_V \leq 1$ is employed, while for the results
including prospective measurements at
ILC250 and ILC500 no assumption on $\kappa_V$ is employed.}
\label{fig:WaescheleineMh125_MA1750}
\end{figure}

We now turn to the effects in the rates of the Higgs boson at $125\,\GeV$
that would arise for the assumed parameter points with
$M_A = 1.75\,\TeV$ and large values of $\tb$.
The predicted light Higgs boson couplings, presented in terms of $\kappa$
scale factors, for the three discussed MSSM scenarios are shown as
\emph{W\"ascheleinen-plots} in \fig{fig:WaescheleineMh125_MA1750}, in
comparison with the precision of the prospective $\kappa$ determination
at the HL-LHC and the ILC. As expected
for such a large value of $M_A$,
corresponding to a scenario that is quite far in the decoupling region,
we find that even for the relatively large chosen values of \tb\ the
coupling deviations from the SM predictions are very small.
Deviations at this level will be
extremely challenging to resolve experimentally.
The ultimate precision of
ILC500 seems necessary in order to experimentally establish
a significant pattern of
deviations. The best chances in this context would occur if a relatively
large \emph{negative} value of $\mu$ was realized, as exemplified
in the $M_h^{125,\mu-}$ scenario in \fig{fig:WaescheleineMh125_MA1750},
where an enhancement of the bottom-quark (and perhaps also
tau-lepton) Yukawa coupling with respect to the SM prediction could
clearly be established at the ILC.
In this case the pattern and size of deviations is driven by genuine
  SUSY effects, i.e.\ beyond the THDM type~II.

\medskip

We summarize this discussion of MSSM points assumed to be realized in
nature by giving the $\chi^2$ values of the SM hypothesis in these
assumed scenarios for the prospective accuracies of the rate
measurements at the HL-LHC and the ILC
in Tab.~\ref{tab:chi2values}. These $\chi^2$ values can be interpreted
either directly as the goodness-of-fit of the SM, i.e.\ how well it fits
the set of observations, or they can be used to perform
a likelihood-ratio test that quantifies
the significance of the tension between the SM prediction and
the MSSM interpretation
(with the assumed parameter point used as null
hypothesis). Here we want to focus on the latter. As the future
measurements will naturally feature statistical fluctuations, we rather
refer to the $\chi^2$ of the SM hypothesis as $\Delta\chi^2_\text{SM}
\equiv \chi^2_\text{SM} - \chi^2_\text{MSSM}$, where in our projection
study with idealized measurements we have $\chi^2_\text{MSSM} = 0$ for
the considered realized MSSM parameter point.
In this likelihood ratio test between two simple hypotheses, with
no adjustable model parameters, the levels $\Delta \chi^2 = 4$ and
$9$ correspond to a $2\sigma$ and $3\sigma$ tension, respectively,
between the SM hypothesis and the MSSM hypothesis. It should be noted
that this
level of sensitivity does not allow one to exclude the SM hypothesis on
grounds of the measurements alone, but instead only allows one to
discriminate between two models. As these tensions are inferred
\emph{indirectly} from the signal rates of the SM-like Higgs boson,
the combination with the information from \emph{direct} collider
searches, both regarding limits and possible hints for signals, will be
crucial to firmly
establish an observed pattern of BSM physics and to narrow down its possible
nature.

\begin{table}
\centering
\small
\begin{tabular}{l c ccc}
\toprule
MSSM scenario	& assumed $(M_A, \tan\beta)$ point   & \multicolumn{3}{c}{$\Delta\chi^2_\text{SM}$} \\
\cmidrule{3-5}
	&	        &		HL-LHC	&	ILC250	&	ILC500	\\
\midrule
\multirow{3}{*}{$M_h^{125}$} 		& $ (700\,\GeV,~8) $ & $22.0$ & $34.3$ &  $50.7$ \\
					& $ (1000\,\GeV,~8) $ & $4.6$ & $7.2$ & $10.6$  \\
					& $ (1750\,\GeV,~50) $ & $0.3$ & $0.5$ & $0.7$ \\
\midrule
\multirow{3}{*}{$M_h^{125,\mu-}$}	& $ (700\,\GeV,~8) $ & $27.6$ & $41.3$ &  $61.2$ \\
					& $ (1000\,\GeV,~8) $ & $5.9$ & $8.8$ & $13.0$ \\
					& $ (1750\,\GeV,~25) $ & $2.5$ & $3.5$ &$4.2$ \\
\midrule
\multirow{2}{*}{$M_{h,\text{EFT}}^{125}(\tilde\chi)$} 	& $ (700\,\GeV,~3) $ & $113.8$ & $130.7$ & $152.0$\\
								& $ (1000\,\GeV,~3) $ & $177.3$ & $182.1$ & $187.9$\\
\midrule
$M_{h}^{125}(\tilde\chi)$	& $ (1750\,\GeV,~50) $ & $1.2$ & $1.3$ & $1.7$\\
\bottomrule
\end{tabular}
\caption{Goodness-of-fit of the SM hypothesis, $\Delta\chi^2_{\text{SM}}$, from a test against future signal rate measurements
of the Higgs boson at $125\,\GeV$ at the
HL-LHC and in combination with ILC250 and ILC500 measurements,
assuming different MSSM realizations in nature.}
\label{tab:chi2values}
\end{table}

From Tab.~\ref{tab:chi2values}
we see that in the $M_h^{125}$  and the $M_h^{125,\mu-}$ scenario the
HL-LHC will reveal a significant tension between the SM and the MSSM
interpretation only if a realization with $M_A=700\,\GeV$ is assumed. For the
larger value of $M_A=1000\,\GeV$, the HL-LHC can only discriminate
the SM at the $2\sigma$ level
from the MSSM hypothesis, while the ILC
measurements would be crucial to
clearly establish a deviation from the SM.
The situation is somewhat different in the
$M_{h,\text{EFT}}^{125}(\tilde\chi)$ scenario at low values of $\tan\beta$, where
the large enhancement of $h\to \gamma\gamma$ from loop contributions of
light charginos leads to a very strong
tension with the SM, already with its precise determination at the
HL-LHC. In fact, such a strong deviation in the $h\to \gamma\gamma$ rate
would clearly exclude the SM as null hypothesis, based solely on
the goodness-of-fit, with very high significance.
For the scenarios with $M_A = 1.75\,\TeV$ shown in Tab.~\ref{tab:chi2values}
the deviations in the properties of the Higgs boson at $125\,\GeV$ are so
small that the HL-LHC accuracy for the Higgs rate measurements will yield
only small shifts in $\chi^2$ between the SM and the MSSM. A significant
distinction between the SM hypothesis and the MSSM hypothesis,
with $\Delta\chi^2 > 4$, can only be achieved for the
$M_h^{125,\mu-}$ scenario with the precision at ILC500.


\section{Conclusions}
\label{sec:conclusions}

In this paper we have assessed the indirect sensitivity of the HL-LHC for
probing extended Higgs sectors via rate measurements of the observed Higgs
boson at $125\,\GeV$, alone%
\footnote{A selection of
our HL-LHC results has already been included in Ref.~\cite{Cepeda:2019klc}.}
and in combination with a future
$e^+e^-$ Higgs factory, and we have compared the indirect sensitivity with
the present LHC reach and the projected HL-LHC reach for direct searches for
additional heavy Higgs bosons.
For the direct searches we have taken into account present results and projected
sensitivities
in the $\tau^+\tau^-$, $b\bar{b}$ and di-Higgs ($hh$) final states.
Concerning the projections of the rate measurements of the detected Higgs
boson at $125\,\GeV$ at a future $e^+e^-$ Higgs factory,
we have considered for
definiteness the case of the ILC since it is the currently most advanced
project, and its projected accuracies are based on full detector simulations.
We have performed our analysis for the specific example of the MSSM, based on
benchmark scenarios that were defined in
Refs.~\cite{Bahl:2018zmf,Bahl:2019ago} and a new benchmark scenario, called
the $M_h^{125,\mu-}$ scenario, which we have defined in the present paper.
The $M_h^{125,\mu-}$ scenario is characterized by a relatively large negative
value of the parameter $\mu$, namely $\mu = -2\,\TeV$, leading to an
enhancement of the Higgs-boson couplings to bottom quarks via SUSY loop
contributions. This new benchmark scenario should be well suited in
particular for the presentation of experimental results for the search for
heavy Higgs bosons in the $H / A \to b \bar b$ decay channel.

Within the $M_h^{125}$ benchmark scenario,
in which all supersymmetric particles have masses above the TeV scale such
that the model at the electroweak scale resembles a THDM with MSSM relations
in the Higgs sector, we find that the rate measurements
at the HL-LHC could set a lower limit of about $M_A \gtrsim 900\,\GeV$ in the
absence of any deviation of the measured results
from the SM predictions. In this scenario the
indirect sensitivity from the rate measurements at the HL-LHC
is not sufficient to access
allowed parts of the parameter space that would not be covered by the direct
searches in the $\tau^+\tau^-$ channel.
This is in contrast to the $M_{h, {\rm EFT}}^{125}$~scenario, where the
region that is compatible with a prediction of about $125\,\GeV$ for the mass
of the SM-like Higgs boson extends to lower values of $\tb$
around $\tb \ge 1$ through a very
heavy sfermion spectrum. At such low values of $\tb$ the indirect HL-LHC
sensitivity via the rate measurements of the state at $125\,\GeV$ surpasses
the reach of the standard search for heavy Higgs bosons
in the di-tau final state. We have emphasized in this context that the
search channel $H \to hh$ has the potential to cover parts of the parameter
space in the low-$\tb$ region, and it is also of interest to exploit
${H/A\to t\bar t}$ and ${A\to Zh}$ searches.

In addition to the $M_h^{125}$ and $M_{h, {\rm EFT}}^{125}$ scenarios, where
all SUSY particles are relatively heavy, we have also considered the specific
case of very light electroweakinos, as defined in the
$M_h^{125}(\tilde\chi)$ and $M_{h, {\rm EFT}}^{125}(\tilde\chi)$ scenarios.
These scenarios would constitute a particularly favorable case for the
HL-LHC, since the light charginos would give rise to large loop contributions
to the $h\to \gamma\gamma$ rate (and could also be probed by direct searches for
electroweakinos). In the context of a scenario of this kind,
the Higgs rate measurements
at the HL-LHC (assuming that they agree with the SM prediction) could be used to
rule out a certain part of the plane of the two parameters $M_2$ and $\mu$,
which together with $\tb$ fix the chargino sector at tree-level.
For instance, at $\tb = 2.5$ the HL-LHC could indirectly probe
mass values of up to $\sim 255\,\GeV$ at the $2\sigma$ level.
We have emphasized in the context of this example of a scenario where some of
the SUSY particles are relatively light that the impact of direct searches
for heavy Higgs bosons at the LHC can be further strengthened by
supplementing the searches in the $\tau^+\tau^-$ and
$b \bar b$ final states with dedicated
searches for the decays of $H$ and $A$ to BSM particles, for instance
charginos, neutralinos and sleptons.

\smallskip

We then extended our investigations to the situation where the results from
the HL-LHC are combined with prospective high-precision measurements of the
Higgs signal rates at a future $e^+e^-$
machine. Specifically, we considered the combination of the HL-LHC results
with the results that could be achieved at the first ILC stage
at $250~\GeV$ and $2~\mathrm{ab}^{-1}$ of data, and furthermore also the
additional combination with the results achievable at
the ILC at $500~\GeV$ with $4~\mathrm{ab}^{-1}$.
In this context we did not take into
account the capabilities of the ILC for detecting new light states, and we
also did not attempt a combination with the prospective results from direct
searches at the HL-LHC, but only commented on the possible interplay between the
indirect information from the rate measurements and the results of the direct
searches.

We first discussed the case where the future rate measurements at the HL-LHC
and the ILC exactly agree with the SM predictions. This corresponds to an
MSSM scenario that is far in the decoupling limit, where the sensitivity to
variations in $M_A$ is small. Accordingly, the additional
measurements from  ILC250 and from ILC500 strengthen the indirect
lower bound on $M_A$ from the HL-LHC measurements
by only a moderate amount of about $+100~\GeV$ each in
this scenario. It should be noted that in a realistic future
analysis not all
measurements would uniformly yield a push into the decoupling region, but even
in case of the absence of any source of BSM physics statistical fluctuations
could mimick hints for new physics. Both the gain in accuracy and the much larger
set of high-precision observables that would be available from the
combination of HL-LHC and ILC measurements would be instrumental to correctly
interpret the observed pattern of the measurements.
The analysis of scenarios with a relatively large negative value of $\mu$
revealed the interesting feature that the Higgs rate measurements in such a
case have the potential to set an \emph{upper bound} on $\tb$ and / or the
absolute value of $\mu$.

We furthermore investigated the question to what extent measurements at the
HL-LHC and the ILC can establish significant deviations in the
properties of the Higgs boson at $125\,\GeV$ from the SM predictions via
the Higgs rate measurements
if a certain parameter point of the MSSM is
realized in nature. In this context we also addressed the issues of
how well the parameters $M_A$ and \tb\ can be indirectly constrained and
to what extent the SM can be discriminated from the MSSM. This information,
obtained on the basis
of just the rate measurements of the state at $125\,\GeV$, would be
important in order to complement it with
the (possibly inconclusive) information from direct
searches for new physics. For this analysis we assumed that a particular
parameter point of the ($M_A$, $\tb$) plane of the considered benchmark
scenario is realized and we used the corresponding predictions as
hypothetical results of the rate measurements performed at the HL-LHC and the
ILC. Specifically we considered the $M_A$ values of $700\,\GeV$, $1\,\TeV$
and $1.75\,\TeV$ with different settings for the accompanying $\tb$ value.

For the optimistic case where a value of $M_A = 700\,\GeV$ was realized
nature, the accuracy of the rate measurememts at the HL-LHC would allow one
to obtain an indirect \emph{upper} bound on $M_A$ of about $1\,\TeV$ at the
$2\sigma$ level, which would further be strengthened with the precision
measurements at the ILC.
In contrast, for the case where the true value would be $M_A = 1\,\TeV$
the prospective accuracy of the signal strength measurements
at the HL-LHC would not be sufficient to place
such an indirect upper bound on $M_A$. This would be of relevance not only
for discriminating between the SM and effects of new physics but also
set a
target for the direct searches for additional heavy Higgs bosons.
The incorporation of the precision measurements at the ILC would not only
have a strong impact regarding the distinction between the SM and
the MSSM. Moreover,
the much larger set of high-precision observables available from the
combination of the HL-LHC with the ILC in comparison to the case of just the
HL-LHC would also be crucial for disentangling the underlying nature of
observed deviations from the SM. For the benchmark scenarios with light
charginos the induced large loop contributions to the
$h\to \gamma\gamma$ rate would make it possible, in addition to the
constraints on $M_A$, to set tight indirect
constraints on $\tb$. This would be possible via the rate
measurements at the HL-LHC if a low value of $\tb$ was
realized in nature (we considered the example of $\tb = 3$).
For the case of $1.75\,\TeV$ we considered the large values of $\tb=50$ and
$\tb = 25$, where for the former value this parameter region will be
accessible via the $H / A \to \tau^+\tau^-$ searches at the LHC in the near
future. As expected for such a scenario that is quite far
in the decoupling region, the deviations in the Higgs rates from the SM
predictions are very small. A significant discrepancy from the SM
could only be established with the ultimate precision of ILC500
for the $M_h^{125,\mu-}$ scenario.

Besides the analysis of indirect constraints in the $(\mA, \tb)$ plane
we also displayed the sensitivities in the different scenarios via
\emph{``W\"ascheleinen-plots''} (washing line plots) showing
the predicted light Higgs couplings (normalized to the SM prediction) at
the assumed MSSM points in the $\kappa$ framework in comparison
with the anticipated precision of the
$\kappa$ determination from the prospective Higgs rate
measurements at the HL-LHC and the ILC.
We summarized the capabilities of the HL-LHC and the two ILC realizations to
discriminate between the SM and the MSSM
by providing the goodnees-of-fit of the SM assuming the MSSM
realizations discussed in this paper.

It should be noted that the analyses in this paper have been carried out
within specific MSSM benchmark scenarios, where besides $M_A$ and $\tb$ all other
SUSY parameters are fixed to specific values by definition.
This setting implies large correlations between the different Higgs rates,
so that precise measurements of just a few observables already have a large
impact on constraining the parameter space. As a consequence, these analyses
cannot demonstrate the full potential of the entire set of high-precision
observables that will be available by combining the information from the
HL-LHC and the ILC. Instead, more realistically
if no assumption is made on the underlying structure
of the physics scenario that is realized in nature, the
full breadth of precision measurements at the HL-LHC and the ILC will be
crucial in order to either determine the nature of observed patterns of
deviations from the SM or to set constraints on wide classes of possible
extensions or alternatives to the SM.
For the scenarios considered in this paper this would mean that the
correlations arising from assuming the MSSM Higgs sector with the considered
parameter settings could actually be tested, and results for the Higgs
couplings could be obtained in a model-independent way.


\subsection*{Acknowledgements}

We thank Michael Peskin, J\"urgen Reuter and Jonas Wittbrodt for interesting
discussions, and Maria Cepeda, Martin Flechl, Andrew Gilbert, Marumi Kado and
Lei Zhang for helpful discussions on the HL-LHC projections. H.B., T.S.~and
G.W.\ acknowledge support
by the Deutsche Forschungsgemeinschaft (DFG, German Research
Foundation) under Germany‘s Excellence Strategy -- EXC 2121 ``Quantum
Universe'' -- 390833306.
The work of S.H.~is supported in part by the MEINCOP Spain under
contract FPA2016-78022-P, in part by the Spanish ``Agencia Estatal de
Investigaci\'on'' (AEI) and the EU ``Fondo Europeo de Desarrollo
Regional'' (FEDER) through the project FPA2016-78022-P, in part by the
``Spanish Red Consolider MultiDark'' FPA2017-90566-REDC, and in part by
the AEI through the grant IFT Centro de Excelencia Severo Ochoa
SEV-2016-0597.

{\footnotesize
\bibliographystyle{utphys}
\bibliography{MSSM_benchmarks_HL-LHC}

\providecommand{\href}[2]{#2}\begingroup\raggedright\begin{thebibliography}{100}

\bibitem{Aad:2012tfa}
{\bf ATLAS} Collaboration, G.~Aad {\em et al.}, {\em {Observation of a new
  particle in the search for the Standard Model Higgs boson with the ATLAS
  detector at the LHC}}.
  \href{http://dx.doi.org/10.1016/j.physletb.2012.08.020}{Phys. Lett. {\bf
  B716} (2012)  1--29},
\href{http://arxiv.org/abs/1207.7214}{{\tt arXiv:1207.7214 [hep-ex]}}.

\bibitem{Chatrchyan:2012xdj}
{\bf CMS} Collaboration, S.~Chatrchyan {\em et al.}, {\em {Observation of a New
  Boson at a Mass of 125 GeV with the CMS Experiment at the LHC}}.
  \href{http://dx.doi.org/10.1016/j.physletb.2012.08.021}{Phys. Lett. {\bf
  B716} (2012)  30--61},
\href{http://arxiv.org/abs/1207.7235}{{\tt arXiv:1207.7235 [hep-ex]}}.

\bibitem{Khachatryan:2016vau}
{\bf ATLAS, CMS} Collaboration, G.~Aad {\em et al.}, {\em {Measurements of the
  Higgs boson production and decay rates and constraints on its couplings from
  a combined ATLAS and CMS analysis of the LHC pp collision data at $
  \sqrt{s}=7 $ and 8 TeV}}.
  \href{http://dx.doi.org/10.1007/JHEP08(2016)045}{JHEP {\bf 08} (2016)  045},
\href{http://arxiv.org/abs/1606.02266}{{\tt arXiv:1606.02266 [hep-ex]}}.

\bibitem{ATLAS:2019slw}
{\bf ATLAS} Collaboration, T.~A. collaboration, {\em {Combined measurements of
  Higgs boson production and decay using up to $80~\mathrm{fb}^{-1}$ of
  proton--proton collision data at $\sqrt{s}=$ 13 TeV collected with the ATLAS
  experiment}}.
\href{http://arxiv.org/abs/ATLAS-CONF-2019-005}{{\tt ATLAS-CONF-2019-005}}.

\bibitem{Sirunyan:2018sgc}
{\bf CMS} Collaboration, A.~M. Sirunyan {\em et al.}, {\em {Measurement and
  interpretation of differential cross sections for Higgs boson production at
  $\sqrt{s} =$ 13 TeV}}.
  \href{http://dx.doi.org/10.1016/j.physletb.2019.03.059}{Phys. Lett. {\bf
  B792} (2019)  369--396},
\href{http://arxiv.org/abs/1812.06504}{{\tt arXiv:1812.06504 [hep-ex]}}.

\bibitem{Sirunyan:2018koj}
{\bf CMS} Collaboration, A.~M. Sirunyan {\em et al.}, {\em {Combined
  measurements of Higgs boson couplings in proton–proton collisions at
  $\sqrt{s}=13\,\text {Te}\text {V} $}}.
  \href{http://dx.doi.org/10.1140/epjc/s10052-019-6909-y}{Eur. Phys. J. {\bf
  C79} (2019) no.~5, 421},
\href{http://arxiv.org/abs/1809.10733}{{\tt arXiv:1809.10733 [hep-ex]}}.

\bibitem{Nilles:1983ge}
H.~P. Nilles, {\em {Supersymmetry, Supergravity and Particle Physics}}.
\href{http://dx.doi.org/10.1016/0370-1573(84)90008-5}{Phys. Rept. {\bf 110}
  (1984)  1--162}.

\bibitem{Haber:1984rc}
H.~E. Haber and G.~L. Kane, {\em {The Search for Supersymmetry: Probing Physics
  Beyond the Standard Model}}.
\href{http://dx.doi.org/10.1016/0370-1573(85)90051-1}{Phys. Rept. {\bf 117}
  (1985)  75--263}.

\bibitem{Gunion:1984yn}
J.~F. Gunion and H.~E. Haber, {\em {Higgs Bosons in Supersymmetric Models. 1.}}
  \href{http://dx.doi.org/10.1016/0550-3213(86)90340-8,
  10.1016/0550-3213(93)90653-7}{Nucl. Phys. {\bf B272} (1986)  1}.
[Erratum: Nucl. Phys.B402,567(1993)].

\bibitem{Bahl:2018zmf}
E.~Bagnaschi {\em et al.}, {\em {MSSM Higgs Boson Searches at the LHC:
  Benchmark Scenarios for Run 2 and Beyond}}.
  \href{http://dx.doi.org/10.1140/epjc/s10052-019-7114-8}{Eur. Phys. J. {\bf
  C79} (2019) no.~7, 617},
\href{http://arxiv.org/abs/1808.07542}{{\tt arXiv:1808.07542 [hep-ph]}}.

\bibitem{Bahl:2019ago}
H.~Bahl, S.~Liebler, and T.~Stefaniak, {\em {MSSM Higgs benchmark scenarios for
  Run 2 and beyond: the low $\tan \beta $ region}}.
  \href{http://dx.doi.org/10.1140/epjc/s10052-019-6770-z}{Eur. Phys. J. {\bf
  C79} (2019) no.~3, 279},
\href{http://arxiv.org/abs/1901.05933}{{\tt arXiv:1901.05933 [hep-ph]}}.

\bibitem{Buchmueller:2013rsa}
O.~Buchmueller {\em et al.}, {\em {The CMSSM and NUHM1 after LHC Run 1}}.
  \href{http://dx.doi.org/10.1140/epjc/s10052-014-2922-3}{Eur. Phys. J. {\bf
  C74} (2014) no.~6, 2922},
\href{http://arxiv.org/abs/1312.5250}{{\tt arXiv:1312.5250 [hep-ph]}}.

\bibitem{Heinemeyer:2019vbc}
S.~Heinemeyer, M.~Mondragón, N.~Tracas, and G.~Zoupanos, {\em {Reduction of
  Couplings and its application in Particle Physics}}.
  \href{http://dx.doi.org/10.1016/j.physrep.2019.04.002}{Phys. Rept. {\bf 814}
  (2019)  1--43},
\href{http://arxiv.org/abs/1904.00410}{{\tt arXiv:1904.00410 [hep-ph]}}.

\bibitem{Bagnaschi:2017tru}
E.~Bagnaschi {\em et al.}, {\em {Likelihood Analysis of the pMSSM11 in Light of
  LHC 13-TeV Data}}.
  \href{http://dx.doi.org/10.1140/epjc/s10052-018-5697-0}{Eur. Phys. J. {\bf
  C78} (2018) no.~3, 256},
\href{http://arxiv.org/abs/1710.11091}{{\tt arXiv:1710.11091 [hep-ph]}}.

\bibitem{Cepeda:2019klc}
M.~Cepeda {\em et al.}, {\em {Report from Working Group 2}}.
  \href{http://dx.doi.org/10.23731/CYRM-2019-007.221}{CERN Yellow Rep. Monogr.
  {\bf 7} (2019)  221--584},
\href{http://arxiv.org/abs/1902.00134}{{\tt arXiv:1902.00134 [hep-ph]}}.

\bibitem{Gunion:2002zf}
J.~F. Gunion and H.~E. Haber, {\em {The CP conserving two Higgs doublet model:
  The Approach to the decoupling limit}}.
  \href{http://dx.doi.org/10.1103/PhysRevD.67.075019}{Phys. Rev. {\bf D67}
  (2003)  075019},
\href{http://arxiv.org/abs/hep-ph/0207010}{{\tt arXiv:hep-ph/0207010
  [hep-ph]}}.

\bibitem{Carena:2013ooa}
M.~Carena, I.~Low, N.~R. Shah, and C.~E.~M. Wagner, {\em {Impersonating the
  Standard Model Higgs Boson: Alignment without Decoupling}}.
  \href{http://dx.doi.org/10.1007/JHEP04(2014)015}{JHEP {\bf 04} (2014)  015},
\href{http://arxiv.org/abs/1310.2248}{{\tt arXiv:1310.2248 [hep-ph]}}.

\bibitem{Carena:2014nza}
M.~Carena, H.~E. Haber, I.~Low, N.~R. Shah, and C.~E.~M. Wagner, {\em
  {Complementarity between Nonstandard Higgs Boson Searches and Precision Higgs
  Boson Measurements in the MSSM}}.
  \href{http://dx.doi.org/10.1103/PhysRevD.91.035003}{Phys. Rev. {\bf D91}
  (2015) no.~3, 035003},
\href{http://arxiv.org/abs/1410.4969}{{\tt arXiv:1410.4969 [hep-ph]}}.

\bibitem{Bechtle:2016kui}
P.~Bechtle, H.~E. Haber, S.~Heinemeyer, O.~Stål, T.~Stefaniak, G.~Weiglein,
  and L.~Zeune, {\em {The Light and Heavy Higgs Interpretation of the MSSM}}.
  \href{http://dx.doi.org/10.1140/epjc/s10052-016-4584-9}{Eur. Phys. J. {\bf
  C77} (2017) no.~2, 67},
\href{http://arxiv.org/abs/1608.00638}{{\tt arXiv:1608.00638 [hep-ph]}}.

\bibitem{Haber:2017erd}
H.~E. Haber, S.~Heinemeyer, and T.~Stefaniak, {\em {The Impact of Two-Loop
  Effects on the Scenario of MSSM Higgs Alignment without Decoupling}}.
  \href{http://dx.doi.org/10.1140/epjc/s10052-017-5243-5}{Eur. Phys. J. {\bf
  C77} (2017) no.~11, 742},
\href{http://arxiv.org/abs/1708.04416}{{\tt arXiv:1708.04416 [hep-ph]}}.

\bibitem{Ellwanger:2009dp}
U.~Ellwanger, C.~Hugonie, and A.~M. Teixeira, {\em {The Next-to-Minimal
  Supersymmetric Standard Model}}.
  \href{http://dx.doi.org/10.1016/j.physrep.2010.07.001}{Phys. Rept. {\bf 496}
  (2010)  1--77}, \href{http://arxiv.org/abs/0910.1785}{{\tt arXiv:0910.1785
  [hep-ph]}}.

\bibitem{Domingo:2015eea}
F.~Domingo and G.~Weiglein, {\em {NMSSM interpretations of the observed Higgs
  signal}}. \href{http://dx.doi.org/10.1007/JHEP04(2016)095}{JHEP {\bf 04}
  (2016)  095},
\href{http://arxiv.org/abs/1509.07283}{{\tt arXiv:1509.07283 [hep-ph]}}.

\bibitem{Drechsel:2016jdg}
P.~Drechsel, L.~Galeta, S.~Heinemeyer, and G.~Weiglein, {\em {Precise
  Predictions for the Higgs-Boson Masses in the NMSSM}}.
  \href{http://dx.doi.org/10.1140/epjc/s10052-017-4595-1}{Eur. Phys. J. {\bf
  C77} (2017) no.~1, 42},
\href{http://arxiv.org/abs/1601.08100}{{\tt arXiv:1601.08100 [hep-ph]}}.

\bibitem{Domingo:2018uim}
F.~Domingo, S.~Heinemeyer, S.~Paßehr, and G.~Weiglein, {\em {Decays of the
  neutral Higgs bosons into SM fermions and gauge bosons in the
  $\mathcal{CP}$-violating NMSSM}}.
  \href{http://dx.doi.org/10.1140/epjc/s10052-018-6400-1}{Eur. Phys. J. {\bf
  C78} (2018) no.~11, 942},
\href{http://arxiv.org/abs/1807.06322}{{\tt arXiv:1807.06322 [hep-ph]}}.

\bibitem{Biekotter:2019kde}
T.~Biekötter, M.~Chakraborti, and S.~Heinemeyer, {\em {A 96 GeV Higgs boson in
  the N2HDM}}. \href{http://dx.doi.org/10.1140/epjc/s10052-019-7561-2}{Eur.
  Phys. J. {\bf C80} (2020) no.~1, 2},
\href{http://arxiv.org/abs/1903.11661}{{\tt arXiv:1903.11661 [hep-ph]}}.

\bibitem{Robens:2015gla}
T.~Robens and T.~Stefaniak, {\em {Status of the Higgs Singlet Extension of the
  Standard Model after LHC Run 1}}.
  \href{http://dx.doi.org/10.1140/epjc/s10052-015-3323-y}{Eur. Phys. J. C {\bf
  75} (2015)  104}, \href{http://arxiv.org/abs/1501.02234}{{\tt
  arXiv:1501.02234 [hep-ph]}}.

\bibitem{Robens:2016xkb}
T.~Robens and T.~Stefaniak, {\em {LHC Benchmark Scenarios for the Real Higgs
  Singlet Extension of the Standard Model}}.
  \href{http://dx.doi.org/10.1140/epjc/s10052-016-4115-8}{Eur. Phys. J. C {\bf
  76} (2016) no.~5, 268}, \href{http://arxiv.org/abs/1601.07880}{{\tt
  arXiv:1601.07880 [hep-ph]}}.

\bibitem{Robens:2019kga}
T.~Robens, T.~Stefaniak, and J.~Wittbrodt, {\em {Two-real-scalar-singlet
  extension of the SM: LHC phenomenology and benchmark scenarios}}.
  \href{http://dx.doi.org/10.1140/epjc/s10052-020-7655-x}{Eur. Phys. J. C {\bf
  80} (2020) no.~2, 151}, \href{http://arxiv.org/abs/1908.08554}{{\tt
  arXiv:1908.08554 [hep-ph]}}.

\bibitem{Heinemeyer:1998yj}
S.~Heinemeyer, W.~Hollik, and G.~Weiglein, {\em {FeynHiggs: A Program for the
  calculation of the masses of the neutral CP even Higgs bosons in the MSSM}}.
  \href{http://dx.doi.org/10.1016/S0010-4655(99)00364-1}{Comput. Phys. Commun.
  {\bf 124} (2000)  76--89},
\href{http://arxiv.org/abs/hep-ph/9812320}{{\tt arXiv:hep-ph/9812320
  [hep-ph]}}.

\bibitem{Heinemeyer:1998np}
S.~Heinemeyer, W.~Hollik, and G.~Weiglein, {\em {The Masses of the neutral CP -
  even Higgs bosons in the MSSM: Accurate analysis at the two loop level}}.
  \href{http://dx.doi.org/10.1007/s100529900006}{Eur. Phys. J. {\bf C9} (1999)
  343--366},
\href{http://arxiv.org/abs/hep-ph/9812472}{{\tt arXiv:hep-ph/9812472
  [hep-ph]}}.

\bibitem{Degrassi:2002fi}
G.~Degrassi, S.~Heinemeyer, W.~Hollik, P.~Slavich, and G.~Weiglein, {\em
  {Towards high precision predictions for the MSSM Higgs sector}}.
  \href{http://dx.doi.org/10.1140/epjc/s2003-01152-2}{Eur. Phys. J. {\bf C28}
  (2003)  133--143},
\href{http://arxiv.org/abs/hep-ph/0212020}{{\tt arXiv:hep-ph/0212020
  [hep-ph]}}.

\bibitem{Frank:2006yh}
M.~Frank, T.~Hahn, S.~Heinemeyer, W.~Hollik, H.~Rzehak, and G.~Weiglein, {\em
  {The Higgs Boson Masses and Mixings of the Complex MSSM in the
  Feynman-Diagrammatic Approach}}.
  \href{http://dx.doi.org/10.1088/1126-6708/2007/02/047}{JHEP {\bf 02} (2007)
  047},
\href{http://arxiv.org/abs/hep-ph/0611326}{{\tt arXiv:hep-ph/0611326
  [hep-ph]}}.

\bibitem{Hahn:2013ria}
T.~Hahn, S.~Heinemeyer, W.~Hollik, H.~Rzehak, and G.~Weiglein, {\em
  {High-Precision Predictions for the Light CP -Even Higgs Boson Mass of the
  Minimal Supersymmetric Standard Model}}.
  \href{http://dx.doi.org/10.1103/PhysRevLett.112.141801}{Phys. Rev. Lett. {\bf
  112} (2014) no.~14, 141801},
\href{http://arxiv.org/abs/1312.4937}{{\tt arXiv:1312.4937 [hep-ph]}}.

\bibitem{Bahl:2016brp}
H.~Bahl and W.~Hollik, {\em {Precise prediction for the light MSSM Higgs boson
  mass combining effective field theory and fixed-order calculations}}.
  \href{http://dx.doi.org/10.1140/epjc/s10052-016-4354-8}{Eur. Phys. J. {\bf
  C76} (2016) no.~9, 499},
\href{http://arxiv.org/abs/1608.01880}{{\tt arXiv:1608.01880 [hep-ph]}}.

\bibitem{Bahl:2017aev}
H.~Bahl, S.~Heinemeyer, W.~Hollik, and G.~Weiglein, {\em {Reconciling EFT and
  hybrid calculations of the light MSSM Higgs-boson mass}}.
  \href{http://dx.doi.org/10.1140/epjc/s10052-018-5544-3}{Eur. Phys. J. {\bf
  C78} (2018) no.~1, 57},
\href{http://arxiv.org/abs/1706.00346}{{\tt arXiv:1706.00346 [hep-ph]}}.

\bibitem{Bahl:2018jom}
H.~Bahl and W.~Hollik, {\em {Precise prediction of the MSSM Higgs boson masses
  for low M$_{A}$}}. \href{http://dx.doi.org/10.1007/JHEP07(2018)182}{JHEP {\bf
  07} (2018)  182},
\href{http://arxiv.org/abs/1805.00867}{{\tt arXiv:1805.00867 [hep-ph]}}.

\bibitem{Bahl:2018qog}
H.~Bahl, T.~Hahn, S.~Heinemeyer, W.~Hollik, S.~Paßehr, H.~Rzehak, and
  G.~Weiglein, {\em {Precision calculations in the MSSM Higgs-boson sector with
  FeynHiggs 2.14}}.
\href{http://arxiv.org/abs/1811.09073}{{\tt arXiv:1811.09073 [hep-ph]}}.

\bibitem{Bahl:2019hmm}
H.~Bahl, S.~Heinemeyer, W.~Hollik, and G.~Weiglein, {\em {Theoretical
  uncertainties in the MSSM Higgs boson mass calculation}}.
\href{http://arxiv.org/abs/1912.04199}{{\tt arXiv:1912.04199 [hep-ph]}}.

\bibitem{Harlander:2012pb}
R.~V. Harlander, S.~Liebler, and H.~Mantler, {\em {SusHi: A program for the
  calculation of Higgs production in gluon fusion and bottom-quark annihilation
  in the Standard Model and the MSSM}}.
  \href{http://dx.doi.org/10.1016/j.cpc.2013.02.006}{Comput. Phys. Commun. {\bf
  184} (2013)  1605--1617},
\href{http://arxiv.org/abs/1212.3249}{{\tt arXiv:1212.3249 [hep-ph]}}.

\bibitem{Harlander:2016hcx}
R.~V. Harlander, S.~Liebler, and H.~Mantler, {\em {SusHi Bento: Beyond NNLO and
  the heavy-top limit}}.
  \href{http://dx.doi.org/10.1016/j.cpc.2016.10.015}{Comput. Phys. Commun. {\bf
  212} (2017)  239--257},
\href{http://arxiv.org/abs/1605.03190}{{\tt arXiv:1605.03190 [hep-ph]}}.

\bibitem{Harlander:2005rq}
R.~Harlander and P.~Kant, {\em {Higgs production and decay: Analytic results at
  next-to-leading order QCD}}.
  \href{http://dx.doi.org/10.1088/1126-6708/2005/12/015}{JHEP {\bf 12} (2005)
  015},
\href{http://arxiv.org/abs/hep-ph/0509189}{{\tt arXiv:hep-ph/0509189
  [hep-ph]}}.

\bibitem{Harlander:2002wh}
R.~V. Harlander and W.~B. Kilgore, {\em {Next-to-next-to-leading order Higgs
  production at hadron colliders}}.
  \href{http://dx.doi.org/10.1103/PhysRevLett.88.201801}{Phys. Rev. Lett. {\bf
  88} (2002)  201801},
\href{http://arxiv.org/abs/hep-ph/0201206}{{\tt arXiv:hep-ph/0201206
  [hep-ph]}}.

\bibitem{Harlander:2002vv}
R.~V. Harlander and W.~B. Kilgore, {\em {Production of a pseudoscalar Higgs
  boson at hadron colliders at next-to-next-to leading order}}.
  \href{http://dx.doi.org/10.1088/1126-6708/2002/10/017}{JHEP {\bf 10} (2002)
  017},
\href{http://arxiv.org/abs/hep-ph/0208096}{{\tt arXiv:hep-ph/0208096
  [hep-ph]}}.

\bibitem{Anastasiou:2014lda}
C.~Anastasiou, C.~Duhr, F.~Dulat, E.~Furlan, T.~Gehrmann, F.~Herzog, and
  B.~Mistlberger, {\em {Higgs Boson GluonFfusion Production Beyond Threshold in
  N$^{3}LO$ QCD}}. \href{http://dx.doi.org/10.1007/JHEP03(2015)091}{JHEP {\bf
  03} (2015)  091},
\href{http://arxiv.org/abs/1411.3584}{{\tt arXiv:1411.3584 [hep-ph]}}.

\bibitem{Anastasiou:2015yha}
C.~Anastasiou, C.~Duhr, F.~Dulat, E.~Furlan, F.~Herzog, and B.~Mistlberger,
  {\em {Soft expansion of double-real-virtual corrections to Higgs production
  at N$^{3}$LO}}. \href{http://dx.doi.org/10.1007/JHEP08(2015)051}{JHEP {\bf
  08} (2015)  051},
\href{http://arxiv.org/abs/1505.04110}{{\tt arXiv:1505.04110 [hep-ph]}}.

\bibitem{Anastasiou:2016cez}
C.~Anastasiou, C.~Duhr, F.~Dulat, E.~Furlan, T.~Gehrmann, F.~Herzog,
  A.~Lazopoulos, and B.~Mistlberger, {\em {High precision determination of the
  gluon fusion Higgs boson cross-section at the LHC}}.
  \href{http://dx.doi.org/10.1007/JHEP05(2016)058}{JHEP {\bf 05} (2016)  058},
\href{http://arxiv.org/abs/1602.00695}{{\tt arXiv:1602.00695 [hep-ph]}}.

\bibitem{Degrassi:2010eu}
G.~Degrassi and P.~Slavich, {\em {NLO QCD bottom corrections to Higgs boson
  production in the MSSM}}.
  \href{http://dx.doi.org/10.1007/JHEP11(2010)044}{JHEP {\bf 11} (2010)  044},
\href{http://arxiv.org/abs/1007.3465}{{\tt arXiv:1007.3465 [hep-ph]}}.

\bibitem{Degrassi:2011vq}
G.~Degrassi, S.~Di~Vita, and P.~Slavich, {\em {NLO QCD corrections to
  pseudoscalar Higgs production in the MSSM}}.
  \href{http://dx.doi.org/10.1007/JHEP08(2011)128}{JHEP {\bf 08} (2011)  128},
\href{http://arxiv.org/abs/1107.0914}{{\tt arXiv:1107.0914 [hep-ph]}}.

\bibitem{Degrassi:2012vt}
G.~Degrassi, S.~Di~Vita, and P.~Slavich, {\em {On the NLO QCD Corrections to
  the Production of the Heaviest Neutral Higgs Scalar in the MSSM}}.
  \href{http://dx.doi.org/10.1140/epjc/s10052-012-2032-z}{Eur. Phys. J. {\bf
  C72} (2012)  2032},
\href{http://arxiv.org/abs/1204.1016}{{\tt arXiv:1204.1016 [hep-ph]}}.

\bibitem{Actis:2008ug}
S.~Actis, G.~Passarino, C.~Sturm, and S.~Uccirati, {\em {NLO Electroweak
  Corrections to Higgs Boson Production at Hadron Colliders}}.
  \href{http://dx.doi.org/10.1016/j.physletb.2008.10.018}{Phys. Lett. {\bf
  B670} (2008)  12--17},
\href{http://arxiv.org/abs/0809.1301}{{\tt arXiv:0809.1301 [hep-ph]}}.

\bibitem{Spira:1995rr}
M.~Spira, A.~Djouadi, D.~Graudenz, and P.~Zerwas, {\em {Higgs boson production
  at the LHC}}. \href{http://dx.doi.org/10.1016/0550-3213(95)00379-7}{Nucl.
  Phys. B {\bf 453} (1995)  17--82},
  \href{http://arxiv.org/abs/hep-ph/9504378}{{\tt arXiv:hep-ph/9504378}}.

\bibitem{Bonvini:2015pxa}
M.~Bonvini, A.~S. Papanastasiou, and F.~J. Tackmann, {\em {Resummation and
  matching of b-quark mass effects in $ b\overline{b}H $ production}}.
  \href{http://dx.doi.org/10.1007/JHEP11(2015)196}{JHEP {\bf 11} (2015)  196},
\href{http://arxiv.org/abs/1508.03288}{{\tt arXiv:1508.03288 [hep-ph]}}.

\bibitem{Bonvini:2016fgf}
M.~Bonvini, A.~S. Papanastasiou, and F.~J. Tackmann, {\em {Matched predictions
  for the $ b\overline{b}H $ cross section at the 13 TeV LHC}}.
  \href{http://dx.doi.org/10.1007/JHEP10(2016)053}{JHEP {\bf 10} (2016)  053},
\href{http://arxiv.org/abs/1605.01733}{{\tt arXiv:1605.01733 [hep-ph]}}.

\bibitem{Forte:2015hba}
S.~Forte, D.~Napoletano, and M.~Ubiali, {\em {Higgs production in bottom-quark
  fusion in a matched scheme}}.
  \href{http://dx.doi.org/10.1016/j.physletb.2015.10.051}{Phys. Lett. {\bf
  B751} (2015)  331--337},
\href{http://arxiv.org/abs/1508.01529}{{\tt arXiv:1508.01529 [hep-ph]}}.

\bibitem{Forte:2016sja}
S.~Forte, D.~Napoletano, and M.~Ubiali, {\em {Higgs production in bottom-quark
  fusion: matching beyond leading order}}.
  \href{http://dx.doi.org/10.1016/j.physletb.2016.10.040}{Phys. Lett. {\bf
  B763} (2016)  190--196},
\href{http://arxiv.org/abs/1607.00389}{{\tt arXiv:1607.00389 [hep-ph]}}.

\bibitem{Banks:1987iu}
T.~Banks, {\em {Supersymmetry and the Quark Mass Matrix}}.
\href{http://dx.doi.org/10.1016/0550-3213(88)90222-2}{Nucl. Phys. {\bf B303}
  (1988)  172--188}.

\bibitem{Hall:1993gn}
L.~J. Hall, R.~Rattazzi, and U.~Sarid, {\em {The Top quark mass in
  supersymmetric SO(10) unification}}.
  \href{http://dx.doi.org/10.1103/PhysRevD.50.7048}{Phys. Rev. {\bf D50} (1994)
   7048--7065},
\href{http://arxiv.org/abs/hep-ph/9306309}{{\tt arXiv:hep-ph/9306309
  [hep-ph]}}.

\bibitem{Hempfling:1993kv}
R.~Hempfling, {\em {Yukawa coupling unification with supersymmetric threshold
  corrections}}.
\href{http://dx.doi.org/10.1103/PhysRevD.49.6168}{Phys. Rev. {\bf D49} (1994)
  6168--6172}.

\bibitem{Carena:1994bv}
M.~Carena, M.~Olechowski, S.~Pokorski, and C.~E.~M. Wagner, {\em {Electroweak
  symmetry breaking and bottom - top Yukawa unification}}.
  \href{http://dx.doi.org/10.1016/0550-3213(94)90313-1}{Nucl. Phys. {\bf B426}
  (1994)  269--300},
\href{http://arxiv.org/abs/hep-ph/9402253}{{\tt arXiv:hep-ph/9402253
  [hep-ph]}}.

\bibitem{Carena:1999py}
M.~Carena, D.~Garcia, U.~Nierste, and C.~E.~M. Wagner, {\em {Effective
  Lagrangian for the $\bar{t} b H^{+}$ interaction in the MSSM and charged
  Higgs phenomenology}}.
  \href{http://dx.doi.org/10.1016/S0550-3213(00)00146-2}{Nucl. Phys. {\bf B577}
  (2000)  88--120},
\href{http://arxiv.org/abs/hep-ph/9912516}{{\tt arXiv:hep-ph/9912516
  [hep-ph]}}.

\bibitem{Carena:2000uj}
M.~Carena, D.~Garcia, U.~Nierste, and C.~E.~M. Wagner, {\em {$b \to s \gamma$
  and supersymmetry with large $\tan\beta$}}.
  \href{http://dx.doi.org/10.1016/S0370-2693(01)00009-0}{Phys. Lett. {\bf B499}
  (2001)  141--146},
\href{http://arxiv.org/abs/hep-ph/0010003}{{\tt arXiv:hep-ph/0010003
  [hep-ph]}}.

\bibitem{Carena:2013ytb}
M.~Carena, S.~Heinemeyer, O.~Stål, C.~E.~M. Wagner, and G.~Weiglein, {\em
  {MSSM Higgs Boson Searches at the LHC: Benchmark Scenarios after the
  Discovery of a Higgs-like Particle}}.
  \href{http://dx.doi.org/10.1140/epjc/s10052-013-2552-1}{Eur. Phys. J. {\bf
  C73} (2013) no.~9, 2552},
\href{http://arxiv.org/abs/1302.7033}{{\tt arXiv:1302.7033 [hep-ph]}}.

\bibitem{deFlorian:2016spz}
{\bf LHC Higgs Cross Section Working Group} Collaboration, D.~de~Florian {\em
  et al.}, {\em {Handbook of LHC Higgs Cross Sections: 4. Deciphering the
  Nature of the Higgs Sector}}.
\href{http://arxiv.org/abs/1610.07922}{{\tt arXiv:1610.07922 [hep-ph]}}.

\bibitem{Noth:2008tw}
D.~Noth and M.~Spira, {\em {Higgs Boson Couplings to Bottom Quarks: Two-Loop
  Supersymmetry-QCD Corrections}}.
  \href{http://dx.doi.org/10.1103/PhysRevLett.101.181801}{Phys. Rev. Lett. {\bf
  101} (2008)  181801},
\href{http://arxiv.org/abs/0808.0087}{{\tt arXiv:0808.0087 [hep-ph]}}.

\bibitem{Noth:2010jy}
D.~Noth and M.~Spira, {\em {Supersymmetric Higgs Yukawa Couplings to Bottom
  Quarks at next-to-next-to-leading Order}}.
  \href{http://dx.doi.org/10.1007/JHEP06(2011)084}{JHEP {\bf 06} (2011)  084},
\href{http://arxiv.org/abs/1001.1935}{{\tt arXiv:1001.1935 [hep-ph]}}.

\bibitem{Mihaila:2010mp}
L.~Mihaila and C.~Reisser, {\em {$\mathcal{O}(alpha_s^2)$ corrections to
  fermionic Higgs decays in the MSSM}}.
  \href{http://dx.doi.org/10.1007/JHEP08(2010)021}{JHEP {\bf 08} (2010)  021},
  \href{http://arxiv.org/abs/1007.0693}{{\tt arXiv:1007.0693 [hep-ph]}}.

\bibitem{Aad:2019zwb}
{\bf ATLAS} Collaboration, G.~Aad {\em et al.}, {\em {Search for heavy neutral
  Higgs bosons produced in association with $b$-quarks and decaying to
  $b$-quarks at $\sqrt{s}=13$ TeV with the ATLAS detector}}.
\href{http://arxiv.org/abs/1907.02749}{{\tt arXiv:1907.02749 [hep-ex]}}.

\bibitem{Sirunyan:2018zut}
{\bf CMS} Collaboration, A.~M. Sirunyan {\em et al.}, {\em {Search for
  additional neutral MSSM Higgs bosons in the $\tau\tau$ final state in
  proton-proton collisions at $\sqrt{s}=$ 13 TeV}}.
  \href{http://dx.doi.org/10.1007/JHEP09(2018)007}{JHEP {\bf 09} (2018)  007},
\href{http://arxiv.org/abs/1803.06553}{{\tt arXiv:1803.06553 [hep-ex]}}.

\bibitem{Bechtle:2008jh}
P.~Bechtle, O.~Brein, S.~Heinemeyer, G.~Weiglein, and K.~E. Williams, {\em
  {HiggsBounds: Confronting Arbitrary Higgs Sectors with Exclusion Bounds from
  LEP and the Tevatron}}.
  \href{http://dx.doi.org/10.1016/j.cpc.2009.09.003}{Comput. Phys. Commun. {\bf
  181} (2010)  138--167},
\href{http://arxiv.org/abs/0811.4169}{{\tt arXiv:0811.4169 [hep-ph]}}.

\bibitem{Bechtle:2011sb}
P.~Bechtle, O.~Brein, S.~Heinemeyer, G.~Weiglein, and K.~E. Williams, {\em
  {HiggsBounds 2.0.0: Confronting Neutral and Charged Higgs Sector Predictions
  with Exclusion Bounds from LEP and the Tevatron}}.
  \href{http://dx.doi.org/10.1016/j.cpc.2011.07.015}{Comput. Phys. Commun. {\bf
  182} (2011)  2605--2631},
\href{http://arxiv.org/abs/1102.1898}{{\tt arXiv:1102.1898 [hep-ph]}}.

\bibitem{Bechtle:2013gu}
P.~Bechtle, O.~Brein, S.~Heinemeyer, O.~Stal, T.~Stefaniak, G.~Weiglein, and
  K.~Williams, {\em {Recent Developments in HiggsBounds and a Preview of
  HiggsSignals}}. \href{http://dx.doi.org/10.22323/1.156.0024}{PoS {\bf
  CHARGED2012} (2012)  024},
\href{http://arxiv.org/abs/1301.2345}{{\tt arXiv:1301.2345 [hep-ph]}}.

\bibitem{Bechtle:2013wla}
P.~Bechtle, O.~Brein, S.~Heinemeyer, O.~Stål, T.~Stefaniak, G.~Weiglein, and
  K.~E. Williams, {\em {$\mathsf{HiggsBounds}-4$: Improved Tests of Extended
  Higgs Sectors against Exclusion Bounds from LEP, the Tevatron and the LHC}}.
  \href{http://dx.doi.org/10.1140/epjc/s10052-013-2693-2}{Eur. Phys. J. {\bf
  C74} (2014) no.~3, 2693},
\href{http://arxiv.org/abs/1311.0055}{{\tt arXiv:1311.0055 [hep-ph]}}.

\bibitem{Bechtle:2015pma}
P.~Bechtle, S.~Heinemeyer, O.~Stal, T.~Stefaniak, and G.~Weiglein, {\em
  {Applying Exclusion Likelihoods from LHC Searches to Extended Higgs
  Sectors}}. \href{http://dx.doi.org/10.1140/epjc/s10052-015-3650-z}{Eur. Phys.
  J. {\bf C75} (2015) no.~9, 421},
\href{http://arxiv.org/abs/1507.06706}{{\tt arXiv:1507.06706 [hep-ph]}}.

\bibitem{HiggsBounds5}
P.~Bechtle, D.~Dercks, S.~Heinemeyer, T.~Klingl, T.~Stefaniak, G.~Weiglein, and
  J.~Wittbrodt, {\em {HiggsBounds-5: Testing Higgs sectors in the LHC 13 TeV
  era}}. {To be published}.

\bibitem{Bechtle:2013xfa}
P.~Bechtle, S.~Heinemeyer, O.~Stål, T.~Stefaniak, and G.~Weiglein, {\em
  {$HiggsSignals$: Confronting arbitrary Higgs sectors with measurements at the
  Tevatron and the LHC}}.
  \href{http://dx.doi.org/10.1140/epjc/s10052-013-2711-4}{Eur. Phys. J. {\bf
  C74} (2014) no.~2, 2711},
\href{http://arxiv.org/abs/1305.1933}{{\tt arXiv:1305.1933 [hep-ph]}}.

\bibitem{HiggsSignals2}
P.~Bechtle, S.~Heinemeyer, T.~Klingl, T.~Stefaniak, G.~Weiglein, and
  J.~Wittbrodt, {\em {HiggsSignals-2: Probing new physics with precision Higgs
  measurements in the LHC 13 TeV era}}. {To be published}.

\bibitem{Aad:2019mbh}
{\bf ATLAS} Collaboration, G.~Aad {\em et al.}, {\em {Combined measurements of
  Higgs boson production and decay using up to $80~\mathrm{fb}^{-1}$ of
  proton-proton collision data at $\sqrt{s}=$ 13 TeV collected with the ATLAS
  experiment}}.
\href{http://arxiv.org/abs/1909.02845}{{\tt arXiv:1909.02845 [hep-ex]}}.

\bibitem{Sirunyan:2018egh}
{\bf CMS} Collaboration, A.~M. Sirunyan {\em et al.}, {\em {Measurements of
  properties of the Higgs boson decaying to a W boson pair in pp collisions at
  $\sqrt{s}=$ 13 TeV}}.
  \href{http://dx.doi.org/10.1016/j.physletb.2018.12.073}{Phys. Lett. {\bf
  B791} (2019)  96},
\href{http://arxiv.org/abs/1806.05246}{{\tt arXiv:1806.05246 [hep-ex]}}.

\bibitem{CMS:2019chr}
{\bf CMS} Collaboration, C.~Collaboration, {\em {Measurements of properties of
  the Higgs boson in the four-lepton final state in proton-proton collisions at
  $\sqrt{s}=13~\mathrm{TeV}$}}.
\href{http://arxiv.org/abs/CMS-PAS-HIG-19-001}{{\tt CMS-PAS-HIG-19-001}}.

\bibitem{CMS:1900lgv}
{\bf CMS} Collaboration, C.~Collaboration,
{\em {Measurements of Higgs boson production via gluon fusion and vector boson
  fusion in the diphoton decay channel at $\sqrt{s} = 13$ TeV}}.

\bibitem{Sirunyan:2018hbu}
{\bf CMS} Collaboration, A.~M. Sirunyan {\em et al.}, {\em {Search for the
  Higgs boson decaying to two muons in proton-proton collisions at $\sqrt{s} =$
  13 TeV}}. \href{http://dx.doi.org/10.1103/PhysRevLett.122.021801}{Phys. Rev.
  Lett. {\bf 122} (2019) no.~2, 021801},
\href{http://arxiv.org/abs/1807.06325}{{\tt arXiv:1807.06325 [hep-ex]}}.

\bibitem{CMS:2019pyn}
{\bf CMS} Collaboration, C.~Collaboration, {\em {Measurement of Higgs boson
  production and decay to the $\tau\tau$ final state}}.
\href{http://arxiv.org/abs/CMS-PAS-HIG-18-032}{{\tt CMS-PAS-HIG-18-032}}.

\bibitem{Sirunyan:2017elk}
{\bf CMS} Collaboration, A.~M. Sirunyan {\em et al.}, {\em {Evidence for the
  Higgs boson decay to a bottom quark–antiquark pair}}.
  \href{http://dx.doi.org/10.1016/j.physletb.2018.02.050}{Phys. Lett. {\bf
  B780} (2018)  501--532},
\href{http://arxiv.org/abs/1709.07497}{{\tt arXiv:1709.07497 [hep-ex]}}.

\bibitem{Sirunyan:2017dgc}
{\bf CMS} Collaboration, A.~M. Sirunyan {\em et al.}, {\em {Inclusive search
  for a highly boosted Higgs boson decaying to a bottom quark-antiquark pair}}.
  \href{http://dx.doi.org/10.1103/PhysRevLett.120.071802}{Phys. Rev. Lett. {\bf
  120} (2018) no.~7, 071802},
\href{http://arxiv.org/abs/1709.05543}{{\tt arXiv:1709.05543 [hep-ex]}}.

\bibitem{CMS:2019lcn}
{\bf CMS} Collaboration, C.~Collaboration, {\em {Measurement of
  $\mathrm{t\overline{t}H}$ production in the $\mathrm{H\rightarrow
  b\overline{b}}$ decay channel in $41.5\,\mathrm{fb}^{-1}$ of proton-proton
  collision data at $\sqrt{s}=13\,\mathrm{TeV}$}}.
\href{http://arxiv.org/abs/CMS-PAS-HIG-18-030}{{\tt CMS-PAS-HIG-18-030}}.

\bibitem{Sirunyan:2018shy}
{\bf CMS} Collaboration, A.~M. Sirunyan {\em et al.}, {\em {Evidence for
  associated production of a Higgs boson with a top quark pair in final states
  with electrons, muons, and hadronically decaying $\tau$ leptons at $\sqrt{s}
  =$ 13 TeV}}. \href{http://dx.doi.org/10.1007/JHEP08(2018)066}{JHEP {\bf 08}
  (2018)  066},
\href{http://arxiv.org/abs/1803.05485}{{\tt arXiv:1803.05485 [hep-ex]}}.

\bibitem{CMS:2018dmv}
{\bf CMS} Collaboration, C.~Collaboration, {\em {Measurement of the associated
  production of a Higgs boson with a top quark pair in final states with
  electrons, muons and hadronically decaying $\tau$ leptons in data recorded in
  2017 at $\sqrt{s} = 13~\mathrm{TeV}$}}.
\href{http://arxiv.org/abs/CMS-PAS-HIG-18-019}{{\tt CMS-PAS-HIG-18-019}}.

\bibitem{Carena:2005ek}
M.~Carena, S.~Heinemeyer, C.~E.~M. Wagner, and G.~Weiglein, {\em {MSSM Higgs
  boson searches at the Tevatron and the LHC: Impact of different benchmark
  scenarios}}. \href{http://dx.doi.org/10.1140/epjc/s2005-02470-y}{Eur. Phys.
  J. {\bf C45} (2006)  797--814},
\href{http://arxiv.org/abs/hep-ph/0511023}{{\tt arXiv:hep-ph/0511023
  [hep-ph]}}.

\bibitem{CMS:2018oxh}
{\bf CMS} Collaboration, C.~Collaboration, {\em {Projection of the Run 2 MSSM
  $\text{H} \to \tau\tau$ limits for the High-Luminosity LHC}}.
\href{http://arxiv.org/abs/CMS-PAS-FTR-18-017}{{\tt CMS-PAS-FTR-18-017}}.

\bibitem{Adhikary:2018ise}
A.~Adhikary, S.~Banerjee, R.~Kumar~Barman, and B.~Bhattacherjee, {\em {Resonant
  heavy Higgs searches at the HL-LHC}}.
  \href{http://dx.doi.org/10.1007/JHEP09(2019)068}{JHEP {\bf 09} (2019)  068},
\href{http://arxiv.org/abs/1812.05640}{{\tt arXiv:1812.05640 [hep-ph]}}.

\bibitem{Sirunyan:2018two}
{\bf CMS} Collaboration, A.~M. Sirunyan {\em et al.}, {\em {Combination of
  searches for Higgs boson pair production in proton-proton collisions at
  $\sqrt{s} = $ 13 TeV}}.
  \href{http://dx.doi.org/10.1103/PhysRevLett.122.121803}{Phys. Rev. Lett. {\bf
  122} (2019) no.~12, 121803},
\href{http://arxiv.org/abs/1811.09689}{{\tt arXiv:1811.09689 [hep-ex]}}.

\bibitem{Djouadi:2016ack}
A.~Djouadi, J.~Ellis, and J.~Quevillon, {\em {Interference effects in the
  decays of spin-zero resonances into $\gamma \gamma$ and $ t\overline{t} $}}.
  \href{http://dx.doi.org/10.1007/JHEP07(2016)105}{JHEP {\bf 07} (2016)  105},
\href{http://arxiv.org/abs/1605.00542}{{\tt arXiv:1605.00542 [hep-ph]}}.

\bibitem{Carena:2016npr}
M.~Carena and Z.~Liu, {\em {Challenges and opportunities for heavy scalar
  searches in the $ t\overline{t} $ channel at the LHC}}.
  \href{http://dx.doi.org/10.1007/JHEP11(2016)159}{JHEP {\bf 11} (2016)  159},
\href{http://arxiv.org/abs/1608.07282}{{\tt arXiv:1608.07282 [hep-ph]}}.

\bibitem{Hespel:2016qaf}
B.~Hespel, F.~Maltoni, and E.~Vryonidou, {\em {Signal background interference
  effects in heavy scalar production and decay to a top-anti-top pair}}.
  \href{http://dx.doi.org/10.1007/JHEP10(2016)016}{JHEP {\bf 10} (2016)  016},
\href{http://arxiv.org/abs/1606.04149}{{\tt arXiv:1606.04149 [hep-ph]}}.

\bibitem{BuarqueFranzosi:2017jrj}
D.~Buarque~Franzosi, E.~Vryonidou, and C.~Zhang, {\em {Scalar production and
  decay to top quarks including interference effects at NLO in QCD in an EFT
  approach}}. \href{http://dx.doi.org/10.1007/JHEP10(2017)096}{JHEP {\bf 10}
  (2017)  096},
\href{http://arxiv.org/abs/1707.06760}{{\tt arXiv:1707.06760 [hep-ph]}}.

\bibitem{Bernreuther:2015fts}
W.~Bernreuther, P.~Galler, C.~Mellein, Z.~G. Si, and P.~Uwer, {\em {Production
  of heavy Higgs bosons and decay into top quarks at the LHC}}.
  \href{http://dx.doi.org/10.1103/PhysRevD.93.034032}{Phys. Rev. {\bf D93}
  (2016) no.~3, 034032},
\href{http://arxiv.org/abs/1511.05584}{{\tt arXiv:1511.05584 [hep-ph]}}.

\bibitem{Bernreuther:2017yhg}
W.~Bernreuther, P.~Galler, Z.-G. Si, and P.~Uwer, {\em {Production of heavy
  Higgs bosons and decay into top quarks at the LHC. II: Top-quark polarization
  and spin correlation effects}}.
  \href{http://dx.doi.org/10.1103/PhysRevD.95.095012}{Phys. Rev. {\bf D95}
  (2017) no.~9, 095012},
\href{http://arxiv.org/abs/1702.06063}{{\tt arXiv:1702.06063 [hep-ph]}}.

\bibitem{Djouadi:2019cbm}
A.~Djouadi, J.~Ellis, A.~Popov, and J.~Quevillon, {\em {Interference effects in
  $ t\overline{t} $ production at the LHC as a window on new physics}}.
  \href{http://dx.doi.org/10.1007/JHEP03(2019)119}{JHEP {\bf 03} (2019)  119},
\href{http://arxiv.org/abs/1901.03417}{{\tt arXiv:1901.03417 [hep-ph]}}.

\bibitem{Aaboud:2017hnm}
{\bf ATLAS} Collaboration, M.~Aaboud {\em et al.}, {\em {Search for Heavy Higgs
  Bosons $A/H$ Decaying to a Top Quark Pair in $pp$ Collisions at
  $\sqrt{s}=8\text{ }\text{ }\mathrm{TeV}$ with the ATLAS Detector}}.
  \href{http://dx.doi.org/10.1103/PhysRevLett.119.191803}{Phys. Rev. Lett. {\bf
  119} (2017) no.~19, 191803},
\href{http://arxiv.org/abs/1707.06025}{{\tt arXiv:1707.06025 [hep-ex]}}.

\bibitem{CMS:2019lei}
{\bf CMS} Collaboration, C.~Collaboration, {\em {Search for heavy Higgs bosons
  decaying to a top quark pair in proton-proton collisions at $\sqrt{s} =
  13\,\mathrm{TeV}$}}.
\href{http://arxiv.org/abs/CMS-PAS-HIG-17-027}{{\tt CMS-PAS-HIG-17-027}}.

\bibitem{Djouadi:2013uqa}
A.~Djouadi, L.~Maiani, G.~Moreau, A.~Polosa, J.~Quevillon, and V.~Riquer, {\em
  {The post-Higgs MSSM scenario: Habemus MSSM?}}
  \href{http://dx.doi.org/10.1140/epjc/s10052-013-2650-0}{Eur. Phys. J. {\bf
  C73} (2013)  2650},
\href{http://arxiv.org/abs/1307.5205}{{\tt arXiv:1307.5205 [hep-ph]}}.

\bibitem{Djouadi:2013vqa}
A.~Djouadi and J.~Quevillon, {\em {The MSSM Higgs sector at a high $M_{SUSY}$:
  reopening the low tan$\beta$ regime and heavy Higgs searches}}.
  \href{http://dx.doi.org/10.1007/JHEP10(2013)028}{JHEP {\bf 10} (2013)  028},
\href{http://arxiv.org/abs/1304.1787}{{\tt arXiv:1304.1787 [hep-ph]}}.

\bibitem{Maiani:2013hud}
L.~Maiani, A.~D. Polosa, and V.~Riquer, {\em {Bounds to the Higgs Sector Masses
  in Minimal Supersymmetry from LHC Data}}.
  \href{http://dx.doi.org/10.1016/j.physletb.2013.06.026}{Phys. Lett. {\bf
  B724} (2013)  274--277},
\href{http://arxiv.org/abs/1305.2172}{{\tt arXiv:1305.2172 [hep-ph]}}.

\bibitem{Djouadi:2015jea}
A.~Djouadi, L.~Maiani, A.~Polosa, J.~Quevillon, and V.~Riquer, {\em {Fully
  covering the MSSM Higgs sector at the LHC}}.
  \href{http://dx.doi.org/10.1007/JHEP06(2015)168}{JHEP {\bf 06} (2015)  168},
\href{http://arxiv.org/abs/1502.05653}{{\tt arXiv:1502.05653 [hep-ph]}}.

\bibitem{Liebler:2018zul}
S.~Liebler, M.~Mühlleitner, M.~Spira, and M.~Stadelmaier, {\em {The hMSSM
  approach for Higgs self-couplings revisited}}.
  \href{http://dx.doi.org/10.1140/epjc/s10052-019-6594-x}{Eur. Phys. J. {\bf
  C79} (2019) no.~1, 65},
\href{http://arxiv.org/abs/1810.10979}{{\tt arXiv:1810.10979 [hep-ph]}}.

\bibitem{Aboubrahim:2018tpf}
A.~Aboubrahim and P.~Nath, {\em {Naturalness, the hyperbolic branch, and
  prospects for the observation of charged Higgs bosons at high luminosity LHC
  and 27 TeV LHC}}. \href{http://dx.doi.org/10.1103/PhysRevD.98.095024}{Phys.
  Rev. {\bf D98} (2018) no.~9, 095024},
\href{http://arxiv.org/abs/1810.12868}{{\tt arXiv:1810.12868 [hep-ph]}}.

\bibitem{deBlas:2019rxi}
J.~de~Blas {\em et al.}, {\em {Higgs Boson Studies at Future Particle
  Colliders}}.
\href{http://arxiv.org/abs/1905.03764}{{\tt arXiv:1905.03764 [hep-ph]}}.

\bibitem{Freitas:2019bre}
A.~Freitas, S.~Heinemeyer, {\em et al.}, {\em {Theoretical uncertainties for
  electroweak and Higgs-boson precision measurements at FCC-ee}}.
\href{http://arxiv.org/abs/1906.05379}{{\tt arXiv:1906.05379 [hep-ph]}}.

\bibitem{Aaboud:2017sjh}
{\bf ATLAS} Collaboration, M.~Aaboud {\em et al.}, {\em {Search for additional
  heavy neutral Higgs and gauge bosons in the ditau final state produced in
  $36~\mathrm{fb}^{-1}$ of pp collisions at $ \sqrt{s}=13 $ TeV with the ATLAS
  detector}}. \href{http://dx.doi.org/10.1007/JHEP01(2018)055}{JHEP {\bf 01}
  (2018)  055},
\href{http://arxiv.org/abs/1709.07242}{{\tt arXiv:1709.07242 [hep-ex]}}.

\bibitem{Aad:2020zxo}
{\bf ATLAS} Collaboration, G.~Aad {\em et al.}, {\em {Search for heavy Higgs
  bosons decaying into two tau leptons with the ATLAS detector using $pp$
  collisions at $\sqrt{s}=13$ TeV}}.
  \href{http://arxiv.org/abs/2002.12223}{{\tt arXiv:2002.12223 [hep-ex]}}.

\bibitem{ATLAS:2009zmv}
{\bf ATLAS} Collaboration, {\em {ATLAS searches for MSSM Higgs bosons decaying
  into SUSY cascades}}. \href{http://arxiv.org/abs/ATL-PHYS-PUB-2009-079,
  ATL-COM-PHYS-2009-086}{{\tt ATL-PHYS-PUB-2009-079, ATL-COM-PHYS-2009-086}}.

\bibitem{Bisset:2007mi}
M.~Bisset, J.~Li, N.~Kersting, R.~Lu, F.~Moortgat, and S.~Moretti, {\em
  {Four-lepton LHC events from MSSM Higgs boson decays into neutralino and
  chargino pairs}}. \href{http://dx.doi.org/10.1088/1126-6708/2009/08/037}{JHEP
  {\bf 08} (2009)  037}, \href{http://arxiv.org/abs/0709.1029}{{\tt
  arXiv:0709.1029 [hep-ph]}}.

\bibitem{Charlot:2006se}
C.~Charlot, R.~Salerno, and Y.~Sirois, {\em {Observability of the heavy neutral
  SUSY Higgs bosons decaying into neutralinos}}.
  \href{http://dx.doi.org/10.1088/0954-3899/34/1/N01}{J. Phys. G {\bf 34}
  (2007)  N1--N12}.

\bibitem{Arbey:2013jla}
A.~Arbey, M.~Battaglia, and F.~Mahmoudi, {\em {Supersymmetric Heavy Higgs
  Bosons at the LHC}}.
  \href{http://dx.doi.org/10.1103/PhysRevD.88.015007}{Phys. Rev. D {\bf 88}
  (2013) no.~1, 015007}, \href{http://arxiv.org/abs/1303.7450}{{\tt
  arXiv:1303.7450 [hep-ph]}}.

\bibitem{Ball:2007zza}
{\bf CMS} Collaboration, G.~Bayatian {\em et al.}, {\em {CMS technical design
  report, volume II: Physics performance}}.
  \href{http://dx.doi.org/10.1088/0954-3899/34/6/S01}{J. Phys. G {\bf 34}
  (2007) no.~6, 995--1579}.

\bibitem{Craig:2015jba}
N.~Craig, F.~D'Eramo, P.~Draper, S.~Thomas, and H.~Zhang, {\em {The Hunt for
  the Rest of the Higgs Bosons}}.
  \href{http://dx.doi.org/10.1007/JHEP06(2015)137}{JHEP {\bf 06} (2015)  137},
  \href{http://arxiv.org/abs/1504.04630}{{\tt arXiv:1504.04630 [hep-ph]}}.

\bibitem{Belanger:2015vwa}
G.~Belanger, D.~Ghosh, R.~Godbole, and S.~Kulkarni, {\em {Light stop in the
  MSSM after LHC Run 1}}. \href{http://dx.doi.org/10.1007/JHEP09(2015)214}{JHEP
  {\bf 09} (2015)  214}, \href{http://arxiv.org/abs/1506.00665}{{\tt
  arXiv:1506.00665 [hep-ph]}}.

\bibitem{Barman:2016kgt}
R.~K. Barman, B.~Bhattacherjee, A.~Chakraborty, and A.~Choudhury, {\em {Study
  of MSSM heavy Higgs bosons decaying into charginos and neutralinos}}.
  \href{http://dx.doi.org/10.1103/PhysRevD.94.075013}{Phys. Rev. D {\bf 94}
  (2016) no.~7, 075013}, \href{http://arxiv.org/abs/1607.00676}{{\tt
  arXiv:1607.00676 [hep-ph]}}.

\bibitem{Baum:2017gbj}
S.~Baum, K.~Freese, N.~R. Shah, and B.~Shakya, {\em {NMSSM Higgs boson search
  strategies at the LHC and the mono-Higgs signature in particular}}.
  \href{http://dx.doi.org/10.1103/PhysRevD.95.115036}{Phys. Rev. D {\bf 95}
  (2017) no.~11, 115036}, \href{http://arxiv.org/abs/1703.07800}{{\tt
  arXiv:1703.07800 [hep-ph]}}.

\bibitem{Profumo:2017ntc}
S.~Profumo, T.~Stefaniak, and L.~Stephenson~Haskins, {\em {The Not-So-Well
  Tempered Neutralino}}.
  \href{http://dx.doi.org/10.1103/PhysRevD.96.055018}{Phys. Rev. D {\bf 96}
  (2017) no.~5, 055018}, \href{http://arxiv.org/abs/1706.08537}{{\tt
  arXiv:1706.08537 [hep-ph]}}.

\bibitem{Kulkarni:2017xtf}
S.~Kulkarni and L.~Lechner, {\em {Characterizing simplified models for heavy
  Higgs decays to supersymmetric particles}}.
  \href{http://arxiv.org/abs/1711.00056}{{\tt arXiv:1711.00056 [hep-ph]}}.

\bibitem{Arganda:2018hdn}
E.~Arganda, V.~Martin-Lozano, A.~D. Medina, and N.~Mileo, {\em {Potential
  discovery of staus through heavy Higgs boson decays at the LHC}}.
  \href{http://dx.doi.org/10.1007/JHEP09(2018)056}{JHEP {\bf 09} (2018)  056},
  \href{http://arxiv.org/abs/1804.10698}{{\tt arXiv:1804.10698 [hep-ph]}}.

\bibitem{Gori:2018pmk}
S.~Gori, Z.~Liu, and B.~Shakya, {\em {Heavy Higgs as a Portal to the
  Supersymmetric Electroweak Sector}}.
  \href{http://dx.doi.org/10.1007/JHEP04(2019)049}{JHEP {\bf 04} (2019)  049},
\href{http://arxiv.org/abs/1811.11918}{{\tt arXiv:1811.11918 [hep-ph]}}.

\bibitem{Adhikary:2020ujn}
A.~Adhikary, B.~Bhattacherjee, R.~M. Godbole, N.~Khan, and S.~Kulkarni, {\em
  {Searching for heavy Higgs in supersymmetric final states at the LHC}}.
  \href{http://arxiv.org/abs/2002.07137}{{\tt arXiv:2002.07137 [hep-ph]}}.

\bibitem{Liu:2020muv}
J.~Liu, N.~McGinnis, C.~E. Wagner, and X.-P. Wang, {\em {Searching for the
  Higgsino-Bino Sector at the LHC}}.
  \href{http://arxiv.org/abs/2006.07389}{{\tt arXiv:2006.07389 [hep-ph]}}.

\bibitem{Heinemeyer:2014yya}
S.~Heinemeyer and C.~Schappacher, {\em {Heavy Higgs Decays into Sfermions in
  the Complex MSSM: A Full One-Loop Analysis}}.
  \href{http://dx.doi.org/10.1140/epjc/s10052-015-3383-z}{Eur. Phys. J. C {\bf
  75} (2015) no.~5, 198}, \href{http://arxiv.org/abs/1410.2787}{{\tt
  arXiv:1410.2787 [hep-ph]}}.

\bibitem{Heinemeyer:2015pfa}
S.~Heinemeyer and C.~Schappacher, {\em {Higgs Decays into Charginos and
  Neutralinos in the Complex MSSM: A Full One-Loop Analysis}}.
  \href{http://dx.doi.org/10.1140/epjc/s10052-015-3442-5}{Eur. Phys. J. C {\bf
  75} (2015) no.~5, 230}, \href{http://arxiv.org/abs/1503.02996}{{\tt
  arXiv:1503.02996 [hep-ph]}}.

\bibitem{Barman:2017swy}
R.~K. Barman, G.~Belanger, B.~Bhattacherjee, R.~Godbole, G.~Mendiratta, and
  D.~Sengupta, {\em {Invisible decay of the Higgs boson in the context of a
  thermal and nonthermal relic in MSSM}}.
  \href{http://dx.doi.org/10.1103/PhysRevD.95.095018}{Phys. Rev. D {\bf 95}
  (2017) no.~9, 095018}, \href{http://arxiv.org/abs/1703.03838}{{\tt
  arXiv:1703.03838 [hep-ph]}}.

\bibitem{Pozzo:2018anw}
G.~Pozzo and Y.~Zhang, {\em {Constraining resonant dark matter with combined
  LHC electroweakino searches}}.
  \href{http://dx.doi.org/10.1016/j.physletb.2018.12.062}{Phys. Lett. B {\bf
  789} (2019)  582--591}, \href{http://arxiv.org/abs/1807.01476}{{\tt
  arXiv:1807.01476 [hep-ph]}}.

\bibitem{Wang:2020dtb}
K.~Wang and J.~Zhu, {\em {Funnel annihilations of light dark matter and the
  invisible decay of the Higgs boson}}.
  \href{http://dx.doi.org/10.1103/PhysRevD.101.095028}{Phys. Rev. D {\bf 101}
  (2020) no.~9, 095028}, \href{http://arxiv.org/abs/2003.01662}{{\tt
  arXiv:2003.01662 [hep-ph]}}.

\bibitem{Barman:2020vzm}
R.~K. Barman, G.~Bélanger, B.~Bhattacherjee, R.~Godbole, D.~Sengupta, and
  X.~Tata, {\em {Current bounds and future prospects of light neutralino dark
  matter in NMSSM}}. \href{http://arxiv.org/abs/2006.07854}{{\tt
  arXiv:2006.07854 [hep-ph]}}.

\bibitem{CidVidal:2018eel}
X.~Cid~Vidal {\em et al.}, {\em {Report from Working Group 3}}.
  \href{http://dx.doi.org/10.23731/CYRM-2019-007.585}{CERN Yellow Rep. Monogr.
  {\bf 7} (2019)  585--865},
\href{http://arxiv.org/abs/1812.07831}{{\tt arXiv:1812.07831 [hep-ph]}}.

\bibitem{Biekotter:2017xmf}
T.~Biekotter, S.~Heinemeyer, and C.~Munoz, {\em {Precise prediction for the
  Higgs-boson masses in the $\mu \nu $ SSM}}.
  \href{http://dx.doi.org/10.1140/epjc/s10052-018-5978-7}{Eur. Phys. J. C {\bf
  78} (2018) no.~6, 504}, \href{http://arxiv.org/abs/1712.07475}{{\tt
  arXiv:1712.07475 [hep-ph]}}.

\bibitem{Drechsel:2018mgd}
P.~Drechsel, G.~Moortgat-Pick, and G.~Weiglein, ``{Sensitivity of the ILC to
  light Higgs masses},'' in {\em {International Workshop on Future Linear
  Collider (LCWS2017) Strasbourg, France, October 23-27, 2017}}.
\newblock 2018.
\newblock
\href{http://arxiv.org/abs/1801.09662}{{\tt arXiv:1801.09662 [hep-ph]}}.
\newblock

\bibitem{Wang:2018awp}
{\bf International Large Detector Concept Group} Collaboration, Y.~Wang, {\em
  {Search for Light Scalars Produced in Association with a Z boson at the 250
  GeV stage of the ILC}}.
\href{http://dx.doi.org/10.22323/1.340.0630}{PoS {\bf ICHEP2018} (2019)  630}.

\bibitem{Wang:2019mzd}
{\bf International Large Detector concept group} Collaboration, Y.~Wang,
  J.~List, and M.~Berggren, ``{Search for Extra Scalars Produced in Association
  with Muon Pairs at the ILC},'' in {\em {International Workshop on Future
  Linear Colliders (LCWS 2018) Arlington, Texas, USA, October 22-26, 2018}}.
\newblock 2019.
\newblock
\href{http://arxiv.org/abs/1902.06118}{{\tt arXiv:1902.06118 [hep-ex]}}.
\newblock

\bibitem{Biekotter:2019gtq}
T.~Biekotter, S.~Heinemeyer, and C.~Munoz, {\em {Precise prediction for the
  Higgs-Boson masses in the $\mu \nu $ SSM with three right-handed neutrino
  superfields}}. \href{http://dx.doi.org/10.1140/epjc/s10052-019-7175-8}{Eur.
  Phys. J. C {\bf 79} (2019) no.~8, 667},
  \href{http://arxiv.org/abs/1906.06173}{{\tt arXiv:1906.06173 [hep-ph]}}.

\bibitem{Bechtle:2014ewa}
P.~Bechtle, S.~Heinemeyer, O.~Stål, T.~Stefaniak, and G.~Weiglein, {\em
  {Probing the Standard Model with Higgs signal rates from the Tevatron, the
  LHC and a future ILC}}. \href{http://dx.doi.org/10.1007/JHEP11(2014)039}{JHEP
  {\bf 11} (2014)  039},
\href{http://arxiv.org/abs/1403.1582}{{\tt arXiv:1403.1582 [hep-ph]}}.

\bibitem{Desch:2004cu}
K.~Desch, E.~Gross, S.~Heinemeyer, G.~Weiglein, and L.~Zivkovic, {\em {LHC / LC
  interplay in the MSSM Higgs sector}}.
  \href{http://dx.doi.org/10.1088/1126-6708/2004/09/062}{JHEP {\bf 09} (2004)
  062},
\href{http://arxiv.org/abs/hep-ph/0406322}{{\tt arXiv:hep-ph/0406322
  [hep-ph]}}.

\bibitem{Weiglein:2004hn}
{\bf LHC/LC Study Group} Collaboration, G.~Weiglein {\em et al.}, {\em {Physics
  interplay of the LHC and the ILC}}.
  \href{http://dx.doi.org/10.1016/j.physrep.2005.12.003}{Phys. Rept. {\bf 426}
  (2006)  47--358},
\href{http://arxiv.org/abs/hep-ph/0410364}{{\tt arXiv:hep-ph/0410364
  [hep-ph]}}.

\bibitem{Barklow:2017suo}
T.~Barklow, K.~Fujii, S.~Jung, R.~Karl, J.~List, T.~Ogawa, M.~E. Peskin, and
  J.~Tian, {\em {Improved Formalism for Precision Higgs Coupling Fits}}.
  \href{http://dx.doi.org/10.1103/PhysRevD.97.053003}{Phys.\ Rev.\ D {\bf 97}
  (2018) no.~5, 053003}, \href{http://arxiv.org/abs/1708.08912}{{\tt
  arXiv:1708.08912 [hep-ph]}}.

\end{thebibliography}\endgroup
}

\end{document}